# Model-Guided Fieldwork: A Practical, Methodological and Philosophical Investigation in the use of Ethnomethodology for Engineering Software Requirements

**Chris Hinds**

Wadham College, Oxford

A thesis submitted for the degree of DPhil

Term of submission: Michaelmas 2007

Date of submission: 11th January 2008

# Abstract

Ethnomethodological fieldwork has long been acknowledged as a potentially valuable way of informing the design of technology. However, there is relatively little methodological support for this activity, particularly in relation to the systematic approaches to development advocated in mainstream software and requirements engineering. This thesis focuses on the use of ethnomethodological fieldwork for the engineering of software requirements. Firstly, it proposes an approach, dubbed "Model Guided Fieldwork," to support a fieldworker in making observations that may contribute to a technological development process. It does this by supplementing the normal debriefing sessions that a fieldworker and a technologist might have, with a lightweight iterative system modelling exercise, in such a way that the fieldwork and modelling can be mutually guiding. Secondly, the thesis presents an application of this approach in a high-profile e-Science project. This case study provides an opportunity to examine the relationship between ethnomethodological ethnography and requirements engineering empirically. Thirdly, the thesis addresses a number of theoretical and philosophical concerns relating to its project. This consists in a number of clarifications and counterarguments which aim to better situate ethnomethodological fieldwork as a method of requirements elicitation. In these three regards the thesis constitutes a practical methodological and philosophical investigation into the topic at hand.

# Contents

# 1. Introduction

This thesis is concerned with the use of ethnomethodological fieldwork in the development of software requirements, a topic which it will tackle practically, methodologically, and philosophically. By way of introduction this chapter will approach the subject in four ways. Firstly, it will outline a summary of the current state-of-the-art in relation to ethnomethodological fieldwork and technological development; secondly, it will present what this thesis proposes to achieve; thirdly, it will propose a motivation for doing this; and finally, it will introduce how this work will be achieved.

Fieldwork is a form of investigation conducted within the natural environment of those being studied, away from the workplace of those actually doing the studying. Ethnomethodology is an orientation under which fieldwork can be conducted. It is a sensibility, or way of understanding the nature of, and ways in which, people create social order within the world. It acts as a focus for the fieldworker by helping to inform the way that they "see" the situations they encounter. One of the critical features of ethnomethodology is that it allows a fieldwork to come to an intrinsic understanding of a domain that approaches the way in which the people within that domain itself see it. This is in stark contrast to many methods which seek to impose extrinsic analytic structures on a domain, imported from the discipline of those who might be conducting the study.

There have been many case studies which illustrate the value that can be gained by conducting fieldwork studies that are designed to inform the development of technology. However, it is acknowledged (see for example [1], p15, p306) that in over a decade, there has been little methodological progress regarding how such studies should be conducted. Typically the role of fieldwork is described as being to "inform design" [1], and following [2], this is generally done through an iterative process where fieldworkers engage in debriefing sessions with technologists. However, such sessions are far from straightforward. As [1] suggests: one must identify precisely the observations that are relevant to provide; one must do this at precisely the time that they are needed; and moreover, one must make their relevance to the process clear. The state-of-the-art provides relatively little methodological support for this kind of analytic work.

This thesis will contribute to the above situation in three ways. The first will be to propose an extension of the iterative debriefing style of working proposed in [2]. It will be designed to support the analytic activity of identifying and developing relevancies between fieldwork observations and ideas about the emerging system, in order to derive requirements for that system. The method will depend on supplementing the iterative debriefing format with a system modelling exercise. These lightweight disposable models will act to focus the fieldwork onto issues of relevance, and conversely, that fieldwork will help to evolve the models. From this focused activity it will be possible to distil system requirements. The second contribution will be to study the application of this proposed method in a real-world case study, details of which will be given below. The third contribution will be to examine the foundations of these activities, and the foundational consequences of their combination. Ethnomethodological fieldwork and requirements engineering are often thought of as very different activities, and the literature has identified a number of puzzles and problems relating to their combination, which this thesis will seek to clarify.

From the above it is possible to conclude that *if* there is a need for fieldwork within software engineering, then there will also be a need for approaches to support the analysis of that fieldwork. It is therefore pertinent to consider whether software engineering ought in the first place to take fieldwork-based approaches seriously, particularly given the apparent difficulties that their application bring. This is an important question and one which chapter 7 will return to in more detail. However, for the present it will be sufficient to motivate an interest in fieldwork by using a case study.

The London Ambulance Service (LAS) Computer Aided Dispatch system is a well known case study, not just because of the consequential nature of its failure but also because many details of the project are publicly available [3]. In what follows some details of this case study will be briefly presented.

Prior to 1992 the dispatch system for the LAS functioned in the following way:

1. The Caller rings "999" and speaks to a Call Taker

2. The Call Taker takes their details and fills out a paper form; a map book is used to verify the caller's location.

3. The paper form is put on a conveyer belt.

4. It is picked up by a member of the Control Staff, who passes it to the Resource Allocator for the appropriate region.

5. The Resource Allocator then despatches an available Ambulance to the incident using a radio.

The number of staff involved in this process led to a perception that the process was slow, error prone, and costly. The proposed replacement LAS despatch system streamlined the process as follows:

1. The Caller rings "999" and speaks to a Call Taker

2. The Call Taker takes their details and enters them onto a Computer System

3. The Computer System, which can track Ambulances using GPS, then electronically despatches the nearest available Ambulance to the incident.

The new system was deployed in a "big bang," without a fall-back system, and proved to be disastrous. It was in full operation for less than a day, but during this time patients suffered significant delays, in some cases waiting 11 hours for an ambulance to arrive; media reports suggested that up to 30 deaths may be attributable to its failings.

There were many problems with the system. Because of the nearest-first ambulance allocation algorithm, ambulance drivers, who had traditionally worked within one particular region, tended to drift further and further away from this area. This was consequential, they were intimately familiar with their own region, and how the traffic conditions would change within it over the day (at rush hour, or when the schools finish). As they drifted away from this familiar region they found it harder to reach incidents quickly. Similarly, nearest-first allocation meant that ambulances would drift toward "hot-spots" where the most incidents were occurring. This left other areas of the region uncovered, and if incidents did occur there, they could not be reached quickly. The GPS coverage was also incomplete, and when an ambulance went out of GPS range, the system would allocate a new ambulance to the incident. As the response time slowed, patients would begin to call back in and re-request ambulances. As a result the system slowly ground to a halt, as all trust in the information it was producing evaporated.

The LAS despatch system was an unmitigated disaster for many reasons. However, for present purposes one lesson stands out. *The developers failed to understand the real work that was involved in allocating an ambulance*. The Regional Resource Allocators take great care in allocating incidents to ambulances. They treat it almost like a game of chess, thinking one move ahead, so that they'd have ambulances positioned across their region to deal with the next incident, no matter where it might occur. The Ambulance Drivers are similarly knowledgeable about their particular region, and how to get around it quickly. If only the developers had properly understood the *real work* involved in being a Resource Allocator, or an Ambulance Driver, they would almost certainly have designed the system differently.

An interest in the mundane working practices of a domain is often reflected in contemporary guidance to software engineers. For example, the British Computer Society's Good Practice Guide [4] recommends that requirements engineers (sec 5.1):

> Strive to understand the organization's business and search for changes that will bring tangible benefits.

> Be aware of the impact of new or changed business solutions on people's working lives and deal sensitively with them.

> When analysing current practices, show respect for people at all levels in the organization.

However, this may not be a simple goal to achieve as the domain of interest may be partly or even wholly unfamiliar to the development team. Fieldwork studies provide a practical way to gain this kind of familiarity. The exact way in which one chooses to understand that domain is a matter of theoretical orientation. Ethnomethodology provides one orientation which is argued powerfully for as a perspective for gaining an adequate understanding of social order [5]. Moreover it has also been routinely drawn on as an orientation for conducting fieldwork for technological development [1].

The foregoing provides at least some motivation for an interest in fieldwork, and specifically in ethnomethodological fieldwork. It is an interest that is shared by a substantial corpus of case studies within the field of Computer Supported Cooperative Work. What is distinct about this thesis it that it chooses to study fieldwork, not in relation to design, but for the engineering of requirements. Of course, there is some literature within the academic field of requirements engineering advocating fieldwork; [6] provides a methodological review, [7] discusses many of the issues, [8] even suggests that it is an approach favoured by expert practitioners. However, others are more sceptical [1] (p156):

> We are doubtful as to whether it is sensible to simply regard ethnography as a method that can be unproblematically incorporated into the requirements capture process. It is more a matter that is best described as a method which can *inform* design by identifying problems and concerns which a system has to accommodate if it is to effectively support work activities.

For this reason it will be important to make clear why this thesis has chosen to study requirements rather than design.

Firstly, design is a somewhat flexible term, it can refer to anything from the formation of concepts, to the development of detailed internal software architecture. It can also be manifest in different ways, from something that is merely in the mind of the designer, to formal diagrams. The term "informing design" consequently inherits this flexibility. This flexibility makes it potentially harder to tie down a methodological relationship between fieldwork and technology. Requirements engineering on the other hand is by definition concerned with expressing precise statements. This concrete deliverable provides a fixed point around which to think about fieldwork's contribution. Secondly, the fact that requirements engineering produces a clear, precise, documented statement provides a resource that may make it easier to study any relationship with fieldwork using empirical methods. However, third and perhaps most importantly, engaging with requirements engineering represents an attempt to take seriously the needs of software engineering itself. By focusing on design, or rather on the wrong aspect of a malleable notion of design, it would be possible to neglect the very real needs of software engineering, which is itself a fraught activity with its own professional commitments. By focusing on requirements engineering, which is itself a sub-discipline of software engineering, this issue can be addressed.

Another strand of the thesis relates to the philosophical and foundational aspects of the proposed programme of research. The notion of using fieldwork as a technique for engineering technology appears to raise a number of puzzles. In the most simplistic sense, engineering and ethnomethodology come from very different traditions. The former is often thought of as the application of scientific principles to practical problems, whereas the latter appears firmly to resist scientific theorising. As a result it would be easy to imagine that their combination could be far from complementary.

This tension is, in philosophical terms, expressed by [9]. It points out that, if this philosophical divide is taken seriously, then it may actually render the use of ethnomethodological fieldwork impossible. Showing that this divide is illusory will therefore be an important task for the thesis.

Meanwhile, there are also puzzles at a methodological and practical level. On the one hand there are a great many case studies that, on the basis of hands-on experience, extol the virtues of doing fieldwork to inform the development of technology. Yet, on the other there is a dearth of methodological advice for conducting such studies. It is a situation that [10] (p251) reflects on.

> Though workplace studies may be useful for highlighting systematic, robust features of work practices, they do not and perhaps cannot conclude either that these features should be preserved or that they will be preserved when new technology is introduced. Yet this is precisely the type of information that is tantalisingly suggested by these studies.

It is an analysis with which this author would, in part, concur. When one comes to a proper understanding of a set of ethnomethodological domain observations, they really do appear to "suggest" something about how technologies for that domain might be best configured. That ethnomethodologists have been unable to articulate in precise terms exactly why this should be so, constitutes another important puzzle for the thesis.

The last task for this introduction is to outline how the thesis as a whole will tackle these topics. Perhaps the most important aspect of the thesis that has yet to be properly introduced is the case study around which much of the rest of the research revolves. The Digital Mammography National Database (eDiaMoND) [11] was a flagship UK e-Science [12] project that sought to develop a collaborative platform for sharing medical mammography data. It formed an excellent case-study for this thesis as, although it was a research project, the core software development work was being conducted by professional, commercial, software engineers who expected to be provided with robust and well-documented requirements.

The project aimed to develop a next generation Grid [13] enabled prototype to demonstrate the potential benefits of a national infrastructure to support digital mammography. It was a large interdisciplinary e-Science project, jointly funded to approximately £4.25m, involving multiple academic and commercial partners[1]. The project received extensive public interest, including articles published by Wired [11], the BBC [14], and even a press statement by then-PM Blair [11].

The context for this project was the UK National Breast Screening Programme, which currently examines women from the ages of 50 to 64 every three years. One in nine women in Britain develops breast cancer in her lifetime [15] and the screening programme is a critical part of the NHS's cancer prevention strategy. It is, however, a massive undertaking: some 1.5 million women are screened each year and work is being done to extend the programme to screen women up to and including the age of 70 (a 50% increase in workload). All of this comes within a context of a national shortage of radiologists [16]. The system proposed by eDiaMoND would allow radiologists to share expertise and reading work between clinics, irrespective of the geographical location of the radiologists involved. In this regard the technology was intended to benefit not just the process of scientific discovery or the training of radiologists, but also directly to improve the care of patients.

---

[1] The project involved a core of 30-35 staff spread over 12 locations, including: 5 universities, 4 NHS trusts, a multi-national company, and a rapidly expanding university spin-out enterprise.

Mammograms are X-rays of the breast and are currently based on film technology. Central to the project's vision was an expected shift from film-based to digital mammography technologies. Once mammograms were in digital form, it was anticipated that a database would be required to store and manage them. eDiaMoND sought to develop such a database on a national level, and then apply Grid technology to manage a series of services to utilize those data. The Grid presented a series of potential benefits for digital mammography: radiologists would be able to share images and expertise in new ways; new Grid enabled Computer Aided Decision algorithms would be developed and used to assist in the reading process; and epidemiologists would be able to use vast stores of accumulated image data to research cancer in new ways.

Breast screening itself is a complex process. Under its current organisation women are asked to attend a screening clinic where they will have a series of these mammograms taken by a radiographer. After the films have been developed, they are displayed on a light-box so that they can be examined for potential signs of cancer by two radiologists. In the vast majority of cases the image will indicate a clear diagnosis and the woman will be asked to return at her next regular appointment. However, if an image is seen to contain a warning sign, the woman will be asked to return to the clinic for a full assessment. This may involve conducting a biopsy, or the taking of further images, perhaps using other modalities. Breast screening therefore shares many common features with other scientific tasks; it is by nature detailed, methodical work that is concerned with using complex equipment to take specific measurements of scientific phenomena.

The radiologists' abilities to identify potential cancers in these mammograms are central to the process. To the untrained eye, a mammogram is a spidery blur, but to the radiologist it is something that can be normal, unusual, or even amazing. Radiologists have a specialized language for talking about different regions of the breast, and a vocabulary for describing features within it (micro-calcifications, spiculative masses, etc). This is a complex skill that is taught through a mix of scientific theory and apprenticeship [17].

Several members of the project, the author included, held the view that in order to develop a Grid-based notion of digital mammography and produce a prototype to demonstrate this concept, it would be important to understand the aforementioned practices.[2] The author's specific involvement in the eDiaMoND project, which forms the empirical core of this thesis, was in developing requirements for the screening application. This was done by drawing on an approach, also developed within this thesis, that extends [2]'s notion of an iterative debriefing meeting.

Consisting as it did of such a large coalition of interests, eDiaMoND was not without friction. However, the project was widely regarded as successful, and led to series of follow-on initiatives and retrospective studies.[3]

---

[2] A view later outlined at length in [18].
[3] eDiaMoND spawned several follow-on projects such as [19], for which the author later acted as a research associate.

This thesis will approach the foregoing topics in the following way. Chapter 2 contains a *literature review*. This will look at three topics that will be essential to the rest of the thesis: firstly, at software and requirements engineering, activities to which the research will contribute; then at different modes for engaging with the stakeholders of a technological development initiative; and finally, at fieldwork itself, and how it has been applied to such projects.

Chapter 3 presents *a methodological introduction to ethnomethodological fieldwork*. Ethnomethodological fieldwork is central to the thesis. This chapter will introduce both fieldwork as a practical activity, and ethnomethodology as a theoretical orientation. It will be necessary to detail a little of the sociological and philosophical background to ethnomethodology, in order to highlight the distinctiveness of its approach, and to lay the ground-work for further foundational discussions later in the thesis.

Chapter 4 considers *the case against a positivist philosophy of requirements engineering*. Arguments against using ethnomethodological fieldwork are sometimes philosophical in nature. [9] for example implies that requirements engineering is a fundamentally positivist activity. This is irreconcilable with an ethnomethodological perspective; if this were so, it would make it very difficult to use ethnomethodological fieldwork as a requirements elicitation method. This chapter argues to the contrary: that positivism is a fundamentally inappropriate view to take of requirements engineering. It thus acts to clear the philosophical ground for the following chapters.

Chapter 5 introduces an approach called *Model Guided Fieldwork*. It proposes the use of fieldwork in iterative combination with informal system modelling. Since modelling and ethnomethodology come from two very different traditions, this proposal will be accompanied by a careful methodological discussion.

Chapter 6 then examines the ***eDiaMoND Case Study***, mentioned above. This was a valuable chance to apply Model Guided Fieldwork in a real software engineering situation. The work done was all recorded and this empirical evidence then provided a means of evaluating the strengths and weakness of the proposed approach. Still, it is as much an investigation into the possibilities afforded by the notion of model guided fieldwork, as it is an evaluation of that approach.

Chapter 7 presents a discussion of the work done. It focuses on some of the difficulties that have traditionally been associated with fieldwork. In particular it examines *approaches to justifying ethnomethodological fieldwork as a requirements elicitation method.* It is proposed that the site at which puzzles, such as those mentioned above, are most appropriately considered, may be found when fieldwork is justified as a method of requirements elicitation.

Chapter 8, gives some *conclusions* from the thesis. It both summarises and draws together the contributions that the thesis has made and looks at some further work which could be use this thesis as its basis.

# 2. Literature Review

This thesis is concerned with the use of fieldwork for the engineering of software requirements. As a consequence three clear areas of interest arise within the existing literature, and this chapter will detail these. Firstly, it is the aim of this thesis to contribute methodologically to requirements engineering, and thus also to software engineering. It will therefore be necessary to establish a clear characterisation of these disciplines, and their relationship. Moreover, as will become apparent in later chapters modelling will play a significant role in this, and so its role within these fields will also be examined. The aim of the thesis is to use fieldwork as a way of engaging with those who might be stakeholders in a potential technology, but it is only one of many approaches to such a task. A second job for this chapter will be to examine such approaches and how they relate to fieldwork techniques. Finally, because the thesis will make specific proposals relating to the use of fieldwork, it will of course be necessary to survey existing methods for its application.

## 2.1. Software and Requirements Engineering

Software engineering has its origins in a desire to professionalize the development of software. This brought with it a series of implications, for example, that there would be a separation of activities, of development from maintenance and use; it implied a specialisation of skills with distinct roles for programmers, testers and analysts; and, it marked the rise of methodology as a distinct and valid topic of research and discussion in itself. Methodology[4] thus became the vehicle for carrying "best practice" from one circumstance of use to another.

For much of software engineering, methodology means outlining the process through which systems should be developed. In its early form the process of software engineering consisted of a linear sequence of activities, since one flowed into the next, it became known as the waterfall model. Classically these phases would include: requirements, where that which is needed from the system is specified; analysis, in which requirements are examined in such a way as to make ready for design; design itself, where the system is fleshed out; implementation where that design is realised; deployment and then ongoing maintenance [20].

Early experiences with this kind of software development revealed problems. Studies showed that requirements was in practice proving to be a problematic part of the software engineering process (summarised by [21]). Moreover, the earlier mistakes in the development process could be detected, the less expensive they were to correct [22]. It was with these concerns in mind, that the spiral model of development was introduced. Instead of engaging in one linear sequence, the spiral model seeks to iterate through the essential software engineering activities. Thus requirements will be revisited and revised in the light of what has been learned about a problem through all the preceding iterations. It was, in Boehm's original work, a method for addressing a perceived risk associated in developing requirements [23].

---

[4] "Methodology" is used here, as it often is in software engineering, to mean an approach or family of methods. In the remainder of the thesis it is used purely in its more traditional sense to indicate the analysis, discussion or production, of methods themselves.

Often referred to now as "iterative and incremental" the spiral model is still a mainstay of software engineering methodology. The Unified Process [24], for example, represents a modern version of it, and adds a notation so that customisations of the process can be systematically documented. This drive toward systemisation is a typical methodological aspiration within software engineering, one that is characterised by the Capability Maturity Model (CMM)[5]. The CMM is not a process in itself, but rather a set of properties that define what a "good" process should be, and in this sense it captures the aspirations of many within software engineering. The CMM presents a stratification of software engineering practice, where superior levels are characterised by activities like systematic requirements management, or the use of quantitative progress measures.

Requirements engineering is a branch of software engineering, and thus inherits many of these properties. The IEEE defines a requirement as (Std 610.12-1990):

> (1) A condition or capability needed by a user to solve a problem or achieve an objective

> (2) A condition or capability that must be possessed by a system or system component to satisfy a contract, standard, specification or other formally imposed document

> (3) A documented representation of a condition or capability as in (1) or (2)

In both its content and style of presentation, this definition echoes many of the qualities of precision alluded to by the CMM.

---

[5] The Capability Maturity Model [25], is now referred to as the Capability Maturity Model Integration [26].

Like its parent discipline, requirements engineering has also been traditionally decomposed into separate activities. Requirements elicitation is concerned with finding information that will be relevant to the requirements process; requirements analysis is the process of examining this information and identifying requirements; requirements negotiation is concerned with gaining agreement between stakeholders over a set of requirements; requirements documentation aims to put them into written form; and requirements verification, is the task of confirming that requirements have been accurately produced [27].

Many of the techniques involved in realizing these activities might initially seem methodologically quite basic. For instance, requirements documents are often based on a template, which defines the various sections that should be completed in order to thoroughly describe a desired system (see for example the Volere template presented in [28]). Similarly, requirements elicitation may simply involve talking to stakeholders in the system and asking them what they need from it; and requirements negotiation might take the form of a simple facilitated discussion. However, despite such apparent simplicity, the task that each of these techniques tries to achieve is in reality extremely complex and difficult to apply in practice. Thus, when examined in more methodological detail, requirements engineering can be seen to be a richly varied discipline.

[29] provides evidence of such diversity. It takes a survey of requirements engineering methods, and then categorises them using Lyotard's taxonomy of social theories [30]. They explain how it is possible to see each requirements method as having a meta-narrative, which would outline the assumptions and "world view" which that method

adopts. The different categories span perspectives from those which take an objective, science-like, attitude (unitary hard), through to those which see the introduction of technology as a democratic process of political persuasion.

Many of the earliest methods discussed within requirements engineering as an independent field of study, were of the former kind. The so-called structured methods sought to replace the traditional requirements document, which was seen as lengthy and difficult to deal with. Instead, using notations such as the data-flow diagram, they allow the requirements engineer to build a model, either of the current domain, or of the problem, or of a potential solution (for example [31]). Diagrams were seen as bringing structure and clarity to the problems of requirements.

Domain modelling played a significant role in many of these early approaches. However, practical experience began to suggest some associated pitfalls. Domain modelling is a time-consuming activity. Because of the almost infinite detail that *could* potentially be expressed in such a model it has become associated with the phrase: "paralysis by analysis." By the mid-80s requirements engineers such as Yourdon [32] were suggesting a much lighter approach to domain representation.

Today, modelling methods are dominated by the UML notation [33]. The Unified Process [24], which provides a reference point for how that notation should be used in practice, still presents domain modelling as a useful tool. However, its major modelling focus in terms of requirements engineering, is on Use Cases. These provide a way of representing a system so as to capture the features which it is required to have. Rather like many of the structured methods, the diagrammatic nature of use cases can provide a valuable overview of a system. Moreover, the simplicity of the notation allows quite clear and accessible diagrams to be created. However, experiences both with using and teaching the technique, show that while simplicity is possible, producing use cases is actually a very complex task. In this respect use cases face similar difficulties to the use of more traditional requirements document templates. Indeed, some have pointed out that use cases, each of which should be specified with a template of its own, should be thought of more as a textual format than a graphical one [34].

Modelling techniques thus provide a useful notation for expressing requirements, but often do less for structuring the process of requirements analysis. Some in the requirements community have sought to provide modelling notations to help with this too. Goal oriented notations, such as [35, 36], provide a way of starting with high level aims and expressing a process of analysis whereby these are decomposed into more specific, detailed, sub-aims. Goal oriented modelling techniques are often sufficiently rich that enough entities from the requirements problem can be expressed to enable formalisation to take place. In this way it is possible to start with informal or semi-formal representations of a problem, and work towards a formal specification of it through analysis.

Formal techniques for requirements engineering have also been a continued strand of research within the field. Many of the earliest academic works on requirements engineering focused on providing formal notations, similar either to those in software engineering or to artificial intelligence. Practical experiences, however, have shown that these alone are not the answer to all requirements engineering's problems. Rather they have revealed the difficulties associated with the process of formalisation itself [37].

Modelling and formalisation are both approaches that mirror strands of software engineering methodology. Since requirements is a sub-field of software engineering, this is in some sense very natural. However, requirements engineering may also contain challenges that are more distinct from those most commonly addressed in software engineering methodology. Requirements elicitation has traditionally been based on techniques such as surveys, interviews, focus groups, and fieldwork, all of which have undergone significant methodological development in the social sciences [38].

Other techniques also overlap with, or have been imported from other fields. The Soft Systems Methodology (SSM) [39] for example, based on Systems Theory, is seen as a general problem solving approach. It recommends a series of activities that are designed to help a group to develop and agree on a system concept, and concomitant change to a domain through its introduction. This is done through facilitation, and the use of group activities which help to make everyone's perspectives clearer to those assembled. SSM is thus useful in complex requirements situations where there is little or no clarity or agreement over what role a new system might fulfil. Similarly, ETHICS (Effective Technical and Human Implementation of Computer Based Systems) [40], is an approach which hails from both the socio-technical and participative traditions. The socio-technical view of systems development considers work and technology to be fundamentally linked, and thus an optimal solution must necessarily involve the joint optimisation of both. Participative methods will be returned to below.

However, this variation is not just a product of importing techniques from outside the field. For example, in recent years scenarios have been a significant focus of attention in requirements engineering research [41]. Scenarios have been used at many different points in the development process, and in many different forms [42] but fundamentally, as [43] points out, they rely on our innate ability to work effectively with narratives.

Requirements engineering is therefore a methodologically varied field. Certainly it draws on, and follows trends within software engineering, but it also draws on other fields, such the social sciences, to provide it with methodology, and moreover, has developed many lines of research which look quite different from mainstream software development techniques.

This gloss, of requirements engineering as a methodologically innovative off-shoot from software engineering, is challenged by more recent initiatives within the latter. Perhaps most obvious is the agile programming movement [44], which has challenged some of the principles that software engineering might traditionally have taken for granted, such as the need for up-front architectural design. Software engineering often proceeds in a hierarchical way, first determining the overall architecture of a system, then decomposing it into separate problems which can each be subjected to more detailed design, often using modelling notations, before coding can begin in earnest. Such an approach is predicated on the assumption that it is possible to gain a good sense of a system's requirements early in a project, something that agile methods also reject. For example, Extreme Programming (XP) insists that it is sufficient to gather detailed requirements *as* one develops [45]. Thus, development is organised into iterations, often of a month, and for each iteration a selection of "user stories," which are much like scenarios, are developed and selected for implementation. The design of the system, rather than being shaped by an overall architectural plan, emerges under the principle that the system should never be more complex than it needs to be to realise the user stories at hand. Although this strategy may sometimes require the design of the system to be "refactored," this is seen as a small price to pay for the overall ease and agility that it affords.

Though both modelling and requirements engineering are mainstays of current software engineering methodology, it is clear from the above that the relationship between them has shifted over the years. Models may be of use either as throw-away sketches, or as permanent documentation. Similarly, requirements may be developed as a formal statement in advance, or established during the process of development itself. However, what all of these approaches depend on is establishing a clear sense of what is needed from the system under development. No matter what kind of methodology is employed, software engineering is therefore deeply committed to engaging with stakeholders in order to elicit these needs.

## 2.2. Approaches to an engagement with stakeholders

Requirements engineering often draws on other disciplines in its methodological treatment of stakeholders. In this regard considering some of the ways in which other technologically focused disciplines engage those who might use it, will be of value in situating fieldwork. Three movements will be considered, Human Computer Interaction (HCI), Participative Design (PD), and Computer Supported Cooperative Work (CSCW). Given their overlapping aims, it is no surprise that these disciplines have similarities. Nevertheless, differences exist and are worth examining.

HCI has its origins in the need to design acceptable user interfaces for computational devices. Traditionally it has conceived of its task as one of designing the interactions between users and computers by determining principles for good interface design [46]. These might be derived through laboratory experiments, where multiple participants are timed or assessed in the completion of a pre-determined task using a particular interface. This provides data which can then be used to support or refute the principle, much like other scientific work. Principles may then form the basis for the evaluation of prototype interfaces, for example by a panel of expert HCI evaluators [47].

HCI often adopts a cognitive perspective [48]. By seeing interactions between man and machine in terms of cognitive processes, and by using existing theories about these processes, HCI has been able to determine further principles for design. Moreover, this orientation forms the basis for other evaluative approaches, such as cognitive walkthroughs [49], or testing by "thinking aloud" [50]. However, critiques have suggested that this orientation may be too narrow to properly understand the nature of work and thus design appropriate systems  [51-54].

Amongst these, Suchman's critique [54] was particularly influential. It presents a study of an "intelligent" interface to a photocopy machine. The interface uses its interactions with the user to build a model of that user's photocopying "plan," and on the basis of this deduce what problems they might be having, and hence provide relevant assistance. The study showed how the assumptions on which this design was based, were flawed. The design assumed that users would have stable, pre-existing plans, whereas observations of usage suggested that their actions related more strongly instead to the local circumstances of a particular situation. This formed a basis for a more wide ranging critique of the cognitive perspective, and for some marked a "turn to the social" [7, 55] in the design of technology.

The rise of mobile devices and more powerful networking technologies has for some heralded a new era in computing. [55] points to 5 phases, from programmer-as-user in the early days of the computer, through to today's ubiquitous and wireless technologies. In this it draws a distinction between the notion of machines as having individual users, in the sense often associated with HCI, and a paradigm shift towards computational devices conceived of as a seamless part of the ways that groups collaborate to produce work.

These new technologies, and this alternative view of work as being inherently collaborative rather than individual, both underpin the field of CSCW. An interest in collaboration has led CSCW toward methods for achieving detailed studies of naturalistically occurring work. Often these studies have been fieldwork based [52]. They have sought to understand work practices by studying those practices individually, and in detail. As a result of such studies, specific designs might be provided, or guidelines might be given so that others may do design work. Alternatively, findings from studies might be used as supporting evidence in academic debates which seek to develop the way in which collaboration and work are conceptualised [56].

Although the fieldwork studies of CSCW are very different from the classic experimental style of HCI, the former being methodologically informed by sociology rather than cognitive science, there is nevertheless similarity. Both involve researchers studying potential users of a technology, and then producing research or recommendations that would guide other design work. The third field of interest takes a slightly different approach. PD [57] seeks to involve those who might use a technology as fully as possible within the process of design. PD has its heritage in a Scandinavian movement of collaboratively involving the trade unions of a workplace in the design of work itself. Participatory methods thus do not just seek to involve users, but often to empower them within the technological development process.

The above thus presents, albeit briefly, three different perspectives on how to engage with stakeholders, and it is possible to see how each has influenced requirements engineering. For example, methods like Rapid Application Development (RAD) and SSM both have participatory elements, and academic research is often influenced by either CSCW or HCI. Though three modes of engagement have been outlined, this is something of a simplification of what are, in reality, methodologically varied fields. While the above emphasises differences, it is equally possible to identify convergence, particularly in recent years, and especially around the adoption of fieldwork. Fieldwork has always been important in CSCW, but initiatives such as Contextual Design [58] have also helped to popularise it within HCI. Moreover, fieldwork and PD represent complementary attempts to explore the stakeholders' own perspectives [59].

## 2.3. Fieldwork and the Production of Technology

Fieldwork, of the ethnographic kind considered in this thesis, is conducted with respect to a particular orientation. However, as [60] points out, the ethnomethodological orientation has become prominent, within Requirement Engineering (RE) and particularly so within CSCW. Nevertheless other approaches are possible, including Actor Network Theory [61], Distributed Cognition [62], and Activity Theory [63]; each is summarised and contrasted against ethnomethodology by [1]. The following chapter will provide a necessarily detailed introduction to ethnomethodology, for the present, and in order to illustrate its relationship to the production of technology, two case studies will be presented.

The first is situated in a London City dealing room [64], it illustrates the kind of value that ethnographic study can bring to the development of technology. The researchers were approached by a telecommunications company, who were interested in constructing technologies to assist dealers in capturing the deals they were making. At this time, deals, which could take place over the phone, electronically, or in person, were recorded by the dealers on paper tickets. These tickets were then collected by administrative staff, who would type up the information and submit it to the market. Doing this quickly and accurately was imperative, and it was felt that technology may improve this. Moreover, an electronic deal capture system might provide a real-time account of the risks that the bank, on whose behalf the dealers were trading, was exposed to.

One of the technologies that the firm was interested in was voice recognition technology. Many of the deals occurred over the phone and were already recorded for legal reasons. Handsets were associated with particular dealers, and thus it was felt that a system could be trained to recognise specific voices. Moreover, it was felt that the content of a deal contained information from a limited set of symbols, for example, the names of a small number of different stocks, and numbers.

With this in mind a team of sociologists were engaged to study the work practices of the dealers, and inform the requirements for this technology. They proceeded by doing ethnographic studies, combined with video recordings of ordinary work. These video recordings were then subjected to repeated analytic viewings, and selected transcription in the style of conversation analysis [65], a branch of the ethnomethodological movement.

The study revealed a number of practices which were of distinct relevance to the project at hand. Dealers would commonly elide the numbers about which they were talking, particularly in relation to market prices. Since the most significant digit of a price would change infrequently it could be left out; for example, "*ninety five nine*" could be given as an abbreviation of "I am prepared to buy this stock from you at a price of six pounds and *ninety five* pence, and I am prepared to sell it to you at a price of six pounds and *ninety nine* pence." The dealers were able to do this because they were able to depend on the competence of the listener, and their knowledge of the marketplace. However, this discovery undermined the assumption that all of the information about a deal would necessarily be mentioned explicitly in the talk of agreeing a deal itself. Other observations revealed more about the process of dealing. For instance, dealers would commonly be simultaneously engaged in multiple deals. Doing this would often involve having multiple callers on several handsets. Although from the context it is always unambiguous which utterance is intended for which caller, this could be quite difficult for a system to establish.

Both of the above observations brought into question the viability of a voice recognition technology. Combined with this, and other observations about the collaborative nature of dealing, the researchers were able to propose alternative technologies for deal capture, such as a pen-based handwriting recognition system. The study also illustrates the powerful role that observation, and in this instance the utility of analysing video recordings, can bring to the development of requirements.

A second case study, and one which represents a seminal attempt to use fieldwork as part of a software engineering project, was the Air Traffic Control study conducted by researchers at Lancaster. Like the example above, fieldwork observations were critical to defining requirements for a system. Their study is made particularly notable by the subsequent methodological reflections expressed on the relationship between software engineering and fieldwork studies.

One of the critical features of fieldwork is that it is often conducted by sociologists, rather than computer scientists; being as it is, a distinct kind of skill, not usually taught as a part of general computer science education. Experiences gained through fieldwork were thus disseminated through debriefings involving the sociologists and computer scientists. Fieldwork was thus iterative, as the meetings would generate new themes for investigation.

[66] outlines some of the questions which seemed helpful in guiding these debriefing sessions (p253).

>What aspects of the work which are currently manual are not important or consequential to system purposes, and need not be supported in a computerised system?

>What are the important manual activities which are characteristic of the system because it is manual? In other words, what activities need not be supported in an electronic system because the system will replace them?

>What aspects of the manual system must be replicated without change in a computerised system?

>What aspects of the work needs to be supported but need not be, or cannot be, replicated in the same way in an electronic version?

Based on experiences from the same study, [2] outlines some of the different modes in which fieldwork might be used during the development of systems. *Concurrent ethnography* is most like that conducted during the ATC study, where ethnography is done iteratively with *regular debriefings*, and at the same time as ongoing technical analysis of the problem. *Quick and dirty ethnography* is done to provide a general but informed sense of the domain to the system development process. In many ways this was led by the observation that the value of further ethnography decreased over time, and that many of the critical observations were made early on in the process. *Evaluative ethnography* would be concerned with looking at a specific proposal, and then doing fieldwork to assess this. Finally, *the examination of previous ethnographies* could provide insight to inform initial design thinking.

Each of the above modes seeks to find a place for fieldwork in systems development, by placing it in a familiar engineering style of process. Other approaches have chosen to target a different part of the fieldwork activity. Ethnography is, in a formal sense, the document that is used to disseminate the findings of a fieldwork study; in its traditional form, when used in sociology or anthropology, this is the primary deliverable. Ethnographies are often very long documents, containing rich detail about the domain. In some regards it is precisely this detail that software engineering may find useful, however, the size of such documents is seen as problematic. For example, [67] relays an anecdote about the ineffective status of an ethnography as medium for communication with designers.

> [The] ethnographers were conducting their observational studies of work, layering on the detail until they were reasonably satisfied that they had adequately described the sociality of work. The study would be so heavy that they would have no need of a brick but, having given the designers ample warning to stand clear, would simply lob it over the great high brick wall with a message attached, 'read this and then build something'.

A number of approaches which have attempted to use fieldwork in software engineering, have focused on transforming this deliverable into something more familiar or accessible. For some this has meant turning ethnographic detail about the domain into models of that domain. [68] shows how UML diagrams can be used to represent observations from a domain. By doing this they use a notion that will be familiar to many software engineers, and in so doing produce an abstract representation which strips away much of the original detail of the observation making it quicker and easier to read. [69] has proceeded along similar lines, only based on the transcription of video data, and using the CSP modelling language, which provides strong facilities for representing the concurrent nature of activities.

[67] represented an attempt to give an accessible form to the ethnographic record, while preserving more ethnographic detail. The Designers' Notepad sought to provide a multimedia tool in which information from the domain could be stored, perhaps linked to diagrams and sketches of the domain. This could then be explored by developers, who would be able to find out information that was more directly relevant to their particular interests, rather than having to read a complete ethnography. Similarly [70] has looked at representing ethnographic observations using virtual reality, to model the flow of work, and spatial qualities of a workplace.

The aforementioned approaches have typically been concerned with communicating details about one particular kind of domain, to one specific project. Patterns [71] attempt to present fieldwork observations in a way that will enable them to be more widely used. Inspired by the work of Christopher Alexander in architecture and urban planning, patterns use a standardised template to help to make the ethnographic record easier to read, much as standard templates for use cases or requirements documents do.

Though the literature contains clear examples that attempt to make fieldwork amenable to a software engineering context, there are many more examples particularly within CSCW where it is talked of in relation to *design*, rather than engineering (for example, see introduction p6). This notion of ethnography "informing design" is a common one, and many have commented on how this might occur. Randall *et al.* suggest that analysis is the key to using ethnography successfully, that fieldwork must be "design relevant but not design laden" (p301), in such a way that it becomes suffused with concerns from other stages in the design process. They suggest that there is currently an "analytic gap" (p144) between observations and their use within the development process. However, closing this gap is far from straightforward (p45), as there can be no general method for doing ethnography-for-design that is applicable to every situation (p13, p16, p148). Perhaps as a result they suggest (p15, p306) that little methodological progress has been made in defining this relationship since the seminal 1994 paper by Hughes *et al.* ([2] discussed above).

A similar picture is provided by other commentators, who emphasise the potential of, and challenges for, fieldwork in design and requirements engineering, rather than provide direct methodological advice regarding its application. For example, [72] has suggested that ethnography, in the sense described by the social sciences, need to be modified before it can be of use in design. [73] goes further and suggests that in fact a hybrid form of methodology may be required to guide this process. As the introductory chapter discusses, both [9] and [10] point to puzzles of a philosophical and foundational nature. Others such as [74] and [75] find fault, not with ethnography, but with the structures and divisions of labour that dominate traditional systems development methodologies. [56] points out, it may be that fieldwork is most valuable in re-specifying our conceptual assumptions, rather than substantively informing individual projects.

Although there are many case studies that show its value, there is a lack of consensus on the precise details of how fieldwork should be applied to the problem of developing technology. Moreover, debate over its role within various fields, and puzzles of a methodological and philosophical nature begin to indicate that there may be issues of a foundational nature yet to resolve.

Finally, one fieldwork-based movement that has sidestepped many of these issues is Contextual Design (CD) [58]. In common with the commentators above, CD draws for inspiration on insights based on ethnomethodologically informed case studies such as Lancaster's Air Traffic Control project, and on sociologists like Goffman. However, as [1] suggests (pp31-9), it does not explicitly engage in any depth with the methodological underpinnings on which these insights are based. CD is thus of interest, less for its theoretical commitments, but because it is a response to the industrial experiences of expert practitioners (predominantly those of its founders: Beyer and Holtzblatt).

At the heart of CD is a form of fieldwork described as "Contextual Inquiry." As [58] puts it (p41): "The core premise of Contextual Inquiry is very simple: go where the customer works, observe the customer as he or she works, and talk to the customer about the work." However in practice this is described as being more like an interview (pp73-6), where the interviewer is only ever with a subject for a couple of hours, than like the kind of longitudinal ethnographic fieldwork referred to above.

Though different both theoretically and practically, from the kind of fieldwork of interest to this thesis, CD nevertheless claims to provide a method for making a successful step from fieldwork into design. One thing CD does is to provide a process by which experiences with customers can be shared within a team. Following a contextual interview it prescribes an "interpretation" session in which a whole team can discuss what the interviewer discovered. Findings about the workplace are represented in diagram form, and implications for design are noted down. Once a corpus of interviews, design implications, and diagrams have been collected, they can be consolidated. Implications are grouped hierarchically on an "affinity diagram," and work diagrams are generalised to provide models of the workplace and its activities. This consolidated information then forms the basis for brainstorming and work practice re-design, which will be captured in further models, and thus lay the way for lightweight prototyping exercises. Though CD gives great emphasis to fieldwork enquiry, in its heavy use of modelling it is somewhat reminiscent of the structured methods of traditional requirements engineering.

## 2.4. Summary

This thesis will draw strongly on a number of the works mentioned in this review. [2] is a seminal description of the different potential relationships that fieldwork might have to systems engineering, and chapter 5 will draw on and extend the notion of the debriefing session it describes. This will be done by drawing on the familiar software engineering notion of modelling. Similarly the thesis will also contain significant parallels with [1], in particular it shares the notion that the primary problem for fieldwork is an analytic one, rather than a representational one. However, the thesis will also take issue with some of the aforementioned positions. Firstly, with the implication expressed in [1], that incorporating fieldwork into the requirements process is in some sense problematic; secondly with the philosophical concerns raised by [9]; and thirdly, with the paradox raised by [10]. In each case examining the foundations of ethnomethodological fieldwork, and requirements engineering more closely, will show that resolutions to these concerns can be found. In particular it will be shown that it is possible to go beyond thinking about fieldwork as merely "informing" design, and instead, with the right kind of analysis, see it as a way of eliciting and justifying precise statements of system requirements, of the kind needed by software engineering.

# 3. A Methodological Introduction to Ethnomethodological Fieldwork.

This thesis seeks to use ethnomethodological fieldwork for specific purposes within an engineering context. It will therefore make methodological proposals involving fieldwork. This chapter is devoted to introducing the approach, its theoretical commitments, and practical properties. Ethnomethodology is often the subject of confusion, because of both the issues it approaches, and the way in which it is written about. This chapter cannot be considered a complete introduction to the topic, but it will cover the aspects of ethnomethodology that are important to later chapters.

"Fieldwork" is a general term used to describe research in the social sciences, archaeology, and so forth, done at the site of interest itself, rather than in the laboratory. This thesis is interested in a particular kind, sometimes called "ethnographic fieldwork" (conducted by "ethnographers"), in reference to the name of the document (an "ethnography") which is traditionally the result of such studies.[6] Specifically, its concern is with a form of fieldwork informed by ethnomethodology.

The approach has much in common with the fieldwork of anthropology, where it is regarded as a more or less essential part of the discipline. In the 19th century these were typically studies of primitive and exotic societies, but groups like the Chicago School in the early 20th century show that mundane urban life is equally valid as a domain of study. The methodological reflections of both contemporary anthropologists and sociologists will thus be of great interest, and it is here that the chapter will begin.

---

6       This informal use of the term "ethnography" is avoided in this thesis, for the simple reason that the method proposed later in the thesis, although deeply involved in fieldwork enquiry, does not advocate actually writing an ethnography of a traditional form.

However, fieldwork is usually additionally informed by a theoretical orientation, which may suggest how the domain of study may be understood, or which points towards the aspects of it that are worthy of the fieldworker's attention. Garfinkel's ethnomethodology [5] is one such orientation which has been adopted by many interested in fieldwork and technological design. It is theoretically well founded, being based on a firm philosophical basis, and possesses a rich set of arguments which establish it in contrast to the theories of mainstream sociology. Presenting this foundation will be essential background to demonstrating its proper use later on in the thesis. This theoretical dimension will therefore form a second focus for the chapter.

## 3.1. Fieldwork and Ethnography as a Practical Activity

Goffman (1922 – 1982), emerged from the aforementioned Chicago School. His subsequent work illuminated sociology, in particular he pioneered a kind of sociology whose findings are often allied with those of ethnomethodology. He describes the practical challenge of fieldwork thus [76][7] (p149):

> It's one of getting data, it seems to me, by subjecting yourself, your own body and your own personality, and your own social situation, to the set of contingencies that play upon a set of individuals, so that you can physically and ecologically penetrate their circle of response to their social situation, or their work situation, or their ethnic situation or whatever. So that you are close to them while they are responding to what life does to them. I feel that the way this is done is to not, of course, just listen to what they talk about, but to pick up on their minor grunts and groans as they respond to their situation. When you do that, it seems to

---

7    Goffman rarely discussed his methodology, the quote provided was recorded covertly by a member of the audience of a panel session in which Goffman spoke about fieldwork. A transcript of the tape was subsequently published with the permission of his widow.

> me, the standard technique is to try and subject yourself, hopefully, to their life circumstances, which means that although, in fact, you can leave at any time, you act as if you can't and you try to accept all of the desirable and undesirable things that are a feature of their life. That "tunes your body up" and with your "tuned-up" body and with the ecological right to be close to them (which you've obtained by one sneaky means or another), you are in a position to note their gestural, visual, bodily response to what's going on around them and you're empathetic enough - because you've been taking the same crap as they've been taking - to sense what it is that they're responding to. To me that's the core of observation.

It is a skill that is best learned through practical experience, and which demands the management of a whole range of concerns. Much of this is a matter of inter-personal skill, of managing relationships with those in the study domain. It is also a personally demanding activity, where the fieldworker must open themselves up to be constantly bemused, even though those around them know exactly what the appropriate next action may be. Geertz, a cultural anthropologist, describes in one of his own fieldwork experiences how Balinese locals initially treated him as if he literally didn't exist, until he was unintentionally caught up with the villagers in a police raid [77] (p416).

> Getting caught, or almost caught, in a police raid is perhaps not a very generalizable recipe for achieving that mysterious necessity of anthropological fieldwork, rapport, but for me it worked very well. It led to a sudden and unusually complete acceptance into a society extremely difficult for outsiders to penetrate. It gave me the kind of immediate, inside-view grasp of an aspect of "peasant mentality" that anthropologists not fortunate enough to flee headlong with their subjects from armed authorities normally do not get.

As Rawls describes, ethnomethodologists are also generally immersed within their domain of study [78] (p6). Moreover, and in common with others, Garfinkel refuses to prescribe any exact methodological details for the conduct of actual ethnomethodological enquiries. Instead he suggests that whatever methods are necessary for the adequate study of a domain, already exist within that domain, and can be learned by becoming a competent participant of it. He calls this the unique adequacy requirement of methods [78] (p175). It may be easier for the fieldworker to acquire competence in some domains more than others. Skills like "polite conversation" may be ones that the fieldworker already innately has; but more specialised domains, such as hospitals, or scientific laboratories, will have their own unique skills which they must learn in order to meet this "unique adequacy" requirement.

Nevertheless, ethnographic fieldwork often takes on a broadly common form. For many, fieldwork is a longitudinal activity which may involve being *in situ* for at least a year. During this time a fieldworker will keep field-notes, which may run into hundreds or even thousands of pages, they may initially be handwritten but should be typed up every evening. Goffman stresses that field-notes are a personal record, which should reflect on the fieldworker's own feelings about events, and relationship with the domain. This means that field-notes are often a private record of events [76] (p153).

Ethnography is literally a written account of a fieldwork study, intended to be a public record of the ethnographer's experiences. Writing an ethnography is thus an analytic activity which may draw on an ethnographer's field-notes as evidence. While student texts may describe ethnography as being a separate analytic step, analysis is nevertheless a continuous part of observation. Although ethnography may literally be done "back in the office", what is written there may have been rehearsed over and over in field-notes.

Perhaps the most famous conception of the ethnography is Geertz's description of it as a process of "thick description" [77]. It is a term first used in the ordinary language philosophy of Ryle. He takes the example of a wink, or more precisely a movement of the eye-lid, and its possible meanings. Such a movement could of course be nothing more than an involuntary muscular spasm, but it could equally well be a conspiratorial signal; the problem is that to a naïve observer the two events could look identical. Furthermore a contraction of the eye-lid could actually signify many other things: it could be a fake-wink designed to fool someone into thinking they are part of a non-existent conspiracy; it could be a burlesque-imitation-of-a-wink designed to mock another winker; it could even be a rehearsal-of-a-burlesque-wink, to perfect future acts of wink-mockery. To describe each of these as merely a movement of the eye-lid would be, in Ryle's words, to give it a "thin" description; a "thick" description is thus to place it within a richer, more meaningful context. Meaning for Geertz is to be found by studying the uses to which these social acts, a wink being but one example, are put. In this regard he also draws on the mature work of Wittgenstein [77] (p17).

Of course such studies must also be conclusive in some regard. Geertz in common with many, is cautious with respect to the relationship between individual observations and general theories. He points to a practical tension between the need to keep analysis "close" to actual events, and the drive to build useful theories based on those events [77] (p24). For Geertz the purpose of such studies is to further our understanding of cultures, however, within this discussion he speaks to issues that are of relevance to anyone who seeks to apply ethnography: "the aim is to draw large conclusions from small, but very densely textured facts; to support broad assertions... by engaging them exactly with complex specifics" [77] (p28). Ultimately small facts speak to large issues only because they are made to by an analyst, for the purposes of that analyst [77] (p23).

Observation of a domain and note taking are essential elements of many classic ethnographies, and many classic studies in ethnomethodology. In more recent years the availability of audio and video recording technologies have introduced new data collection opportunities. Recordings overcome the limitations of recollection; they allow events to be repeatedly examined; they permit other researchers to access the data, for analysis, or for the public scrutiny of analysis; and, they can be reused and re-examined in the light of new themes or issues [79] (p238). The studies of recorded talk conducted by Sacks *et al.* ultimately gave rise to Conversation Analysis, one of ethnomethodology's most successful achievements. Later, the availability of video equipment led to parallel studies in how talk and visual conduct were related, for example detailing the organisation of gaze and gesture [80, 81]. Though fieldwork may be important to such studies, they often focus on close transcription from recordings, of talk and visual conduct. These form the basis for repeated viewings, analysis, and may provide impetus for further data gathering [82].

Early conceptions of both anthropology and sociology were scientific in nature. Indeed the two fields were often theoretically informed by the same sociological figures, such as Durkheim (1858 – 1917) and Weber (1864 – 1920). The late 20[th] century saw moves away from this scientific conception for both fields. It is within the sociological part of this tradition that ethnomethodology is located. Because of the way in which ethnomethodology was established, the following section will introduce it against this backdrop.

## 3.2. Ethnomethodology and Sociological Theory

For some, the methods of the social sciences are deeply related to those of the natural sciences. Durkheim, one of sociology's founding fathers, conceived of the discipline as a science of social facts. Methodologically, this notion of sociology as a science was perhaps most distinctively developed by Lazarsfeld (1901-76). An occasional member of the Vienna Circle, a group devoted to the widest possible use of the scientific method, Lazarsfeld developed a conception of social science as the definition and manipulation of variables. Research would often use surveys or interviews, which could be coded in terms of such variables to provide data for the analysis of statistical covariance between them. It is an approach which has become virtually orthodox within social research [83] (p52). At its core this conception allows social science to create similar tabular data, do similar mathematically based analysis, and use the same method of hypothesis and test that is familiar in the natural sciences. It is, in Garfinkel's terms, an example of "Formal Analysis" [78].

The nature of social life and the methods that should be used to investigate it, have long been the subject of debate. Much of this conflict can be traced back to the Methodenstreit (dispute over method) in late 19[th] century Germany. From this arose an intellectual movement, of which Weber (1864-1920) was a part, that sought to distinguish either their phenomena of interest, or their method of enquiry, or both, from those of the natural sciences, on the grounds that they were irreducibly unique. It is on this presumption, that human action is meaningful and is worthy of study in its own right, that qualitative methods in the social sciences are based [38] (p13).

While the positivists were interested in working nomothetically, to define laws or generalisations through abstraction and the definition of variables, Weber and others were interested in working ideographically, by trying to understand individual, concrete, unique cases of social action. It is argued that, although we may find it useful to deal with phenomena of the natural world in abstract terms, when it comes to the people around us, we are much more concerned with the unique characteristics of any particular interpersonal circumstance. For Weber this detail would come from "interpretive understanding." Indeed, the possibility for such understanding provided the opportunity to study human action at a much greater depth than could ever be achieved in the study of mere inanimate objects. Weber is important as he represents the beginnings of a form of sociology that would become interested in the "microscopic" detail of social life and interaction.

Schutz (1899 – 1959) and Parsons (1902 – 1979) both drew, in different ways, on the writings of Weber (see [84]). The philosophy of the former exemplifies the kind of phenomenology on which ethnomethodology is based. The sociological writings of the latter, a major force within American sociology at the time, provided Garfinkel with a foil [5].

Parsons' theory of social action had been a synthesis of the works of many of the great sociologists, including Durkheim and Weber. For Parsons social actions were constrained and determined (although not caused) by norms. In common with the structural-functionalist view, by following such norms social actors create a self-stabilizing societal system. Such norms are adhered to because they are seen to be the 'proper' way of behaving; they may relate to a value that must be maintained in order for the actor not to suffer internal pain, loss or conflict; or because other actors may punish them for breaking their expectation of adherence. It is a conception that is taken from both Freud and Durkheim, whereby norms are internalized by actors.

Parsons' framework thus rejects a positivistic conception of actors' actions. His notion is of social action determined by subjectively held norms, rather than as a product of objectively knowable biology. However, he maintained a positivistic view of sociology, as a weaving together of empirical observation and abstract theoretical concepts; "he is adamant that, although social sciences deal with 'subjective' phenomena, they are not on this account to be excluded from the general pattern of scientific development" [79] (p19). So, despite the proposed subjective nature of the phenomena of interest, observations of these phenomena may be analysed, decomposed, and then explained in terms of universal laws, of the analyst's making.

Parsons' sociology was one in which the analysis of social action was to occur in terms of concepts wholly external to the point of view of the actor. He thus gives priority to findings of sociologists over those of the actor (there is, of course, no reason to imagine that the two would naturally agree); after all, the sociologist is engaged in a methodologically superior, more rational, more scientific, form of analysis. He suggests that only rarely do actors become aware of their own motivational forces. Garfinkel argues to the contrary, proposing that the actors' own common-sense is severely neglected within the Parsonsian framework, and further, that this common-sense is in fact a topic worth of study in its own right [5] (p36).

Garfinkel argues that Parsons' norm following model turns the social actor into a "cultural dope," whose behaviour is unthinkingly coerced by external rules. It denigrates actors' autonomy and fails to correspond with our own lived experience of the social world. Garfinkel points out that, far from being ruled by norms, social conventions are in fact not only known by actors, but may be used reflexively by them. For example, in many situations it is polite to provide an excuse if one is unable to attend an event to which one has been invited; when refusing such an invitation one could thus choose to be deliberately snide by failing to provide one; for this to occur, both the insulter and insulted must knowingly understand that convention, and what may be meant by an instance of non-compliance. Such reflexivity is deeply problematic for Parsons' sociology.

Another critical difference between Garfinkel and Parsons' sociology was the status of the sociologist. For Parsons the sociologist is engaged in revealing things about society to which those within may not be able to attend. By contrast, Garfinkel sees no difference in principle between the investigations of the professional sociologist and the mundane investigations of the members of society. They are themselves "lay sociologists", using and examining the same social structures that are of interest to the "professional sociologist".

In response to the decline of Parsons' influence on sociology, Garfinkel has in later work re-situated ethnomethodology in contrast to a broader notion of formal sociological analysis [78]. Using theories and methods, formal sociological analysis imposes its own order on a domain; in so doing it orders that domain using structures that are entirely external to it. By contrast ethnomethodology emphasises the existing internal orderliness of a domain as the primary topic of study. Because of this, Garfinkel argues that ethnomethodology is able to study phenomena that systematically escape such formal analysis [78] (p133, p122). From a methodological perspective ethnomethodology is, according to Garfinkel, more an approach to avoiding formal analysis than it is a method of research in itself [78] (p170).

Ethnomethodology places great emphasis on the direct lived experience of the fieldworker. As a basis for such enquiries it draws on the work of phenomenologists such as Gurwitsch, Mearleau-Ponty [78] (p167), and Schutz [5] (p37). During the post war years when Garfinkel was establishing ethnomethodology, this was one of the few major sources of insight into the organisation of everyday experience [79] (p36). Schutz's writings on the philosophical foundations of sociology, and especially his notion of inter-subjective understanding, were of particular importance.

Schutz's first and best known work [85] was an attempt to provide a new foundation, based on the phenomenology of Husserl, for the sociology of Weber. Schutz was very much in favour of Weber's interpretative approach, but felt that some of the more foundational concepts lacked clarity and consistency. What Schutz provided throughout his work was a subtle and profound analysis of lived experience and inter-subjective understanding [85-87]. It provides, amongst other things, a foundation for the observation and analysis of social actions.

Schutz's phenomenological construction begins with what one may introspectively see within one's own stream of consciousness. He notes that while actually immersed in one's stream of consciousness, the individual does not clearly differentiate experiences as being separate. Rather, and drawing on Husserl, he notes that: "[t]hrough the attending directed glance of attention and comprehension, the lived experience acquires a new mode of being. It comes to be 'differentiated', 'thrown into relief'" [85] (p50). Experiences are thus the result of reflecting on one's own stream of consciousness, and as a consequence: "only a past experience can be called meaningful, that is, one that is present to the retrospective glance as already finished and done with" [85] (p52).

Experiences and actions may be meaningful, and these meanings may change over time. Moreover, what seems insignificant at one time, may become important when reflected upon later, and with a different set of practical interests [85] (p74).

> The taken-for-granted is always that particular level of experience which presents itself as not in need of further analysis. Whether a level of experience is thus taken for granted depends on the pragmatic interest of the reflective glance which is directed upon it and thereby upon the particular Here and Now from which that glance is operating. ... a change of attention can transform something that is taken for granted into something problematical.

Of course these experiences may include perceptions of another actor, and their actions. But, the real problem is how one could use the above building blocks to describe a theory of inter-subjective understanding: how two or more people can manage to share and communicate common experiences of the social world. For Schutz, this is a very practical matter. Although actors will not have exactly the same experiences, or attribute the same meanings to those experiences, this is not important because in practice they are able to assume that those experiences are for all practical purposes the same. Schutz calls this the: "general thesis of reciprocal perspectives," which is outlined as follows [86] (v1, pp11-2).

Common-sense thinking overcomes the differences in individual perspectives resulting from these factors by two basic idealizations:

i) The idealization of the interchangeability of the standpoints: I take it for granted – and assume my fellow-man does the same – that if I change places with him so that his "here" becomes mine, I shall be a the same distance from things and see them with the same typicality as he actually does; moreover the same things would be in my reach which are actually in his. (The reverse is also true.)

ii) The idealization of the congruency of the system of relevances: Until counterevidence I take it for granted – and assume my fellow-man does the same – that the differences in perspectives originating in our unique biographical situations are irrelevant for the purpose at hand of either of us and that he and I, that "We" assume that both of us have selected and interpreted the actually or potentially common objects and their features in an identical manner or at least an "empirically identical" manner, i.e., one sufficient for all practical purposes.

Inter-subjective understanding is in Schutz's construction an achievement of the actors themselves, it is a practical matter, accomplished from within society and not enforced, for example, from outside or by prior agreements. Moreover, it provides the basis for the construction of sociological knowledge, through observing, interviewing, and so forth. However, Schutz's work was predominantly theoretical, rather than practical, in nature.

Building on Schutz's foundation, Garfinkel began a programme of empirical studies which would yield a rich and detailed knowledge of the practical social world. Schutz talks of the possibility of inter-subjective understanding. Garfinkel is concerned with the empirical study of the practical work that members do to structure the societies of which they are a part, something that he often describes as its "moral order".

This might include explicitly moral topics, such as what a society believes in relation to corporal punishment. But they may also include more subtle, tacit structures. For example, a small child who repeatedly asks "why" will rapidly discover from the exasperated reaction of an adult, that many kinds of conversation have limits of reasonableness, that there is a moral order or structure to ordinary conversation. Having discovered this about their world, a child might use it to knowingly annoy a parent, or perhaps mock the childishness of a sibling. Parents, being wise to such childish ploys, may act to restore ordinary (tolerable) dialogue, and in so doing attempt to sustain the moral order of reasonable conversation.[8] Unlike the norms of Parsons, which extrinsically exert themselves on members of a society, the moral order is an intrinsic achievement of the members themselves, in this case one which may depend on the practices of parenting and the practical skill of the particular parent involved.

Garfinkel suggests that the professional sociologist (like the child above) can also discover things about the moral order of a society. Though this may be possible through observation, he also proposes the "breaching experiment", as an "aid to the sluggish imagination" [5] (p32). These are intended to highlight the ordinary organisation of social affairs, by disrupting them in some way.

In the following breaching experiment, he highlights the kind of "common sense knowledge of social structures" [5] (p76) that members use in order to understand one another. Students were told they were participating in a trial of a new kind of therapy: they were to ask a question into a microphone, and a therapist in another room would issue a "yes" or "no" response; following each response they were to orally record their reactions; and each participant was advised to ask around ten questions in total. The therapy session was of course, a deception. After each question, the experimenter left a short pause of a random duration, then gave a "yes" or "no" response also at random. Although this sometimes left the participants with puzzling, or sometimes even contradictory responses, in every case they were able to complete the apparent therapy session.

---

[8] By coincidence [1] (ch. 9) also discusses the notion of "moral order" in relation to a domestic example, though the notion applies equally to the workplace (p278).

Despite the random answers, the recorded reflections of participants showed a concern for, and perception of, an underlying pattern. Where there were incomplete, contradictory, or inappropriate answers, these were handled by: a willingness to wait for future answers to provide clarification; or by assigning the trouble to the restrictive character of the "yes-no" format; or that the therapist had changed her mind; or by speculating that the therapist had misunderstood their question in a particular way; and so-forth. Despite the lack of detail, participants tried to assign what they perceived to be its "normal sense;" what any reasonable person in their situation would interpret an answer as referring to. The participants would interpret the meaning of an answer by using their own common-sense knowledge of the social structures that any bona-fide member of that particular society ought to know.

Drawing on Mannheim, Garfinkel calls this the "documentary method of interpretation." Actual appearances are treated as a "document of", as "pointing to" or as "standing on behalf of" an underlying pattern [5] (p78). Document and pattern elaborate each other; the pattern is derived from documentary evidence, while knowledge of this pattern informs the interpretation of the document. For Garfinkel, this form of interpretation is an unavoidable part of all acts of cognition [79] (p85).

Ethnomethodology is a form of empirical study which pays: "the most commonplace activities of daily life the attention usually accorded to extraordinary events," and thus: "seek[s] to learn about them as phenomena in their own right" [5] (p1). At the heart of such study is the notion of "accounts." Garfinkel insists that social actions are accountable, that social acts are accessible to, and observable by, other actors. It is this accountability which allows for the organisation of ordinary affairs. Furthermore, social actions are *reflexively* accountable. The accounts: "are made to happen as events in the same ordinary affairs that in organizing they describe" [5] (p1); the act, the account of that act, and the organizing qualities of that account, are inseparable, they are one and the same thing. Actors may be aware of these qualities and are able to knowingly use them in the organisation of their affairs; in this way reflexivity is an essential part of social order, rather than an inconvenience, as Parsons would have it. These accounts are a situated, ongoing, accomplishment, and their production is dependent on the competence of the members involved. This competence may well be taken for granted by other members in the domain, and moreover the fact that it is give that particular setting its distinguishing and particular features [5]. As Garfinkel puts it [5] (pp 3-4):

> In short, *recognizable* sense, or fact, or methodic character, or objectivity of accounts are not independent of the socially organised occasions of their use. Their rational feature *consist* of what members do with, what they "make of" the accounts in the socially organized actual occasions of their use. Members' accounts are reflexively and essentially tied for their rational feature to the socially organized occasions of their use for they are *features* of the socially organized occasions of their use.

Accounts are essential to ordinary social action, they are essential to the possibility of professionally studying social action, and furthermore are essential to the reports which result from such studies (by being in themselves an account). Because there is no difference in principle between lay and professional accounts Garfinkel is able to introduce the concept by discussing a number of studies of lay sociological "accounting" at work. These include the work of a suicide prevention centre in officially classifying deaths, and the clinical skills involved in reading and writing medical records.

The studies suggest some properties of accounts. They suggest that accounts are given using a knowledge of the practical purposes to which they may be put. So, a title of "suicide" may be assigned by the SPC to a death, where on the evidence available, a professionally defensible case for this title may be made. To properly assign a title it is necessary to draw on a working knowledge of the practical purposes to which a title may be put, and its adequacy for those purposes. Accounts are hence given to be sufficient for practical purposes, and for the time being. But in light of new or unforeseen circumstances they are apt for adjustment to show what those circumstances had really meant all along. For example, new evidence may lead the SPC to reopen a case and perhaps revise their account of that death [5] (p17).

Garfinkel also discusses how accounts are read and used. The way an account is understood, and what the account may mean, is determined by the documentary interpretation work of the reader. As the study of medical records suggests [5] (pp201-2):

> The documents in the case folder had the further feature that what they could be read to be *really* talking about did not remain and was not required to remain identical in meaning over the various occasions of their use. Both actually and by intent, their meanings are variable with respect to circumstances. To appreciate what the documents were talking about, specific reference to the circumstances of their use was required:

emphatically *not* the circumstances that accompanied the original writing, *but the present circumstances of the reader* in deciding their appropriate *present* use.

These studies help to show what Garfinkel means by "accountability"; they show how the practices of accountability are central to understanding the work of a particular domain; and, they indicate some properties of those accounts. However, while such studies are informative, their generalisability is seen as limited. For ethnomethodology, the practices of a particular domain are ultimately a matter for empirical investigation within that domain itself.

Ethnomethodology's study policy is further illuminated by comparing Garfinkel's study of the SPC, with the kind of statistical analysis more typical in sociology. Although both may begin their investigations with very much the same notion of "suicide", as an officially determined label for a death, their subsequent enquiries could hardly be more different. For statistical sociology, showing what suicide is really about involves finding correlations with other factors such as the religion of the deceased (as Durkheim's classic study does). For Garfinkel, showing what suicide is really about involves unpacking the social organisation of the work which led to an event being given the title of a "suicide" in the first place. Ethnomethodology's study policy is thus: to investigate social structures by seeking out the people who maintain, sustain, or construct them, and study the practices through which they do this [78] (p181-2). It is a policy which makes ethnomethodology distinctive within sociology. Indeed the theories of mainstream sociology are something to which ethnomethodology is programmatically indifferent. Ethnomethodology shows a phenomenological concern for the "gritty details" of social organisation, and ethnomethodologists see themselves as using these details to re-specify the accepted treatment of topics within mainstream sociology.

## 3.3. Conclusion

Fieldwork, of the kind considered in this thesis, is an intense activity in which the fieldworker subjects themselves to the experience of living within a particular community. Critical to the adequacy of such studies is the extent to which that fieldworker gains competence as a member of that community. They must learn the skills and terminology of their study environment, not in the sense laid out in textbooks, but in the sense in which they are actually used and made relevant to activities within that environment. This kind of study has its roots in both sociology and anthropology, where it is typically a lengthy affair, the result of which is often a lengthy written ethnography, based on hundreds of pages of field-notes. More recently the availability of audio and video recording technologies have led to additional kinds of materials being gathered from the domain.

Ethnomethodology is an orientation to the study of social order, for which fieldwork has traditionally been a primary tool. It is based on a phenomenological foundation, which focuses attention on the lived, gritty, specific details of each individual domain of study; although many have subsequently drawn parallels with the late works of Wittgenstein ([78] p2 provides discussion of this). Ethnomethodology is interested in the intrinsic orders of ordinary action within a community, and resists the imposition of extrinsic orders on observations of it, for example through use of the theories and methods of professional sociology. This methodological resistance to traditional formal sociological analysis, makes ethnomethodology distinctive within its field. It is perhaps this exclusive concern with the haecceities of each domain (the specific, particular and individual qualities which make it what it is), as opposed to topics which may be of relevance only to a community of sociologists, that motivates an enthusiasm for its use in situations of technological development.

Outlining these details is important to the thesis for two reasons. The primary aim of this thesis is to present an approach to the use of ethnomethodological fieldwork for the engineering of requirements. To do this it will evidently be necessary to show that what is proposed is based on an adequate understanding of the discipline, and furthermore that it constitutes a "proper" application of it. However, this thesis is not just concerned with the theoretical aspects of this problem, in chapter 6 it will also present a case-study in which the proposed approach will be applied to a practical example. This case-study will be the subject of empirical study and will adopt an ethnomethodological orientation. The second reason for the close methodological discussion presented here, is therefore to serve as a basis for this later empirical work.

Finally, this chapter helps to establish one of the central themes of the thesis. Ethnomethodology is methodologically opposed to the use of formal sociological analysis, and the development of generalised theories about social order. By contrast, engineering has traditionally been associated with the application of generalised theories. Evidently the two disciplines have very different concerns, and of course, just because one chooses to theorise about the behaviour of the physical world, does not imply that one must also theorise about social action. Nevertheless, this tension between generalisation and haecceity is something which will be addressed with care in subsequent chapters.

# 4. The Case Against a Positivist Philosophy of Requirements Engineering[9]

## 4.1. Introduction

The practical relationship between requirements engineering and ethnomethodology is taken by this thesis to be of great importance. To this end, chapters 5 and 6 will propose an approach that combines these two disciplines, and then show how such an approach might be applied in practice. However, practicalities alone are not sufficient to assure the role of ethnomethodology within requirements engineering. While practicalities are important, academic discussion is also sometimes theoretical in nature. Discussions that develop or draw on theory are often felt to be powerful because they will have implications for a wide range of potential future practical actions. To do this, theoretical discussion could be seen as drawing on what is felt to be fundamental or essential to the circumstances of practical actions.

To assure the place of ethnomethodology within requirements engineering it will therefore be necessary to engage in dialogues about both theory and practice. For example, chapter 5 will engage with the specific theoretical implications of its practical recommendations. However, it will also be appropriate to consider the existing theoretical discussions that relate to the use of ethnomethodology for requirements engineering in general.

As chapter 1 suggests, there has been relatively little theoretical discussion about the place of ethnomethodology in requirements engineering. Supporters of the approach,

---

such as [10] occasionally raise puzzles relating to its application, and some of these will be discussed in chapter 7. Although anecdotal evidence suggests that others may be less favourably disposed toward ethnomethodology[10], there is relatively little in the way of explicit theoretical dialogue.

One exception to this is [9] which concludes that philosophical concerns may make it impossible to consider using ethnomethodology in requirements engineering. This chapter will seek to refute this conclusion. In part this is motivated by a clear need to defend the central aim of the thesis as a whole. However, also as important will be to take [9] as the strongest example of a position which may be taken by others within requirements engineering, albeit from time to time, perhaps tacitly, and in a less clearly expressed form. Refuting [9] is thus a way of informing similar but more nebulous positions within requirements engineering.

[9] claims that: "the traditional RE research paradigm, in common with most engineering research and practice is founded on the philosophical tradition of positivism." It goes on to justify this by suggesting that: "requirements engineers being members of the culture of science, often fall back on and operate from a *tacit* positivism". It is a stance that has a certain intuitive appeal; the role of abstractions, the status of science, and the lack of formal philosophical reflection, makes its suggestion that positivism may exist as a dogma of requirements engineering seem quite plausible.

---

[10] For example, the author was once told by a Requirements Engineering Doctoral Consortium that there were many in the requirements researchers who considered research into the use of ethnomethodology to be a waste of time.

Requirements engineering is of course a practical endeavour, and there is hence reason to be sceptical toward philosophical discourse which may not at first seem to contribute directly to this. Nevertheless, as Popper puts it: "we all have our philosophies, whether or not we are aware of this fact, and our philosophies are not worth very much. But the impact of our philosophies upon our actions and our lives is often devastating." (p33, [89]) Philosophical arguments need to be taken seriously, but equally well their presence needs to be carefully justified. In this instance accepting the claim that positivism is fundamental to requirements, could have significant methodological consequences for the field as a whole.

[9] asserts that positivism is a fundamental, and hence unavoidable part of both research and practice in requirements engineering (although they point out that these views may be only tacitly held). The assertion is made clear by the paper's central sections, in which they discuss the status of naturalistic enquiry (in which they include ethnomethodological fieldwork) as a requirements method. A comparison is made between the principles of naturalistic enquiry, and those of positivism. This highlights a philosophical tension between the two, which is proposed as significant because: "accepting that the two approaches are irreconcilable would mean that they could not coexist effectively and that the RE community must repudiate one or the other" ([9], p125). The implication of this is clear: positivism is foundational, and any technique one might wish to use in the engineering of requirements must first be reconciled with its commitments. If positivism *were* foundational to requirements engineering, then this would certainly be a major difficulty for the application of ethnomethodological fieldwork.

Part of the justification given in [9] is that positivism, whether tacit or otherwise, is *believed* to be fundamental by requirements engineers. This proposition may be difficult to substantiate. The fact that requirements researchers do not perhaps understand requirements practice as well as they ought, has already been noted [90] (indeed some have even gone so far as to accuse researchers of "armchair engineering" [91]). Nevertheless, irrespective of its actual credibility, the very fact that such a claim is being made by researchers motivates an interest in its philosophical plausibility.

This chapter will therefore argue to the contrary. Based on a more careful examination it will show that positivism does not in fact turn out to form an acceptable philosophical foundation for requirements engineering. The first consequence of this will be that the implications for using fieldwork presented in [9] may be safely disregarded. The second will be to clear the philosophical ground for future attempts to engage with the foundations of the field.

The chapter is structured as follows. Of necessity, it begins with a clear and thorough overview of positivism as a form of philosophy. From this, four particular stances are identified and their relevance to requirements engineering explained. Sections 4.3-10 then present the case against positivism. One of the justifications that [9] gives for positivism, is that the perspective has been inherited from engineering as a whole. Section 4.3 shows that, although engineering has traditionally been seen as an applied science, more modern understandings reveal science to be only one of many different kinds of essential engineering knowledge. As a consequence, it would be wrong to see engineering as fundamentally positivistic. Sections 4.4 and 4.5 both examine potential similarities between the philosophy of positivism and some superficial characteristics of requirements engineering. Section 4.4 shows that, although the demands of software engineering may appear to suggest a positivistic view, in reality the practices of modern software engineering fail to correspond, either ontologically or epistemologically, with this stance. Section 4.5 contrasts formal methods for software engineering, against those for requirements engineering. It shows that while formal methods for requirements engineering might initially seem to be one of the easiest techniques to associate with positivism, under scrutiny this association also turns out to be misplaced. Sections 4.6 and 4.7 both consider how positivism relates to the activities and opinions of requirements researchers. Section 4.6 suggests that the orientation of naïve positivism fails to allow engagement with, what leading requirements researchers see as one of the most fundamental challenges in requirements engineering today. Both sections 4.7 and 4.8 consider scientism. The former examines scientific approaches to requirements research and shows that while the status of science is sometimes unnecessarily privileged, in general, "scientific" and "non-scientific" forms of research are taken to be equally valid. The latter looks at the role of science in requirements practice and concludes that, at best, it would be premature to privilege "scientific" over "non-scientific" approaches. Section 4.9 considers how arguments from the field of Science and Technology Studies might be used to refute a positivist stance in requirements engineering. Finally, section 4.10 examines a critical weakness in the positivist position: its inability to fully engage with claims of value, something that is an essential part of engineering requirements.

## 4.2. The Philosophy of Positivism

Halfpenny identifies no less than 12 separate but related forms of positivism ([92], p114). Thus, in order to arrive at a proper understanding of this philosophy, it will be necessary to locate it within a broader historical context before identifying those aspects of relevance to RE. The term has its origins in the writings of Comte (1789-1857). He rejected the negative critiques of the enlightenment and instead sought a constructive, scientific era where all phenomena could be understood in terms of invariable natural laws. It became a philosophy of wide-ranging influence, not least in the philosophy of science.

Links with science and mathematics makes it unsurprising that some have associated positivism with software engineering [93-95], requirements engineering [9], and related forms of research [96]. So, in order to be more precise about the ways in which a movement as broad as positivism relates to requirements engineering, and to better understand what Potts and Newstetter might mean by the term, four specific views associated with positivism are outlined below.

### 4.2.1. Logical Positivism

Logical positivism is one positivist stance which has been associated by some with the activities of software engineering [93-95]. Philosophically, it characterizes the standpoint of a group of philosophers, scientists and mathematicians, who in the early 1920s became known as the Vienna Circle [97]. In broad terms the group's position was allied to the work of Hume (1711-1776), Mach (1838-1916), Frege (1848-1925), Whitehead and Russell's Principia Mathematica (1910-13), and Wittgenstein's Tractatus (1922). Logical positivism is perhaps best described as an epistemological movement ([98], p29), which sought a new foundation for a variety of fields. Of these, Carnap's (1891-1970) attempts to find a new foundation for physics were particularly influential. Central to Logical Positivism is the notion that statements are only meaningful if they are capable of verification. Statements may be either "analytic" or "empirical"; the former, when expressed in a logical calculus, can be verified by inspection; the latter by experiment, measurement and evidence.

Popper's (1902-1994) view of scientific theories owes much to the work of the Vienna Circle, although Popper counted himself as one of logical positivism's greatest critics. The hypothetico-deductive method of scientific enquiry which he helped to popularise became perhaps the best-known of all those in the positivist project. Rather than verifying laws Popper's view of scientific enquiry was one of attempting to falsify hypotheses [99]. However, the origin of these hypotheses is sidestepped as an irrelevancy: "Their source is a psychological or sociological matter, beyond the bounds of the philosophy of logic of science, which restricts itself to analysing the justification for rejecting or retaining whatever hypotheses scientists entertain, regardless of how they come to entertain them" ([92], p101).

The philosophy of science was, of course, not the only field of interest to the Vienna Circle. It was felt, for example, that philosophy itself should: "be replaced by the logic of science" ([97], p24); economics, sociology, historical analysis, and even poetry, all came to be the subject of similar foundational re-specifications. Carnap used the term "scientific empiricism" to refer to this wider positivist movement with sympathies for the empiricism of Hume, an admiration for the scientific method of 19th century physics, and a willingness, like Russell's and Wittgenstein's, to apply logical analysis such as that pioneered by Frege [100].

Within the philosophy of science from the 1920s until around 1950, it would be difficult to understate the influence of logical positivism. However, during the 1950s alternative views began to surface from prominent figures including Toulmin, Kuhn and Feyerabend [101-103]. Logical positivism is also regarded to have failed as a wider philosophical movement. One of its difficulties rested with the principle of verification. This suggested that only verifiable statements are meaningful, yet since the verification principle is not itself verifiable, it is by its own standards meaningless. Similarly, while Wittgenstein's Tractatus had initially been seen as a shining example of the ideals of logical positivism, close reading by the Vienna Circle revealed significant differences, and his later work famously repudiated even this position.

### 4.2.2. Scientism

Logical positivism is perhaps best regarded as an historical movement. Nevertheless, many of the themes it raised are of relevance in contemporary academic debates. For example, the logical positivists gave great value to scientific knowledge, intending to "lift" other fields like economics, sociology, philosophy, and so on, up to the same level. The relationship between art and science is still very much a matter of discussion, and as a consequence the neologism "scientism" has been used to describe: "the belief that science, especially natural science, is the most valuable part of human learning", or "the belief that science is the only valuable part of human learning, or the view that it is always good for subjects that do not belong to science to be placed on a scientific footing" ([100], p1).

Science is undoubtedly afforded an important status in requirements research. Certainly much of the literature seems to draw on science in some way; [104] is a good example, having been awarded best paper at RE'05. It introduces a case study where requirements are gathered for software that is to be personalised for cognitively disabled users; the users are consequently: "assessed by psychology-based questionnaires and tests to measure cognitive, physical and perceptual abilities" (p21). Furthermore, there is also a sense in which scientific evidence is preferred in the validation of requirements research. For example, the RE'06 call for papers specifically requests: "scientific evaluation papers", which should: "evaluate existing problem situations or validate proposed solutions with scientific means." A scientistic account would suggest that requirements engineering *ought* to be a scientific activity; it would privilege science over non-science; it would urge researchers to go further in finding applications for science, and developing new science to capture the phenomena of relevance to the field; it would suggest that it is simply the immaturity of the current state-of-the-art that prevents requirements from being a science.

### 4.2.3. Naïve Positivism

While science may be considered important, many requirements activities seem to occur without it, and yet still appear positivist in nature. One term which may help to describe this is "naïve positivism", which is perhaps best understood as a label for the tacit positivist standpoints that still remain in a number of fields. For example, in the social sciences it is often used to refer to a belief in the objectivity of theories: "in the social sciences today there is no longer a God's-eye view that guarantees absolute methodological certainty. All inquiry reflects the standpoint of the inquirer. All observation is theory-laden. There is no possibility of theory- or value-free knowledge. The days of naive realism and naive positivism are over" ([105], p3; for examples of its use in other fields see: [106], p6; [107], p57; [108], p9).

In requirements, the term 'naïve positivism' might be most relevant to the activity of modelling. For example, it is often the case that a model is made of the problem which is to be solved. This usually involves the use of typical engineering skills like abstraction and decomposition. Naturally, the fidelity of these diagrams will be dependent on the skill of the analyst – what is particular to naïve positivism is a belief that this activity is unproblematic. The identification of abstractions within the world is neither one that is worthy of special methodological support, nor of investigation as a topic of research in its own right. Naïve positivism in this context holds that formal structures and theories can be unproblematically identified within the world, by the analyst. Scientific justification of such structures is thus unnecessary.

### 4.2.4. Realism

Realism is an ontological position which takes things in the world to exist, and allows those things to have properties, independent of the observer and irrespective of their beliefs, conceptual schemes, or linguistic practices. It is a stance that is relevant to any number of areas. For example, science is concerned with investigating the properties of the natural world; a realist stance would take the objects under study within the world to exist independently of the individual doing the investigation. Similarly, ethics is concerned with the study of value or quality; a realist stance on ethics would take moral facts to actually exist, and be objectively true independently of the beliefs of any particular person. Realism remains the subject of much debate within philosophy, and it is accepted that one may adopt a more-or-less realist stance toward individual issues; for example, it would be regarded as philosophically consistent to be a realist for the purposes of scientific investigation, yet anti-realist with respect to moral values [109].

The relationship between positivism and realism is not simple, as the terms are sometimes conflated, or used together (as the citation from the previous section does). Moreover, logical positivism, one of the best developed forms of positivism, is generally anti-realist ([110] provides an explanation of this tension). Despite huge differences, Popper and Durkheim are both thought of as positivists [92], and both adopt aspects of realism. Popper argued for a realist view of both the objects of scientific study, and the scientific facts that such study produces. Scientific facts exist in the world, ready to be discovered, independent of the enquirer, through a process of conjecture and refutation. Knowledge is therefore really "out there" in the sense that: "we can make theoretical discoveries in a similar way to that in which we can make geographical discoveries" (p74) [89]. Similarly, for Durkheim (1858 - 1917) one of sociology's founding fathers, social facts were objectively and ontologically real, and able to exert their influence on the behaviour of a society. It is relevant to consider realism as part of this discussion on positivism because it represents an important strand in the general positivist attitude, and especially the naïve positivism considered above.

For requirements engineering the realist stance may be relevant in predominantly two respects. Firstly, one may choose to be a realist with respect to domains. For example, a requirements domain model might be taken to be a reflection of an ontologically real set of objects, which exist independently of the author; thus any model of that domain will for all practical purposes be the same, irrespective of the author.

Secondly, one may choose to take a realist stance toward requirements themselves. Under this view, interviews and conversations with a stakeholder are to be considered as clues as to the true nature of the requirements, which are really "out there" independently of the particular stakeholder or requirements engineer ([7], p1).

## 4.3. Modern Understandings of Engineering

As Rogers suggests, engineering is an overwhelmingly practical field, traditionally much more concerned with worldly outcomes than philosophical reflection ([111], p3). In this respect it is difficult to assign its activities a philosophical orientation. However, traditional views of engineering often give it a scientistic rendering. Such views take engineering to be a derivative field, where the production of new knowledge is the preserve of science, and engineering is hence concerned with realizing the "implications" of such discoveries ([112], p148; but also pp 177-185, and by definition in [113]). Commentators too highlight a relationship with science. Petroski, for example, describes how: "Engineers hypothesise about assemblages of concrete and steel that they arrange in the world of their own making. Thus each new building or bridge may be considered to be a hypothesis in its own right" [114]; his descriptions have clear parallels with the positivism of Popper's falsificationism.

This simplistic characterization is challenged by more recent academic work such as [115-117]. Through historical case studies, they present a more realistic picture of technological development. Vincenti in particular begins to describe an independent epistemology for engineering; he shows engineering to be a distinct field of knowledge, neither solely dependent on science to provide it with theories, nor on the scientific method as a means of generating its own. This research shatters the scientistic view of engineering.

For example, research methods like destructive testing provide vital information to engineers, yet would have little standing within the scientific community (p232). Another source of knowledge is to be found in the analogies of engineering design; by way of illustration, he describes how engineers think about the "directional stability" of an aircraft by analogy with a weathercock; so called "weathercock stability" is then a concept that can help both engineering design and engineering education (p221). To choose one final example, he also highlights the importance of practical knowledge (p217):

> Theoretical tools and quantitative data are by definition, precise and codifiable; they come mostly from deliberate research. They are not, however, by themselves sufficient. Designers also needed for their work an array of less sharply defined considerations derived from experience in practice, considerations that frequently do not lend themselves to theorizing, tabulation, or programming into a computer...  [a] seemingly simple but actually complex example is knowing the clearance that must be allowed for tools and hands in putting together assemblies for which the worker must reach inside. Such knowledge is especially difficult to define, since the requirement depends in part on the spatial relationship of the parts of the assembly and the position required of the worker. Knowledge of this kind defies codification, and a mock-up or prototype must often be built to check the designer's work.

Vincenti shows that "real" engineering work consists of much more than simply applying science.

A naïve positivist view of engineering is also challenged by recent research. Bucciarelli's anthropological study describes how engineers work with "object worlds" [118]. These structured, hierarchical worlds (p86) often involve representations that give engineers a view of the world with which they can work (p67). The parallels between naïve positivism and "object worlds" are deliberate, enabling him to demonstrate the contrary. Far from being naïve positivists, he describes the care with which engineers construct their abstractions, and the deeply social nature of this work.

[9] takes positivism to have a foundational role in engineering; the kind of accounts presented above give significant reason to doubt this assumption. But more than this, they suggest the tantalizing possibility that similarly sophisticated accounts of requirements engineering may be possible.

## 4.4. Software Engineering and Requirements

[9] claims that requirements engineering is founded in positivism because: "it is compatible with the need to capture and tie down clear-cut features that can be modelled and turned into artefacts." It is a claim that seems to imply that the philosophy of requirements should be driven by the needs of those who use such requirements; that software engineering should be allowed to exert a coercive influence. In some respects this seems quite reasonable, after all, requirements are only ultimately valuable if they are of utility to software developers. The positivist view of requirements might therefore be justified by reference to two critical properties of software engineering. Development, first of all, is often conducted at a distance from the users or stakeholders of the system under development. This means that requirements models or documents often stand in place of stakeholders themselves. Software engineering is thus deeply dependent on representations of requirements, rather than having those requirements articulated more directly. Second, and fundamental to mainstream software engineering methods, is the idea that these

requirements representations must remain stable, at the very least for the duration of a development iteration, and preferably for the duration of the project as a whole. For software engineers, the ideal representation is one that has a stable and objective relationship with a stakeholder requirement. On the surface it may seem as if software engineering is positivistic in its treatment of requirements. However, on closer examination any resemblance can be shown to be entirely superficial. Of course software engineering may see it as *desirable* to have stable objective requirements, what is at issue is whether they *actually* view them in this way. To further unpack this concern two specific resemblances will be examined. The first considers whether software engineering is positivist in its epistemic treatment of requirements. The second considers whether software engineering is realist in its ontological treatment of requirements.

Software engineering is often iterative in nature, and so requirements may be revisited repeatedly during the course of a project. Requirements might therefore be seen as parallel to scientific theories, which under a falsificationist paradigm appear to develop in a similar way. The resemblance is hence as follows: software engineering appears to demand stable objective knowledge; the positivist view treats scientific knowledge as objective and stable; and at a practical level, software engineering appears to treat requirements in a similar way to a positivist view of scientific theories. As a consequence one might easily conclude that software engineering is positivist in its epistemic view of requirements. However, a more careful examination reveals this kind of positivist ascription to be misplaced.

In Popper's conception, scientific knowledge must be falsifiable. The fact that current knowledge has not yet been falsified, and that there exists a community of scientists whose careers would benefit from so doing if it were possible, means that for the moment at least, such knowledge can be trusted; and indeed, for long periods of time, this may be very much the case. On first inspection modern iterative software development, which affords multiple opportunities to revisit requirements activities, may seem similar to Popper's view of science as a sequence of failed theories. However, the way in which this sequence is envisioned in software engineering, is radically different. Unlike scientific theories, iterative software development explicitly plans at the outset for the revision of the requirements it produces. This expectation, that the revision of requirements is inherently necessary, independent of any attempt to investigate their truth or falsehood, renders claims of epistemic similarity mistaken. Rather than ever seeing them as objectively true, it would seem that software engineering treats its requirements more pragmatically, as perhaps a succession of working approximations. This pragmatic orientation is not at all the same thing as a positivist stance, and the proposed resemblance is therefore illusory.

The second concern which will be investigated is ontological in nature. Realism is arguably part of the natural scientific attitude, and is also often associated with a naïve positivist position. A realist perspective would assert that scientific theories are representations of phenomena that are really out there in the world. It is essential to its status as "real" that each phenomenon is independent of other phenomena, and of any individual who may attempt to observe it [89]. Once again it is easy to see how parallels between the theories of science and the requirements of software engineering might be drawn. A naïve reading of the software engineering process suggests that:

requirements elicitation provides information about a domain, which is then subject to analysis in order to find the requirements expressed within it. The requirements engineer is thus akin to a "butterfly catcher". This resemblance might lead one to conclude that software engineering adopts a realist view of its requirements. But once again, a closer examination reveals this realist ascription to be misplaced.

[24] provides a mainstream view of software engineering. It describes the way in which requirements are produced in iterative software development, suggesting that after inception they become dependent on the system's emerging architecture (pp 65-69). So, stakeholders inform requirements, requirements inform architecture, and then architecture and stakeholders both further inform requirements. [24] describes how architecture is based on a whole raft of concerns, constraints, enablers, and the experience of the engineers involved. These skilled professional judgements then inform the ongoing evolution of the requirements. It is therefore clear that software engineering does not see requirements as being independent from the individuals who are investigating them, and as a consequence, that a realist stance with respect to requirements cannot uniformly be ascribed to it.

In sum it is wrong to ascribe positivist and realist positions to software engineering. The simple fact that requirements are frozen for a development iteration seems more likely to be driven by practical necessity than a belief in their objective truth or ontologically real character. Moreover, the close examination of software engineering provided above, shows that aspects of it contradict such claims. A positivist view of requirements thus contradicts aspects of that which is considered by many to be best practice within software engineering. The idea that one should take a positivist view of requirements because of supposed constraints imposed by its parent discipline is thus wholly misplaced.

## 4.5. Formal Methods and Requirements

Given their shared emphasis on mathematical deduction, it is perhaps unsurprising that some have associated formal methods for software engineering with logical positivism [94]. Moreover, given the obvious similarity between formal methods for software engineering and formal methods for requirements engineering, it is easy to see how some could assume that the latter might "inherit" this positivist foundation. However, a more careful comparison between the two shows this to be an unwarranted assumption.

Although formal methodists sometimes view their craft as being an abstract mathematical one, at a philosophical level they are nevertheless concerned with reasoning about a physical phenomenon – the behaviour of computing hardware.[11] They are often concerned with proving that one formal model is "equivalent" to another, a task which therefore aims ultimately to prove that the two models represent two equivalent sets of physical computational behaviours. Each step of such a proof might consist of an application of a calculus rule, each rule might relate to the behaviour of a programming language element, whose behaviour ultimately depends on the behaviour of a physical computer processor. The validity of the proof step is dependent on the validity of the calculus rule, whose verification therefore ultimately depends on the science of semiconductors. The justification of proofs in formal methods for software engineering is thus epistemically founded in science. This combination of empirical science and logical calculus, makes it easy to see why some have associated formal methods for software engineering with logical positivism.

---

[11] It is a point notably discussed by Fetzer [119], and although many disagreed with his conclusions, even his critics acknowledged that formal methods for software engineering are a form of applied mathematics concerned with reasoning about physical systems; [120] provides an excellent summary.

The story of formal methods for requirements engineering is, however, quite different. The goal-oriented work of van Lamsweerde *et al.* is particularly interesting as a contemporary approach to developing formal requirements models [35]. Although such techniques have similarities with their counterparts in software engineering, sharing for example the use of predicate calculus, the types of worldly claim that may be made against these models are quite different. For example one might wish to "prove" that a set of goals are sufficient to meet the needs of a group of stakeholders. To do this one might proceed in a structural-induction-like fashion over the goal hierarchy, so that at each step of the proof, a justification is given showing that a set of sub-goals is sufficient to meet its parent. Take the train control system case study [121] (p231), and consider the goal:

Maintain[worst-case-stopping-distance-between-trains-on-same-track]

and its three sub-goals:

Maintain[safe-accelerate-command-of-following-train],

Maintain[worst-case-response-of-following-train-to-accelerate-command], and

Avoid[backward-train].

Evidently the rail network as a whole does much more than these three things to ensure that worst-case-stopping-distances are maintained; closing down the rail network during extreme weather conditions being just one. Of course, what is important here is that these three goals are considered sufficient within the context of *this* particular rail safety system; justifying the sufficiency of these goals is therefore tantamount to justifying this system's boundary. A justification of sufficiency is consequently likely to involve: what the project can afford to develop, what the current technologies can provide, what the rail network should reasonably be expected to do to maintain stopping distances, and how this system might relate to the other safety mechanisms that are, or could in the future be, in place within the domain. Justifications of this kind are therefore likely to depend on a complex and interrelated mix of domain, technological and project management expertise.

Of course, it is not proposed that the above style of "proof" reflects real requirements practice; rather, the comparison is intended to highlight the epistemological differences between formal methods in software engineering and requirements engineering. In the former, epistemic claims are justified with reference to a stable, mature branch of science, which is effectively reused time-after-time. In the latter, epistemic claims are justified by drawing on the products of requirements elicitation, which provide a once-off foundation that must be constructed uniquely for each and every project. This distinctive epistemological difference makes the idea that formal methods for requirements might simply "inherit" a positivist foundation from their counterparts in software engineering seem quite inappropriate.[12]

---

[12] Of course, even the basic association between logical positivism and software engineering raises concerns. As section 3 indicates, there are many aspects to real engineering that a positivist perspective may find hard to account for.

## 4.6. Naïve Positivism and Requirements Research

Consider again the goal-oriented example presented above. For the naïve positivist the process of eliciting stakeholders' needs, which may at first be presented informally, and representing these in terms of a goal hierarchy, which may subsequently lead to the production of a formal specification, would be seen as a relatively unproblematic task. For the naïve positivist formal structures and abstractions can be straightforwardly identified by the requirements analyst, without the need, for example, of scientific justification. However, this unproblematic transition from informal to formal would seem to deny much of the literature in requirements that problematizes the topic.

Reconciling the formal and the informal would seem to be one of the critical challenges in the practical production of formal requirements specifications. Indeed, as many leading figures suggest, reconciling the formal and the informal may also be one of the critical theoretical challenges for requirements engineering as a field. For some it is a matter of when this formalization should occur. As Parnas [122] suggests:

> formal descriptions techniques have received considerable attention in RE research, but have not yet been widely adopted into RE practice. Since RE must span the gap between the formal world and software behaviour, the key question over the use of formal methods is not whether to formalize but when to formalize.

A call that is echoed by van Lamsweerde [37]

> constructive techniques are needed to guide requirements engineers in the incremental elaboration and assessment of requirements. In particular, one should clarify when and where to shift from informal, through semi-formal, to formal.

For others the matter is more fundamental. Goguen for example, in emphasising the balance between formality and informality, describes the very activity of requirements itself as a process of formalization [123]. For Nuseibeh and Easterbrook, bridging the divide between the formal and the informal is in fact one of their six major research challenges for the next 25 years of requirements engineering [122]. To adopt the position of naïve positivism would be to dismiss these challenges as a triviality. Yet, for requirements engineering it would seem that it is this stage of the process that is both most critical and of the greatest interest. Naïve positivism simply fails to account for an engagement with such concerns.

## 4.7. Scientism and Requirements Research

Science has an obvious and important role in requirements engineering. However, scientism is an over-valuing of science, of assigning too high a status to science at the expense of other kinds of knowledge; clearly one may still value science without being scientistic. To investigate this form of positivism it will be worth treating the role of science in the activity of engineering requirements, and the role of science in requirements engineering research, separately.

A recent paper [124], discusses the status of science within requirements engineering research. The authors favour a "broad church" view of scientific research, which they define as: "a social practice concerned with the production of claims of knowledge through a process of inquiry in a way that is: 1. relevant… 2. systematic… 3. transparent…" It is a view that would certainly allow one to count a wide variety of requirements research as scientific, and would thus give the field a reassuringly scientific aura. However, this "broad church" definition is so broad that it could arguably include research in medieval history or theology, which seems unsatisfactory. This kind of gerrymandering seems indicative of a scientistic sentiment amongst at least some within the requirements engineering research community.

However, this is by no means universal. Many papers in requirements engineering combine "science" and "non-science" in a very unproblematic way. [125] for example presents techniques for creativity in requirements and cites models from cognitive and social psychology as the basis of its approach. These models are then used to inform the organisation of facilitated stakeholder workshop sessions. Science is therefore being used to inform the design of a requirements method. However, the contributions (discussions, ideas, requirements, and so forth) that stakeholders make as a result of such a workshop may be many and varied. Some contributions may take the form of scientific facts, such as safety parameters or physical properties that are specific to the domain; of course, other contributions may not have such a scientific basis. However, even when a scientific fact is produced by a stakeholder, this is not scientific proof that it necessarily forms a system requirement. Rather the relevance of such a contribution is in practice likely to be dependent on that participant's own understandings of the requirements process. Not that this is cited as a problem, on the contrary, it would naturally be the job of subsequent requirements analysis to verify the results of such a workshop and deal with any resulting inconsistencies. The point is simply that this requirements process, although itself informed by scientific principles, appears to hinge on the facilitation, production, structuring, and management of qualitative material. Science hence informs the method, but following the method is still a pragmatically oriented activity, perhaps not dissimilar to that revealed by the studies of engineering presented in section 4.3. Given the unproblematic way that science and non-science are used, and the fact that this paper is quite typical of requirements research, it seems difficult to build a case for a universal scientistic movement within requirement research.

Indeed, although the RE'06 conference call for papers requested scientific evaluation papers, there is an increasing recognition both in requirements engineering and related fields, of the value and rigour of both quantitative and qualitative research methodologies. [42] provides a classification of the variety of different sorts of research that makes up requirements engineering, and the different orientations, both scientific and non-scientific, that this may entail. Similarly, while [126] calls for a more systematic approach to requirements evaluation, non-scientific approaches like ethnography are proposed as being a part of the solution. More widely, computer science [127], and information systems [128, 129], have both noted a shift during the nineties toward a less dogmatically positivistic view of research method. In sum, although manifest from time-to-time, the rhetoric of scientism seems increasingly irrelevant to requirements engineering research.

## 4.8. Scientism and Requirements Practice

Although section 4.6 seems to neglect the role of science in requirements engineering, this is not to deny its importance. Clearly the rail safety example in section 4.5 would have made use of data whose origins might be found in science. However, as was outlined, these take their place amongst the complex array of opinions that justify any statement of requirements. This places science at the periphery of the requirements process, as just another form of justification, rather than as part of the very methods of engineering requirements itself. Literature by expert practitioners, such as [28] for example, give little or no attention to science and the scientific method. Nevertheless, there is requirements research which places science right at the heart of requirement practice.

[104] uses psychiatric research to measure the cognitive abilities of disabled users, and hence develop a scientifically-based, personalized user profile. Personalized system requirements are selected for each user on the basis of their profile, and this is described as a process of "satisficing" (p23). It is a term which may certainly be given an informal interpretation. In this instance the task of finding an absolutely "optimal" statement of requirements may be hard or impossible to do, however, it may still be possible for the skilful requirements engineer to find a satisfactory and sufficient solution. In this informal sense, the method presented uses scientific measurement to establish certain properties about the current state of the world, but then the skill and creativity of an engineer are necessary in order to formulate a statement of requirements, and thus make a step into the future.

However, the original sense of the term "satisficing" is somewhat more formal, and is often used as a reference to design science. This movement, drawing on [130] seeks: "an explicitly organised, rational and wholly systematic approach to design: not just the utilisation of scientific knowledge of artefacts, but design also in some sense as a scientific activity itself" [131]. [130] presents a science of design, which Simon intends to be taught as part of the discipline of engineering. His aim is to remove design's dependence on experience and judgement (p135), and make it a scientific discipline. Critical to this is his reductionist view of the human condition, very much aligned to his pioneering work in the field of AI, that human intelligence is comparable with a physical symbol processing mechanism (p23). Reductionism is often associated with positivism, and its unity of science agenda. However, there is also an obvious scientism to Simon's aims, viewing a scientific formulation of design as inherently superior to a non-scientific one.

It is a movement that has been controversial even within design itself. [132] (quoted in [131]) suggests: "most opinion among design methodologists and among designers holds that the act of designing itself is not and will not ever be a scientific activity; that is, that designing is itself a nonscientific or ascientific activity." [133] also presents a view of professional design which is intractable to an applied positivist science. In [130] the claim that design can be a science rests on a stated assumption that the designing human mind can be considered as a computing machine. This assumption is critical to the view of design as science because it means that the process of design can be reduced to a set of logical rules, which may be discovered through scientific research, and then used like scientific theories in the production of further designs. But the idea that the conscious mind can be compared to a system of formal logic is contentious and has been argued against, for example by the thought experiments of [134], using physics and mathematics in [135], and by drawing on later Wittgenstein and Ryle in [136]. The status of Simon's positivist design science should therefore also be considered at the very least to be contentious. Even if one abandons a reductionist conception of the activity of design, the notion of design as a science still seems problematic. At a practical level some have suggested that the activities of science and design are actually quite dissimilar [137]. Of course it may be possible to see some parallels with science, however, these seem no more compelling than equally plausible parallels with other rigorous professional disciplines (such as law, for example). Thus, without the contentious positivist conception of design, one is forced into adopting an absurdly broad view of science in order that design might be considered a part of it. Of course this is not to suggest that design science isn't capable of formulating useful and practical guidelines, merely, to highlight the scientistic aspirations of its programme.

To conclude, the above alludes to three possible formulations of requirements engineering practice as science. The first takes the aforementioned "broad church" interpretation of science, thus ensuring that everything necessary to engineer requirements counted as a science. But this seems unsatisfactory. The second would be to mould requirements practice into a strict view of the scientific method, such as the hypothetico-deductive model, and then deny the importance of any methodological knowledge that fails to fit. However, studies such as [117] suggest that this may ultimately be to the field's detriment. The third would adopt a reductionist view of the activity of engineering requirements itself, much as design science does toward the activity of design. However, such a move is at best contentious. In sum, it would be decidedly premature to privilege scientific conceptions of requirements engineering practice over "non-scientific" alternatives.

## 4.9. Science and Technology Studies

Though influential at one time, Positivism is no longer taken seriously in the philosophy of science. Particularly influential in the turn away from positivism was [102]. Kuhn argues that, rather than seeing science as engaged in a steady progress toward truth, it may be better conceived of as a sequence of paradigms. Thus, while there are certainly periods of "normal science," these are interrupted by episodes of revolutionary change. Kuhn's view is of a science organised around communities of ideas and practices, rather than around an idealised view of the scientific theory.

The differences between Kuhn's paradigms, and Popper's falsification, are put forward particularly clearly in chapter 4 of [138]. However, [138] is concerned with much more than comparing two philosophies of science; it suggests that these epistemological positions cannot be fully understood without looking at the ideological, cultural, and social concerns, in which they are embedded. This so-called "strong programme in the sociology of knowledge" argues that the content of scientific knowledge, far from effortlessly transcending the circumstances of its production, can in fact be legitimately understood in sociological terms. One of the critical arguments within the strong programme relates to the role of logic and reason in science and mathematics (for instance, see [138] ch 6-7, or [139]). The following examples (which draw on [140], [141] pp45-48, and [142]) illustrate this concern.

Consider the following hypothetical discussion between a mathematician and a theologian. The theologian is asked to accept that: "all men are mortal" and that: "Socrates is a man;" which he does. The mathematician then urges him to accept by deduction that: "Socrates is mortal." Suppose the theologian doubts this conclusion.

The mathematician could use predicate calculus to reformulate the issue, and point to the rules of a logic which necessitate her conclusion (such as those presented in [143]). Further doubt might result in the calling of a third party, to verify her working. If the theologian persists in his scepticism, he is likely to irritate the mathematician; his intelligence or good character may even be brought into question. The point is that although a rule of logic may *allow* one to draw a particular conclusion, it does not in and of itself necessitate action. Rather, it seems more reasonable to say that it is the social context of the occasion of that logic's use, which impels a particular conclusion.[13]

Suppose the theologian instead engages in a different kind of scepticism, suggesting not that the rules aren't sufficiently persuasive in themselves, but that they are not sufficiently complete for him to apply. Before the theologian may conclude that X is mortal, he must be able to discern that X is in fact a man, and since the mathematician has provided no procedure for him to discern this, as far as the theologian is concerned, her rules are incomplete. Indeed, even if the mathematician provides further rules intended to unambiguously establish whether or not X is a man, these too are susceptible to similar scepticism. Whenever a rule is supplied it is always possible to ask for a further rule which outlines exactly how that rule is to be applied, forming a potentially infinite regress. Once again, the mathematician will have cause for exasperation: the theologian's claim not to be able to identify a man seems ridiculous, surely any reasonable member of society should be able to do this? While an infinite regress is certainly possible, in practice any chain of rules eventually reaches an end, usually with something that is deemed to be incontrovertibly commonsense to the users of that rule. So, just as in the previous example, the logic is not in itself complete: to fully understand what is meant by a rule, one must also understand what is deemed by the users of that rule to be incontrovertibly commonsense.

---

[13] Garfinkel's breaching experiments ([5] ch 2) provide an powerful illustration of the way in which such persistent denial is ordinarily held to moral account.

Returning to the example for the last time, suppose the theologian does eventually accept that "Socrates is mortal." The mathematician might then list other men that the theologian ought to accept as mortal. Suppose this list includes Jesus Christ; he was a man, does the logic therefore compel the theologian to accept that he is also mortal?[14] Evidently the *logic* compels no such thing, and in retrospect the pair might agree that the rule did not apply in this circumstance, or that the rule itself should be revised to account for this case. Thus, while it may initially seem as if the logic compels a certain conclusion, in fact the reverse seems more reasonable: that given a particular instance, the pair first discern a "correct" conclusion, and then account for how the rule relates to it. In their actual everyday use, rules consequently have a *post hoc* character and might be more profitably seen as a resource for action, rather like [54]'s understanding of a "plan".

This kind of scepticism is often taken to be derivative of Wittgenstein's work on rules and rule following. Informed by these kinds of insight, Kuhn, the strong programme in the sociology of knowledge, and Science and Technology Studies more widely, generally take rules, logic, knowledge, and language, as phenomena which may only be understood properly when seen as embedded within a community of use.[15]

---

[14] For example, under the Christological doctrine of the hypostatic union (following the Council of Chalcedon, in 451), Christ has two united natures: one divine and one human.

Rules are obviously of importance in requirements engineering. Software systems are not only rule based constructions, but are introduced into communities which have their own rules and sense of orderliness. The implications for requirements engineering of the insights outlined above, may therefore be significant, and are evidently well beyond the scope of this chapter. However, it will be worth emphasising two conclusions. Firstly, these arguments in no sense make formalisation impossible; the fact that they often focus on logic or mathematics should not be interpreted as asserting the impossibility of these activities. Rather, (and secondly) they are provided here in order to refute positivism's indifference to the context, origin and practical usage of such formalisations.

Science and Technology Studies is a theoretically rich discipline, of which the aforementioned strong programme in the sociology of knowledge forms only one part. For example, social constructivism [145], the Social Construction of Technology [146], and Actor Network Theory [61], may all, in ways that fall well beyond the scope of this chapter, be of relevance to requirements engineering. Indeed, some have already explored such possibilities [147]. However, one does not necessarily have to reject foundationalist accounts, such as realism or essentialism, in order to reject positivism. Even without such theoretical contributions, Science and Technology

---

[15] The examples given here draw in part on the so-called 'community view' of language and rule use supported by [142]. However, although incisive, its relationship to Wittgenstein's original work has been the subject of significant debate [144].

Studies contains a body of empirical studies which may in their own right, be of value to requirements engineering.

The studies of classification in [148] present just one example. Domain modelling is in many respects similar to the way one might build a classification. However, as [148]'s study of the International Disease Classification illustrates, such activities are far from unproblematic. Through detailed study, the history of the International Classification of Diseases is traced from the 17[th] century to the present day where it forms a vital part of medical governance and technology. It would be easy to imagine that advances in modern medicine have brought stability to this classification, however, the authors conclude otherwise (p21):

> what we found was not a record of gradually increasing consensus, but a panoply of tangled and crisscrossing classification schemes held together by an increasingly harassed and sprawling international public health bureaucracy.

The study illustrates the complex human work that goes into building and sustaining classifications, and the richly textured moral properties they consequently possess. Such studies should be at the very least, an aid to the sluggish imagination.

## 4.10. "Value" and Requirements

In seeking a philosophical foundation, requirements engineering is often compared with science. This may be due to a perceived relationship between science and engineering, as discussed in section 4.3; it may be due to a perception that the methods of science hold an epistemic strength which requirements engineers should replicate, discussed in section 4.4; or, it may be because science already has a rich philosophy on which to draw. This section will, by contrast, present the view that *if* requirements engineering is to have a philosophy then it may end up looking more like a branch of ethics.

Essential to this view is the notion that requirements engineering is concerned firstly with thinking about and investigating claims of value, and secondly with concluding how the world ought to be on the basis of such claims. For example, suppose a requirements engineer must make a trade-off between two proposed requirements in order to produce a requirements document. Firstly, in order that any trade-off may be made, the requirements engineer must fully understand the "value" possessed by each proposal. Secondly, in proposing a solution in the form of a new set of requirements that requirements engineer implicitly makes a new value claim. In so far as such claims relate to the "rightness" or "wrongness" of various properties of a system, the requirements engineer's problem is a moral one. In so far as the field of requirements engineering provides approaches to resolving such problems, it is akin to a branch of normative ethics.

The relationship between value, science, and positivism was clearly delineated by the logical positivists. Hume famously highlighted the problematic status of writers who provide statements about what "is", yet conclude something about what "ought to be" as a result ([149] book III, part I, section I). This distinction between "facts" and "values," also described by G. E. Moore (1873 – 1958) and the early Wittgenstein [150], was one that the logical positivists strongly subscribed to. Furthermore, as Ayer describes [151], because matters of value or what "ought to be" could not be given a scientific treatment, they were to be regarded as literally meaningless as their expression was said to add nothing to a statement's factual claims. Since requirements engineering is concerned with investigating and proposing claims of value, it is an activity that the logical positivists would have discarded as meaningless. In this respect logical positivism is clearly not a natural candidate for a philosophy of requirements engineering.

Despite their scientistic ideals the logical positivists didn't think that value-based reasoning could be a scientific activity. Of course, the philosophy of ethics has developed substantially since logical positivism. [152], for example, shows how the later work of Wittgenstein was influential in this development. Whether or not it is philosophically appropriate to treat value, or at least the kind of value that requirements engineering is concerned with, as a topic of scientific enquiry is a meta-ethical question which lies beyond the scope of this chapter. However, one particular stance that some [79] have allied to the later Wittgenstein, will be considered.

Ethnomethodology [5] is of course familiar to requirements engineering [153]. Influenced by the phenomenology of [154], ethnomethodology is an orientation for the study of the locally produced, situated, moral order of everyday mundane human conduct. For example, the moral order of the medical domain may include decisions about whether or not to treat a particular patient; in this sense "moral order" may be quite recognisable as a form of applied ethics. But, this need not be the case, "moral order" equally well includes more mundane matters. During a lecture, for example, it is taken as a basic part of the moral duty of the audience to be quiet and allow the lecturer to speak. Should a member of the audience choose to talk in an inappropriate way, this becomes morally sanctionable, perhaps by fellow members of the audience, perhaps by the lecturer, or by those organising the event. Of course, there may be occasions when it is considered appropriate to abandon such a moral order, for example, if the lecturer were to show himself to be grossly incompetent as a speaker. For ethnomethodology, such moral orders may be widely understood, such as how to attend a lecture, or be quite specific to a particular community, such as how to make medically ethical judgements; they may be explicitly taught, as in organised religion, or only tacitly known, as in the organisation of ordinary conversation [65]. Either way, ethnomethodology is concerned with how they may be uncovered, often using careful observation and fieldwork.

The ethnomethodological view of a domain as a moral order, is thus a richer one than that provided by a scientific conception, which might describe in terms of rules, processes and structures. Garfinkel suggests that a rule-based conception of society makes its members into "dopes" (p68 [5]). By contrast he shows how members quite mundanely work reflexively with any rule, and how society is as a result, better seen as a locally and intrinsically sustained moral order. Ethnomethodology thus represents a strongly practical conception of value, which is capable of identifying issues that may be of practical relevance to requirements engineers.

For example, to understand a claim of value, such as one that a stakeholder might make when asserting (through a requirement) that a system ought to have a particular property, is to appreciate the senseful character of its justification. Certainly it may be necessary to understand the scientific facts that such a justification might use, but this is not in itself a sufficient condition for understanding that justification as a whole. To achieve this one must also understand the local moral order which makes that claim meaningful. In this regard, ethnomethodology is an analytic orientation capable of producing the sort of rigorous empirical backdrop which would allow the sense of value-claims made during requirements to be adequately determined. More than merely understanding the value-claims of others, requirements engineers must also make claims of their own. Analysing, synthesising, and writing requirements are all activities where a requirements engineer must assert value in a statement of their own making; they are bound, so to speak, to get their moral hands dirty. Naturally, the acceptability of such value-claims, and any justifications which might accompany them, will also relate to this moral order. Ethnomethodological studies can therefore act, not just as background for understanding the value-claims of others, but also as a resource for proposing new ones.

The above of course sketches only one view. It suggests a particular understanding of what value *is*; it suggests methods of empirical enquiry; and, suggests how the requirements engineer *ought* to use the results. For example, this perspective proposes that observing participants in their natural environment provides a resource that is distinct from merely listening to descriptions provided during interviews. Similarly, it proposes that stakeholders' moral orders ought to be accounted for in the production of requirements. Other orientations may propose different views or make different assumptions. Requirements engineering is in part concerned with the exploration and advocacy of such perspectives, and ultimately with considering what adequate requirements practice *ought* to consist in. In so far as a philosophical treatment of this task may be necessary, it is therefore likely to be closer to a branch of normative ethics, than to a philosophy of science.

## 4.11. Conclusions

This chapter has considered various forms of positivism: logical positivism, naive positivism, and more broadly, scientism and elements of realism. It has also considered the different ways in which these forms might be relevant to both research and practice. It has, in particular, considered these in relation to a variety of specific examples, from formal methods for requirements engineering through to the importance of "value". When considered together, they more than adequately refute any broad claim that positivism is foundational to requirements engineering as a whole. Furthermore, they suggest that the positivist perspective is at best detrimental, and at worst antithetical to the activity of engineering requirements. As a consequence, if positivism is not in fact fundamental to requirements engineering, then the problems regarding the use of fieldwork raised by [9] can be safely disregarded.[16] However, the arguments presented in this chapter can also be considered individually. When this is done they provide a catalogue of prompts that may help to inform not just the discussion of a particular paper, but a whole range of conceptual and theoretical discussions within the field.

---

[16] Interestingly, although [9] clearly asserts positivism's foundational status, it not clear that the authors are themselves positivists. On the contrary, they appear to be quite dissatisfied with the state of requirements engineering as they perceive it. The authors go to some lengths to highlight the value that ethnographic studies might bring to the engineering of requirements. That such methods are irreconcilable with what they take to be the field's fundamental principles, seems for them to be an unhappy conclusion.

# 5. Model Guided Fieldwork

The preceding chapters have achieved a number of tasks. Chapter 2 has developed a sense of how requirements engineering relates to the wider aims of software engineering, and detailed some of the existing approaches to using fieldwork for technological development. Chapter 3 has introduced ethnomethodological fieldwork and outlined some of its practical and philosophical characteristics. Finally, chapter 4 has addressed some philosophical concerns within the literature relating to the use of fieldwork for requirements engineering. So, with the philosophical ground cleared, and with a knowledge of fieldwork itself, and the previous attempts to apply it, this chapter will propose an approach of its own: Model Guided Fieldwork (henceforth, MGF).

The chapter will be composed of four main sections. The first will be concerned with establishing a definition of requirements elicitation. Since this is the activity that MGF must contribute to, it will be essential to begin with a clear sense of what this task is intended to achieve. The second section will then propose MGF itself and outline the approach (which is crudely summarised in Figure 1). As motivation, it will draw briefly on some previous experiences with fieldwork. Two discussion sections will then be provided. The first will provide a thorough methodological examination of MGF. Its aim will be to show that MGF is a methodologically consistent proposition. This thesis is concerned with one particular kind of fieldwork, one informed by ethnomethodology. It will therefore be necessary to show that using modelling, a technique sometimes associated with a positivist position, does not compromise the ethnomethodological principles of fieldwork. The second discussion section will examine the relationship between MGF and other approaches to the use of fieldwork in technological development. This chapter will close with a brief summary in preparation for an empirical case study of MGF in the next.

## 5.1. A Definition of Requirements Elicitation

The literature review has outlined the established view of requirements engineering. This typically consists of activities which may include elicitation, analysis, negotiation, and documentation, often arranged in an iterative form. Fieldwork is a form of empirical enquiry, in this respect it is a potential requirements elicitation tool and it is the aim of MGF to harness this potential. However, in order that MGF can legitimately claim to be a requirements engineering method, it will be necessary to establish a clear picture of requirements elicitation, and its purpose. Doing this will provide a firm basis for the subsequent introduction of MGF. A critical step toward this will be to show that *requirements have an inherent rhetorical quality*. Ultimately this will be done by considering some of the different kinds of rhetoric that typically surround them, however, the initial move will be to present some definitions of requirements itself.

One definition is given by Goguen [7]: "requirements are the properties that a system needs to possess to be successful in its target environment." This emphasises the technical nature of requirements, but suggests that they may imply more than this. Requirements thus express properties of a system; they exist to foreshadow the construction of that system; they should consequently describe something that is *technically viable* as a system. For example, suppose one was given a requirement for a system that entailed solving "the halting problem". Since this is a theoretical impossibility for conventional computing, no such software system could ever be constructed to fulfil this demand. Assessing the technical viability of any requirement is consequently of intense interest to the requirements engineer, whose professional duties align with the success of the project.

Although requirements are clearly technical statements, they are also more than this. Requirements must represent something of *value* to a project's stakeholders. As the IEEE definition implies (cited in the Literature Review, p16), they represent properties that will solve a problem or achieve an objective for a user; or they are a necessary consequence of an imposed rule of some kind. In this respect the statements are important because they represent something that will either be of use to those people, or alternatively, because its absence from the finished system would be considered detrimental in some way. The idea that value is inherent to the requirement seems very reasonable. Suppose otherwise, that an argument could be found to cast doubt on the value of a proposed system property, or similarly that no argument could be given to justify that property's value. In these cases that property's status as a requirement would be brought into question, and the requirements engineer might even be accused of engaging in "premature design."

Requirements also imply a mobilisation of resources. The realisation of a requirement will need resources, be it hardware, processing time, software licences, or development effort. Requirements are thus also important because they carry implications of cost and effort. Each project will have constraints on the resources that are available for the realisation of its requirements, and will certainly be judged a failure if it attempts to construct something outside of its capacity so to do. It is therefore not enough for the requirements engineer to propose requirements that are technically viable, they must also be *technically viable given the available resources*. For example, suppose a requirement implies that a system must have a petabyte of online storage. This is certainly a technically viable proposition.[17] However, given the likely cost of such a system, the requirement may well not be technically viable given the available resources.

Requirements engineering, as a professional activity as well as by Goguen's definition, is interested in the success of its projects. It is consequently interested in ensuring that requirements are *sufficiently valuable to warrant their construction cost*. Suppose otherwise, if a system was judged not to be worth what it had cost, that project would certainly be judged a failure. This kind of argument may not be critical to a statement's status as a requirement. Indeed there may be many occasions, such as during brainstorming activities, where it is proper to propose requirements without considering the difficulty of their construction. However, it is an argument that will at some point be essential to almost every requirements engineering situation.

Critical to requirements engineering is the fact that in practice, resources for the realisation of requirements are often limited. This may motivate a number of different kinds of argument relating to their value. For example, it may be necessary to demonstrate the value of a proposed set of requirements before the resources necessary for their construction can be acquired. Or alternatively, fixed resource constraints may make it impossible to implement every requirement, even though each of them may be valued by a project stakeholder. In such a situation it may be necessary to establish which combination of requirements would provide the *greatest value under the given constraints*.

---

[17]     A petabyte is presently a vast amount of storage, today only generated from projects like www.climatepredication.net, businesses like ebay, or by the scientific experiments of CERN.

The above shows that requirements are more than just system properties, they are system properties which can be argued for in particular ways. The first two classes of argument, that a requirement must be both technically viable and valued by its stakeholders, are by definition essential to the very notion of "requirement" itself. A statement cannot be a requirement unless it can be legitimately argued that it is both valued and a viable system property. However, the above also showed that requirements might be involved in a variety of other kinds of argument connected with the potential success of a particular project. For example, to show that they are worth their implementation cost, or that one requirement is more valuable than another. These arguments may not be essential to a statement's status as a requirement but they are nevertheless central to the activity of engineering requirements. These kinds of argument are an essential part of the work of establishing what a project is to build, and so engaging with such arguments is an essential task for the requirements engineer.

It is now possible to return to the notion of requirements' rhetorical qualities, alluded to above. The foregoing pages, which may almost seem to state the obvious, are intended to show that requirements possess an inherent rhetorical property.[18] It is proposed that, although requirements are technical statements, they must additionally be justifiable in a variety of ways. For example, a justification may be given in relation to a requirement's value to its stakeholders, its technical viability, or regarding the resources that may be consumed in its construction. Various examples are given above, but this list should by no means be considered exhaustive. In sum, such arguments can be seen as providing a rhetorical structure in which requirements are embedded; a structure which is essential both to their very status as requirements, and to their relationship with a particular project; a structure which they make meaningful, and are made meaningful by.

---

18      "Rhetoric" is also often used as a pejorative, perhaps to indicate words that lack substance, which is certainly not the sense in which it is used here, where it is used to indicate "persuasiveness".

Of course it is not suggested that it is always necessary to explicitly *give* such justifications, only that a requirement must always be *potentially* justifiable. Neither is it suggested that such justifications always need to be formal, absolute, or final. Justifications need only ever be strong enough for practical purposes; they need only be sufficiently convincing to proceed with the task at hand, for example, to the next step in a development process. Nor is it suggested that these justifications will always be obvious or uncontroversial. It is simply that without the possibility of such a rhetorical case, a requirement's status would be brought into question.

This rhetorical property provides an opportunity to revisit the traditional activities of negotiation, documentation, analysis, and elicitation. It is in the first of these that rhetoric is perhaps most observable. In requirements negotiation justifications will be quite explicitly aired, evaluated, and their convincingness considered. It is an opportunity for competing views on what properties the system should have, to be explored and settled. Negotiation is consequently an explicitly rhetorical task. However, while it is possible to argue for its importance in theory, this rhetorical property may be much harder to practically identify elsewhere in the requirements process.

For example, in its documented form, a requirement may visibly display no rhetorical property whatsoever, the traditional requirements document being a simple numbered list of technical properties, perhaps ordered thematically. The rhetorical character of these requirements would thus only become apparent if one of them was challenged; only then might justifications be given. Of course explicit justifications are unlikely to be provided for every requirement. It may be the case that everyone already understands why a particular system property is necessary. Or it may be that a justification is considered so trivial as to be superfluous. However, by contrast the justification of a requirement may also be complex. Omitting such a justification from a written form may therefore be a reflection of the difficulty of adequately capturing it. Justification might thus be left deliberately implicit until an explicit negotiation activity can take place.

Sometimes of course, a justification is in fact documented along with a requirement. Indeed some requirements templates mandate it, such as the "rationale" field of the Volere requirements shell [28]. This proposed rhetorical property also seems very closely related to the concept of traceability. This is commonly regarded as an important aspect of requirements methodology [155]. According to [156], traceability is concerned with linking requirements either back to the source from which they were elicited, or forward to the designs that later realise them. When the former is achieved, it is possible to see the origin of each particular requirement. This is thought to be important because it makes clear why each requirement is important to the system. Traceability therefore provides a way of making an argument for the importance of a requirement. Arguably, it is a management technique which facilitates the potential formation of a requirement's essential rhetoric. The methodological importance afforded to traceability within requirements engineering is cited here as further evidence of the methodological importance of requirements' rhetorical property.

Analysis is another vital requirements activity, and another place where the notion of requirements rhetoric is difficult to locate. Its role is in broad terms to take elicited information and propose requirements as a result. [24] (ch. 8) suggests that elicitation activities may yield requirements that are set in the language of the domain. For them, analysis is in part concerned with translating these into the language of the system's developers, so that they may form the basis for more technical activities. Following the above, any requirement which such analysis proposes must be potentially justifiable. In the case of rewriting "user requirements" as "system requirements," naturally such a justification would need to show that the latter would entail the former. Of course analysis is a complex activity, it may be creative, it may involve problem solving. Nevertheless, regardless of the exact form it takes, analysis can still be seen as a process of exploring the rhetorical possibilities of the information it is given.

This proposal, that requirements and requirements engineering have an essential rhetorical characteristic, might initially seem alien when considered in the context of mainstream requirements methods; particularly when the justifications which it insists on may be tacit, or perhaps only partially manifest. One topic in particular which may help to bring them together is the role of logic in requirements engineering. Logic, formal, and semi-formal notations form the backbone of many requirements and software engineering methods. Drawing on arguments presented in the chapter 3, it will be worth emphasising three aspects of logic here.

The first is that logic can be regarded as a kind of rhetoric in itself, and a very powerful one at that. Under certain circumstances, the presentation of a derivation based on the careful application of a logical calculus is more than just persuasive, it constitutes a proof of its conclusions. The second is that logic alone is not sufficient. Any such proof is obviously dependent on the veracity of its premises. In requirements engineering, some of these premises will include statements that must stand in some sort of relationship with stakeholders' needs or values. At this point something other than logic is necessary, even if this is simply an appeal to common sense.[19] The third is that logic is not the only form of persuasion. Aristotle, for example, considers that since a speech consists of three elements, a speaker, the content of the speech, and an audience, there are consequently three kinds of persuasion. One may persuade using the structure, logic and content of the argument presented. Alternatively, one may persuade by drawing on the good character of the speaker; for instance, a requirements engineer may use their professional status to persuade stakeholders that a requested feature will be too difficult or costly to implement from a technical perspective. Or, one may persuade by attempting to manipulate the emotional state of the audience; this might be relevant when trying to persuade a sponsor to fund development of a system feature which could positively benefit the working lives of its users [157]. This is not to suggest that Aristotle should necessarily have the last word on rhetoric. On the contrary, this rhetorical property should be a topic of empirical study in its own right.[20]

---

19      Of course to do this it would be necessary to agree on what common sense is. The ordinary and common sense is very much a topic that concerns ethnomethodology, and this is perhaps another reason why it may be useful to requirements engineering.

However, for the task at hand, it is requirements elicitation which is perhaps of greatest interest. As suggested above, elicitation might be seen as being concerned with establishing "user requirements". But in practice its locus of interest may be significantly broader. For example, it is often hard to fully understand what it is that stakeholders mean when they request a feature. Elicitation can therefore provide vital background information which can help in the analysis of their requests. It is consequently concerned, not just with eliciting requirements, but also with information that will support the analysis and interpretation of those requirements. In the discussion above it was noted that analysis is concerned, not just with the production of system requirements, but with the production of *justifiable* system requirements.[21] Moreover, it noted that these justifications were to be based, in part, on elicited information. In this regard elicitation must be concerned with providing information which will be useful in the analysis of requirements. Elicitation is therefore concerned with providing information which may be *rhetorically valuable* to the rest of the requirements process.

---

20      Through study of the work of requirements engineers it may be possible to draw out more specific rhetorical forms. However, it is equally likely that the work of persuasion may be quite particular to each individual project domain. In this regard it may be necessary to empirically establish what constitutes persuasiveness within a domain, as an element of the requirements elicitation activities.

21      Indeed, even if an elicitation activity yielded fully formed system requirements, it would still be necessary to verify this through analysis, by ensuring that suitable justifications for it may be given.

Thinking about the rhetorical properties of requirements is a way of opening up its definition to better account for the variety of activities which it typically involves. This is especially valuable for the present task. By understanding requirements elicitation as the gathering of evidence which may be of rhetorical value during requirements analysis and negotiation, a task is established to which fieldwork can contribute, and thus a foundation is laid for the role of MGF.

## 5.2. An Overview of Model Guided Fieldwork

The use of fieldwork as a requirements elicitation method creates challenges, and in order to set up the MGF approach some of these will be examined. Fieldwork can be an intensely detailed form of enquiry capable of providing a practically infinite supply of observations from a domain of interest. Of course in some senses this closeness is one of the reasons for taking an interest in the technique. But, it is equally possible to be overwhelmed by the volume of potential observations. It may therefore initially seem as if the challenge to using fieldwork effectively, is one of selecting the relevant observations from a vast pool of possibilities. However, on closer examination "selection" is only half the issue.

Blythin, Rouncefield, and Hughes [158], report on their experiences with using fieldwork in an industrial setting. When presenting their ethnography, they were often greeted with a response along the lines of: "never mind all the ethno stuff – what does all this mean and what do we do now?" Crabtree too has reported receiving similar reactions [159]. Such reactions seem to point toward a number of issues with fieldwork. The fieldworkers provide observations, but perhaps these are not adequately interpreted for the audience. It may be that the relevance of these observations is not strong enough, not made clear enough, or not sufficiently well developed. Ultimately they speak to a very practical concern for action – fieldwork observations may be valuable, but only if they indicate an action of some kind. Observations, therefore, do not by themselves seem sufficient. So, although it is clearly possible to describe the use of fieldwork as a matter of "selection" this glosses the analytic work necessary to develop relevancies between observations and the project.

Framing the problem in this way, as an analytic one, is consistent with [1] (as discussed in the literature review p32). Moreover, it also resonates with the definition of requirements elicitation given above which suggests that the aim of elicitation is to find information which may be of rhetorical value to the requirements process. The value of a fieldwork observation is thus in its rhetorical utility for suggesting and justifying possible requirements. A requirements elicitation method which seeks to use fieldwork must therefore be concerned, not just with observations themselves, but with developing arguments which link those observations to issues of importance within the requirements process. This is what MGF aims to achieve.

Of course in order to select, or even in the first instance gather, observations of relevance, it will be necessary to know exactly what they are supposed to be of relevance to. Requirements elicitation often occurs at a stage where ideas within a project are still relatively fluid. At its early stages, a great many different ideas or proposals may be entertained, even its aims may begin in quite a diffuse form. Moreover, even where clear ideas do exist, these may be in flux. Ideas and proposals from a project are thus taken and represented in model form. A potentially diffuse situation is thereby made more concrete. These models can then act as a guide to a fieldwork enquiry, whose resulting observations can be used to evolve those models in an iterative process. The models are therefore throw-away constructs; they are intended to act like a scaffolding to the process of fieldwork; they need not represent ideas that are certain or final, but rather constitute exploratory, proto-models.

The proto-models essentially inform, and are informed by, three different locales: the stakeholders, the project, and the fieldwork domain. They are laid out in the following diagram, and expanded on in the text below it.

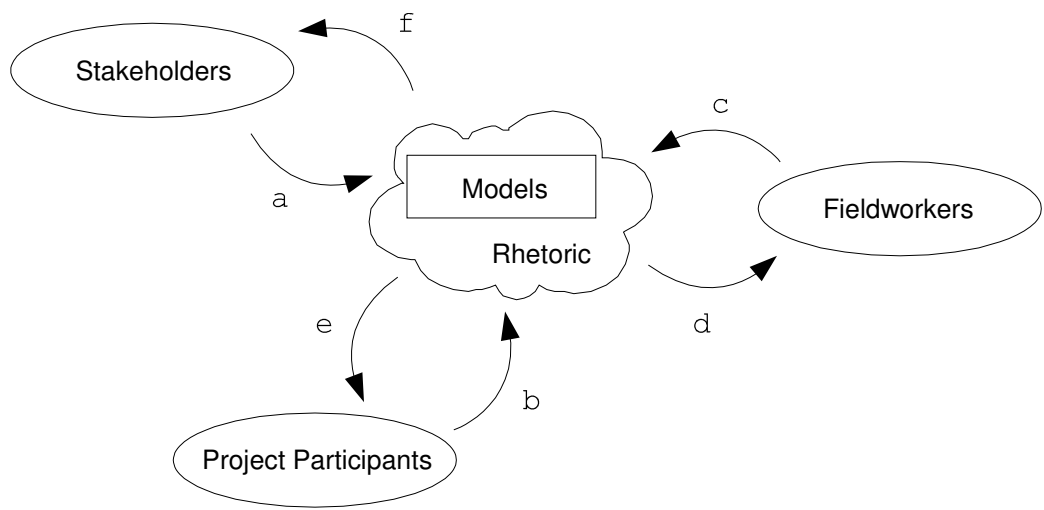

*Figure 1: An Overview of MGF*

One of the most obvious sources for these proto-models will be the ideas or demands of stakeholders (fig.1, a). This is the traditional focus of requirements elicitation activities, and MGF is no exception. It can only focus on producing relevant observations from the fieldwork domain, if the models that are being used to guide it are able to express the current ideas about the system. In this way MGF interlocks with traditional requirements elicitation techniques. These might include interviews or focus groups, to try and elicit the existing views of what the system should achieve. In the first instance it may be the case that a project has already done this kind of elicitation work independently, in which case MGF can simply use these resources for the development of models. If not, then these activities will need to be undertaken as a part of the MGF task.

Another less obvious source of insight for the proto-models are the ideas and proposals of the development project itself (fig.1, b). These may be technically or commercially motivated proposals, and driven by either the ideas of the development team, or aims of the project's management. As discussed above, there is no purpose laying out requirements that cannot be satisfied, either because they aren't technically feasible, or because the resources do not exist to realise them. Because these arguments are critical to the requirements, they form yet another topic which it may be relevant for fieldwork insights to address. For this reason they too should be captured in the modelling exercise. In the first instance, it may be the case that a project has documented its internal ideas and aims, in which case these can feed directly into the MGF modelling work. If not then it may be necessary to undertake some kind of investigation in order to draw out these aims and views, which may be more or less formal depending on the circumstances. If the analysts undertaking MGF are already very much a part of the project team, and if they already fully understand the project's thinking, then this may be more akin to documentation. However, if not then this activity may be a full investigation in its own right. It is a kind of "reverse" elicitation, where the assumptions, themes and issues of the project need to be drawn out and represented.

The third source of insight for the modelling exercise is the fieldwork study domain (fig.1, c). Typically this has been the workplace into which the proposed technology will be deployed.[22] As the thesis introduction has shown, there is a community of practitioners who feel that knowledge of the current practices of a community is a vital resource in developing new technology for them. Evidently the fieldwork study domain will overlap with the stakeholder domain. However, fieldwork provides a different kind of information. It may be used to supplement what has been claimed by the stakeholders via other elicitation techniques. It may indeed replace other techniques where access to stakeholders has proven problematic. Or, it may provide background information which illuminates, or makes comprehensible, the claims of stakeholders. In all these cases, the aim of doing fieldwork is to reveal observations which can be used to influence and develop the models. The exact nature of this relationship, and the way in which fieldwork may influence the models is a critical issue for this research, and one that will be returned to in the following section.

Of course the relationship between models and fieldwork isn't just one way. Part of the purpose of developing such models is so that they can be used to inform the process of fieldwork enquiry (fig.1, d). It is expected that the modelling activity will be carried out in close collaboration with the fieldwork. If both fieldwork and modelling are to be done by the same person then this collaboration may be trivially achieved, but if the tasks are split, then frequent meetings would be necessary. It may be that modelling activities throw up explicit questions which can be answered through fieldwork. Alternatively it may be that by reading and understanding such models, a fieldworker is able to identify matters of relevance in their subsequent observation. When observations are felt to be relevant to the existing models, these relevancies can be examined and discussed and then, if appropriate, models can be changed to represent these new insights. The observations which led to these changes should then be documented, perhaps as a link into a set of field-notes, so that the evolution of the models is made traceable. In this way, through the activities (c) and (d), proto-models help to guide the fieldwork and fieldwork similarly informs the models.

---

22      Although there is no reason why it could not be the workplace of any of a project's other stakeholders, if a deeper understanding of their practices would also help to inform the requirements process.

Each of these three domains (the stakeholders, the project and the fieldwork) is intended to influence the modelling process. But the models that are produced are not to be in any sense influenced in an arbitrary way. Rather the systems they represent should be grounded by a rationale, they are to represent realistic possibilities. They represent system properties which can be argued for. In this sense the models are surrounded by a rhetoric, in just the same sense that the previous section suggests that requirements are. However, it need not be the case that there be only one model, or that every model represents a complete system, or that there is only one system being represented. Neither is it necessary that all the justifications of those models agree.[23] It is simply important that the models represent systems that could be realistically argued for. So, when any of the three aforementioned domains of study yield new insights, not only are these used to develop models, they are simultaneously used to develop the rhetoric that surrounds them. However, to a large extent the exact relationship between modelling and fieldwork activities remains, at this stage, a research question, to be investigated in the following chapter through empirical study.

---

23    Inconsistency between requirements, specifications, or views expressed during requirements is for good reasons a well recognised phenomenon, see for example [160].

However it is envisaged that the MGF activities will proceed iteratively. During this process the models will change and evolve. This is not to suggest that such evolution will be of a linear nature or will converge naturally on a single solution. However, in the general case it seems reasonable to imagine that initial models may start at a relatively high level of abstraction, and become increasingly detailed as the process continues. Observations of the domain may be made early on, which only become relevant to proceedings much later, as the models grow and develop. Over time, the arguments that have been developed and evidence that has been gathered will become stronger.

As MGF proceeds it is expected that the proto-models will evolve. For this reason it will be necessary to maintain an ongoing link between the project, the stakeholders, and the models under consideration within the MGF. Any change that is motivated by fieldwork may evidently have implications for other parts of the project; for example, it may have resourcing implications, or there may be issues of technical viability to assess. For this reason it may be necessary to return to the project domain regularly to discuss such implications (fig.1, e). Similarly, the fieldwork may stimulate ideas and proposals for the system, and the acceptability of these to the stakeholders should also be routinely considered (fig.1, f).

The purpose of MGF is naturally to influence a project's requirements. Once models and rhetoric have stabilised it will be necessary to provide the project with some kind of MGF deliverable. In this second stage, models are distilled to provide some kind of requirements document. The MGF activities will have changed and developed the models with which they began. Where a change has been made, there must be some reason for this change. The models are themselves non-arbitrary, they are initially the result of the project's early views which must themselves be justifiable, and each subsequent change must also be justifiable in some regard. Each change, with its justification, will suggest that the models, and hence the system, ought to be a certain way. Each may therefore imply requirements for the system. The process of distillation is to identify these requirements.

MGF aims to propose properties that a system should possess, arguments for why those properties are important, and provide evidence to support those arguments. The iterative process of doing fieldwork and refining models could be seen as essentially a means of building, rehearsing and refining such arguments. As ultimately it is these arguments that are the main deliverable, the models which supported their development are to be considered disposable.

The format in which these arguments should be presented is an open issue. Evidently, the purpose of such arguments is to influence the project's activities. The exact content of those arguments, and way in which they are presented are thus most likely to be determined by the local circumstances of each individual project. Their details will depend on who the arguments need to convince, the difficulty of this task, and the level of documentation and traceability that is required. For this reason, the exact format of the MGF deliverable will also constitute a topic for subsequent empirical study.

The foregoing has described a problem and the kernel of an approach which seeks to address it. It has done this largely in the abstract and without being prescriptive, because it is felt that the step-by-step design of action must ultimately be a response to the practical circumstances at hand. The practical details of applying this approach to a case study are described in the following chapter, and in particular the reader's attention is drawn to a five-point summary on p164 which describes a typical relationship between a fieldwork observation, a model, and a requirement.

## 5.3. A Methodological Discussion of Model Guided Fieldwork

MGF seeks to apply ethnomethodological fieldwork in a way that will contribute to a process of requirements engineering. In so doing it combines ethnomethodological fieldwork with system modelling, methods from two very distinct traditions. It is therefore reasonable to subject MGF to a more sustained methodological examination, in order to ensure that it does in fact represent a coherent proposal.

In a naïve sense, MGF's combination of modelling and ethnomethodological fieldwork seems like a potentially contradictory proposal, the two activities coming as they do, from very different disciplines. Modelling is often done within a nomological mindset, seeking to establish an abstract, law-like representation of a phenomenon. By contrast ethnomethodological fieldwork, in common with an interpretative tradition which stretches back to Weber, is concerned with understanding the unique details of individual instances rather than seeking to create abstract representation of the general case. Rather than develop laws or models it is ideographic and hence concerned with producing rich, detailed, "thick" descriptions of the social. For the ethnomethodologist the work of developing an abstract model might be a fascinating phenomenon for study, however, as an ethnomethodological description the model itself is likely to be seen as inadequate.

Though tension does exist between these perspectives, MGF is arranged in such a way as to combine modelling and ethnomethodolgical fieldwork in a complementary fashion. Though ethnomethodology may be opposed to modelling of the social, MGF uses models purely for the representation of ideas relating to a system, on which topic ethnomethodology would be indifferent. Similarly illusory is the distinction between ideographic and nomothetic approaches, which makes little sense when applied to MGF's system modelling which seeks to project a possible future technology rather than represent an existing worldly phenomenon.

Indeed the idea of using models to represent possible aspects of a system seems generally uncontroversial. Systems can be seen as logical structures for the manipulation of symbols, it therefore seems appropriate for aspects of their behaviour to be represented and reasoned about using other kinds of symbolic logical structures. In practice too, modelling is (as the literature review discusses) a well recognised technique within software engineering. Interestingly, MGF seeks to use modelling to streamline the use of fieldwork, in a way that is resonant with the way that agile methods like XP streamline the gathering of requirements. XP, for instance, elicits requirements continuously, as the software is being written; this is quite similar to the way that MGF elicits fieldwork observations as its models evolve.

The use of ethnomethodological fieldwork is a central part of MGF. For ethnomethodology, the thick description that Geertz presents is given a specific emphasis. Ethnomethodology is concerned with the reflexively accountable nature of ordinary social action. It is concerned both with the reflexive accounts that participants' actions provide, and the way in which professional sociologists might make these accounts available to their colleagues (this is described in greater depth in the methodology chapter). In using an ethnomethodological style of fieldwork it is inevitable that MGF will be engaging with this notion of account. As the above suggests, the fact that such accounts are used to create abstract models is not in itself contradictory. Rather, what is important is the rhetorical value of those accounts. Nevertheless, ethnomethodological accounts have a number of properties which may make them, from an engineering perspective, an unfamiliar resource.

It will therefore be appropriate to consider some of the properties of these accounts. MGF describes a situation where fieldwork activities will result in accounts being provided for the purposes of influencing an iterative modelling exercise. In this regard, the process of fieldwork and reporting is much like that done in professional sociology, where instead of models, fieldwork is made to speak to issues of academic interest. Much has been written about the nature of fieldwork accounts developed for this purpose, and so this will be a good starting point. Ethnomethodology, however, considers lay and professional sociology to be essentially equivalent. Garfinkel thus chooses to introduce the ethnomethodological notion of "account," by providing a detailed description of essentially equivalent accounting practices within particular domains. Possible parallels with these studies will thus provide a further way to consider the kind of account that might drive MGF. Ultimately, however, the nature of these accounts and the practices with which they are handled must really be a matter for empirical study itself (see following chapter).

The first of these properties is that accounts are, in comparison to models, detailed rather than abstract. One of the concerns with the use of the traditional form of ethnography in systems design was the overwhelming amount of detail which they contained. By contrast the abstraction of software engineering seeks to do exactly the opposite, to exclude unnecessary detail and thus manage complexity. The overhead involved in handing excessive detail can be problematic, particularly where that extra information isn't relevant to proceedings. However, when working within a fieldwork enquiry it would seem much more likely that the relevance of particular details is not initially completely clear. Since working out the relevance of particular observations is central to MGF, it seems reasonable to think that details should be dealt with as fully as possible. While the burden of dealing with this detail is potentially considerable, MGF mitigates this risk by avoiding giving this detail a written expression until much later in the process, when its relevance has been determined. Thus much of the detail produced by fieldwork can be dealt with through informal discussion, without the need to provide it with a formal representation.

Though an ethnomethodologist's accounts are typically detailed in nature, this is not their only significant property. This detail is often based in the language and terminology of the domain itself, which is also very different from the kind of information provided in a domain model. These are typically supported by some kind of modelling language which provides a framework of constructs that can be chosen and applied by an analyst for different purposes. For example, Object Oriented

Analysis (OOA, [161]) presents a framework which consists of "classes," "services," and "associations," which are then chosen by an analyst to provide information about a domain to a set of software developers. There is consequently a dual act of selection; clearly the analyst plays an active role in selecting aspects of the domain that should be communicated; however, the framework provided by OOA is also part of that selection process. The fact that the framework stresses classes and associations often means that certain information, like the organisational structure of a workplace, is selected for presentation at the expense of other details, such as its physical, spatial or aural properties. The former act of selection occurs at the time of analysis, and in direct response to the needs of a project; the latter act of selection occurs long before the time of analysis, when the method itself was created, with no specific knowledge of that project. For requirements engineering the notion of domain modelling is one whose popularity has subsided over the years (see literature review for a discussion).

Instead of the prior selection and externally imposed language of a domain model, ethnomethodology stresses the use of a domain's own terminology and structure for providing accounts of itself. Garfinkel insists that a domain already contains whatever terms or constructs are necessary to describe it. However, one of the advantages of a standardised modelling language is that it produces an account which will be familiar to the technologist, which is something that cannot always be said for the ethnomethodological account. Accounts in MGF may be provided in an interactive way, during discussion. Because of this, whatever explanation is required to flesh out those accounts can be given as and where it is needed, and be tailored to the exact demands of the audience.

As previously mentioned, an ethnomethodological account is concerned with particular instances of social action rather than generalised theories. This makes it distinct from much of the knowledge that is traditionally associated with engineering, which is powerful precisely because of its predictive power over many instances. Ethnomethodology, however, views rules and theories as an inadequate means of conveying the rich nature of social action. Garfinkel, for example, shows how attempts to do this end up making the participants into "dopes". He also shows how rules alone are inadequate as a description, as rules can, and often are, treated reflexively by those who participate in them. For the ethnomethodologist, rules, laws and models are not considered sufficient as a means for describing the social.

Instead, ethnomethodological accounts provide detailed insight into particular instances of social action. These accounts are provided by a fieldworker who has begun to observe and understand that domain in a way that approaches the domain participants themselves. By making such accounts available, the fieldworker provides a resource for others within the development process to begin to understand the domain as those within it do. It may be possible to understand the domain in a great many ways, however the way in which the participants themselves understand it (they gain a "vulgar competence" in the practices of that domain) is clearly a valuable one. For the requirements analyst such understandings may have a significant rhetorical utility.

Ethnomethodological accounts also differ in other ways from the more familiar laws and theories that might be used by engineering. Laws and theories are often developed in relation to one set of instances, but are then held to apply to all other instances of a sufficient similarity. They are consequently fixed and transportable, at least insofar as they should not change, and may be widely applied. The kind of thick description that fieldwork provides is quite different. Accounts remain by contrast "close" to the data from which they are taken. Goffman [76], for example, talks about the closeness of the analyst to the data, and their personal connection with the field-notes that they write, both in the sense of ensuring they are a part of those field-notes, and in the sense of those notes being a personal record which would not be suitable for general use. Heath and Hindmarsh [162] also talk about the role of transcripts; these are held to be simply an aide-memoir or a sketch, and not a replacement for the data itself. In this sense, data and analysed forms of that data, are kept in close proximity. Indeed, whether in a written paper or a conference presentation, ethnomethodologists generally show as much data as possible.

Relatedly, ethnomethodological accounts are also open and contingent in nature. As Garfinkel's study of the suicide prevention centre suggests ([5], ch. 1), accounts may be revised in the light of subsequent circumstances. Accounts only ever need to be good enough for the practical purposes to which those accounts may be put. Although the open and close nature of accounts may be unfamiliar to some, MGF is nevertheless arranged in such a way as to acknowledge these properties. It is an iterative approach. This means that accounts may be given and revised as circumstances develop, and stabilise; as the practical purposes to which the accounts may be put become clearer. Similarly, in MGF an account can remain "close" to the fieldworker whose observations it personally relates to. If a fieldworker gives an account from the domain which is of sustained interest to the modelling activities, then MGF provides the opportunity for their involvement in that analysis to be equally sustained.

Accounts are then a very different kind of resource from the kind of scientific theories or tables of data that engineering is traditionally associated with. They cannot be treated as if they were law-like or as an objective record. Despite this accounts are central to MGF. One of the critical moves within ethnomethodology is the use of empirical examples to introduce the methodological notion of "account." It may thus be valuable to consider how some of these studies relate to the use of accounts in MGF. One such study considered the reading and writing of medical records ([5], ch.6). These are accounts of medically consequential interactions between patients and clinicians. He shows that these records are not, for example, a simple "objective" representation of what occurred during a consultation. Instead they are in fact written with specific concerns in mind, in this instance, with concern for a sense of medico-legal responsibility. Moreover, the ways in which a record could be read were not at all determined by the way it was written. Rather, the proper meanings which could be given to that document were only determined at the time of reading and in light of the circumstances of that reading. The practices of reading and writing were thus quite particular and localized.

The study might, by analogy, reveal some of the properties that accounts within MGF may possess. By analogy then, the meaning of an account would be determined by the circumstances of its use. Thus, what an observation from the domain can be said to say will be determined by what is usefully sayable for the project. Although this is given by analogy, it does also stand to reason: if an observation can be made to say nothing of use, then from the project's perspective it effectively says nothing at all. This certainly suggests that MGF must ensure that what is usefully sayable about a project is a vital resource. But at an even more basic level, it suggests that accounts do not simply speak for themselves. An account is made to speak for a particular purpose, and this takes practical work, which MGF is designed to facilitate.

However, there are obviously limits to what an account may be made to say. It is evidently possible to misrepresent or misuse an account. Accounts should only be used to say things that are legitimate for them so to do, and therefore, to use them clearly requires knowledge of what an account may be legitimately made to say. Perhaps this is more straightforward for the medical record than it is for MGF. Often a medical record might be both written and read by the same kind of professional, somebody with a similar training and background. By contrast, in MGF the fieldwork domain and the project domain might be very different. If a fieldwork observation is being presented by a fieldworker to an analyst, it may be non-trivial for that analyst to know exactly what that observation may be legitimately made to say. So while the written medical record is a fixed resource, accounts in MGF are dynamic and interactive in a way that allows their legitimate use to be negotiated and discussed.

Medical records are, according to Garfinkel, written with an overriding sense of "medico-legal responsibility". They are in this sense written with respect for the possible uses to which they may be put. By analogy, it seems similarly reasonable to expect that accounts in MGF will also be given with a respect for the rhetorical possibilities which they may offer. Observations then may be chosen with a sense of their possible utility to the project. However, there are again obvious respects in which "medico-legal responsibility" differs from the "rhetorical possibilities" of MGF. The former is perhaps a more stable or fixed notion, insomuch as it is perhaps recognisably similar between instances; while the latter may not only be radically different between projects but is also likely to evolve over the course of a single project. Providing a constant sense of a project's rhetorical possibilities is one of the central challenges for MGF which it addresses through the development and use of models, and concern for the rhetoric that surrounds them.

## 5.4. A Discussion of Model Guided Fieldwork in Relation to other Fieldwork-Based Approaches

In many respects MGF draws most strongly on the descriptions given by Hughes *et al.* [2]. They describe four modes in which ethnography could be used in the system design process, and to a certain extent this framework uses elements of each. The ethnographic work is very much intended to be *quick and dirty*, inasmuch as the studies attempt to gather as many details from the domain in as brief a time as possible. To a certain extent the fieldwork studies could be seen as being *concurrent,* at least with the process of technical modelling; however, given that this is a requirements framework, the fieldwork is clearly not intended to be concurrent with the whole development process, which may be the original sense in which the term was meant. The framework is also *evaluative* inasmuch as the ethnography is constantly seeking to evaluate the models, as they are produced. Finally, where *existing studies* are available it is envisaged that these should be brought into the process (indeed, these played an important role in the case study described in the following chapter).

The claim that fieldwork can be effective when done in a quick and dirty fashion is quite significant. Fieldwork is often held to be a lengthy, time consuming activity where an ethnographer may be in situ for years at a time. Ethnomethodologists talk about the need to establish a 'vulgar competence' within the domain. Hughes *et al.*, many of whom are informed by ethnomethodology, suggest that in fact it is possible to get valuable results from relatively short spells in a domain. It stands to reason that the details of interest to sociology are not necessarily the same as those of interest to technologists, which may perhaps be too grossly mundane to be newsworthy elsewhere.

Another similarity between the proposed framework and the work of Hughes *et al.* is in the notion of debriefing sessions. They describe how ethnographers and system designers undertook lengthy meetings where fieldwork and technology was discussed. Debriefings are also a critical part of MGF, however the format is slightly different. Instead of having meeting between fieldworkers and the rest of the project, MGF introduces an intermediate step. By having debriefing-style meetings internally before presenting to the project as a whole, an opportunity is created for arguments to be more carefully refined. Also, by having meetings that are internal to the MGF exercise it is possible to have more of them, and thus increase the amount of time in which ideas may be refined. The ATC study was based on 4 weeks of fieldwork, which was matched by 4 debriefing sessions. It is envisaged, using MGF's intermediate proto-model-centric debriefing sessions, that MGF should make it possible to increase the amount of time spent synthesising fieldwork findings with technology-based arguments by having more meetings and by working in more rapid iterations.

Another issue raised by this work is that of who should actually conduct the prescribed activities. Fieldwork and modelling have both traditionally been specialised activities. Hughes *et al*. embrace this divide, describing experiences in a collaboration between computer and social scientists. Though disciplinary boundaries have been blurred in more recent years, there is certainly not a ready supply of hybrid computer-science-anthropologists. Developing such a hybrid worker would undoubtedly be beneficial for the field, and MGF would certainly be a useful technique for such a team. Meanwhile, it seems likely that MGF will also be carried out by interdisciplinary teams, where there are computer scientists or software engineers who have had little exposure to fieldwork, and ethnographers who may perhaps have had some exposure to software development but are probably not developers themselves. This kind of situation is reasonably common in CSCW research, and indeed in design ethnography circles. In an interdisciplinary environment, MGF allows the responsibility for modelling and fieldwork to be split along disciplinary boundaries. Evidently the greater the mutual appreciation between these two distinct roles, the more rapid progress with MGF is likely to be.

Another notion which stemmed from Lancaster's ATC study are 'guiding questions'. After some time of working together, the computer scientist proposed four guiding questions (cited in the literature review, p29) "intended to provide a framework for directing the ethnographic research toward issues of system design." Both sociologists and computer scientists came to recognise that these were very difficult questions to answer. For Randall et al. understanding how to answer these questions was crucial to understanding the role of ethnography in systems design. In many respects MGF tries to address the same challenge. Each of the above questions is complex because each is concerned with detail, both of the domain and of the proposed system, and these properties seem to be mutually relevising. For example, it is impossible to know what aspects of current work will be relevant to the system, without a firm idea of how that system is going to function; yet conversely, knowledge of those aspects of work practice may in fact suggest that the system ought to function in a certain way. The use of MGF's proto-models provides concrete system details for discussion and an iterative way to build answers to these questions.

In the anthropological tradition, the fieldwork deliverable, an ethnography, is a written document often of substantial length. It is typically thought to be an inappropriately weighty tool for conveying the value of a fieldwork study within systems development. As the literature review shows, there are consequently a number of approaches to the use of fieldwork which focus on changing the representation of fieldwork findings in an attempt to tackle this problem.

The literature review (p30) also discusses the use of modelling for representing fieldwork observations [68, 69]. By using a modelling language these approaches reduce the amount of information that is presented. While for the purposes of ethnomethodology this kind of representation may be seen as inadequate, this is not to say that such models are necessarily inadequate for the purposes of requirements engineering. However, practical experiences with this approach appear mixed. On the basis of their case study [69] suggests that (p129):

> If the modelling exercise is merely intended to convey information to
> system developers, it is unclear why a formal (or semi-formal) notation is
> necessary, particularly because natural language is quite an effective way
> of revealing the ambiguous, complex, and rich nature of everyday
> conduct.

Both of the above approaches, in common with the Designers Notepad [67], seek to change the format of the fieldwork's deliverable. None of them deals specifically with the issue of which observations or details should in the first instance be presented, or how such findings may be made relevant to the task of engineering requirements. In this regard MGF provides an entirely complementary approach, as its focus is on the production of relevant observations, and the arguments that make them so, rather than the format of how those observations should be presented. It would be possible to couple MGF with any of the above methods. For MGF, the format of the deliverable is a matter of local concern, to be decided on a per-project basis.

Another method that combines fieldwork with modelling is Contextual Design (CD, discussed in the literature review, p33). Although CD is a much larger in its scope and more complex than MGF, which mainly focuses on supporting the analytic aspects of fieldwork, some comparisons between the two may be instructive. Rather like [68, 69], CD uses models to represent activities within the domain, and does so as a way of recording the experiences of a fieldworker. Rather like MGF, these models are to be made immediately after a fieldwork visit, perhaps before the next. However, unlike CD, the models in MGF are system models, not domain models; and unlike CD, their main purpose is to guide and focus the fieldwork, rather than record its outcome. There is a further sense in which MGF is distinct from CD, which relates to the kind of fieldwork that each advocates. For the latter fieldwork is about gathering and recording the scenic details of work in a particular place, for example, which steps happen, and in what order.[24] For the former, fieldwork may be something more than this: to come to an understanding of work which approaches how that work is understood by the people who ordinarily do it. This theme, of what it is that ethnomethodological fieldwork is really providing access to and why this might be a worthwhile contribution to the requirements process, is one that will be returned to in the discussion chapter.

## 5.5. Conclusions

This chapter has introduced and discussed the MGF approach. A number of moves have been necessary to achieve this. The first was to consider the nature of requirements engineering itself, so as to be clear about what it is that MGF will contribute to. It was suggested that requirements had an inherently rhetorical property, and that fieldwork may be useful as an evidential source for building and evolving both requirements and the rhetoric that surrounds them. The next was to consider a critical problem with existing experiences of the use of fieldwork in systems development. The problem of identifying relevant observations was considered, and the relationship between relevance and rhetoric was discussed. MGF was then introduced. At the centre of the approach are representations of possible systems. These models both inform and are informed by the fieldwork activities. In this way fieldwork observations are focused on matters that remain of relevance to the ongoing requirements process. Following this, a substantial section was devoted to showing that MGF's combination of system modelling and ethnomethodological fieldwork is methodologically consistent.

---

[24] Of course it is difficult to summarise CD's orientation to the domain as little of its methodological background is available for study.

However, a number of aspects of the approach still remain open for consideration. Firstly, it was observed that a number of aspects of MGF will require tailoring to meet the particular needs of its circumstances of application, in particular, the format of the modelling to be used during the approach, and the kind of deliverable that is to be produced as a result of it. Secondly, although much methodological discussion has occurred, it remains to be seen whether MGF is actually a practical proposition. Indeed, it was noted earlier that the exact nature of the practices which will be used in relation to models and fieldwork remains a matter for empirical study; methodological analysis being at this stage no more than informed speculation. This will be a subject for the following chapter. Finally, there is the issue of fieldwork's status as a method of requirements elicitation. Following the evidence presented in the thesis introduction, this chapter has been premised on a notion that fieldwork can be of value to requirements engineering. This chapter has gone further and suggested that to be a requirements elicitation technique it must provide some rhetorical value to the requirements process. However, no argument has yet been given for why this should be the case; this will form a task for chapter 7.

# 6. The eDiaMoND Case Study

In the previous chapter, Model Guided Fieldwork (MGF), an approach to eliciting relevant observations from the domain was outlined. The approach proposed the use of a system modelling exercise, to be completed in close iteration with fieldwork observation. This chapter presents a study of that approach as applied to a real world situation: the eDiaMoND project. This chapter will show how the approach might be applied in practice; it will examine this application and provide some evaluation of it and the approach; and furthermore, will act as an investigation into the relationship between fieldwork and requirements engineering. In many respects eDiaMoND provides an ideal case study. There was within the project a strong interest in applying fieldwork methods, which meant there were certainly some people who would be receptive to the results the approach might yield. Its aim was to engineer a prototype system, and as such there were professional software developers who took the notion of proper requirements very seriously. It therefore provided an ideal environment in which to study the relationship between fieldwork and requirements engineering.

Chapter 1 has already outlined the general shape and objectives of the eDiaMoND project (p8), so this chapter will skip forward to describing the approach itself. It will begin by considering the tailoring of MGF to the circumstances of the eDiaMoND project. It will then outline the approach that was taken to studying the case study itself. Following this, some empirical description of the work that was done will be provided. It will attempt to describe the development of one particular requirement, tracing a statement made in the final requirements document back through our working materials to documents that were in circulation within the project before the activity in question began. This will end in a presentation of some detailed transcripts of one particular incident, and a close analysis of it, which will then form the basis for discussion and an evaluation of the approach.

## 6.1. Applying MGF to the eDiaMoND project

The first aspect to setting up the eDiaMoND case study was to establish who would be involved. It seemed as if, by good fortune, a number of people simultaneously converged upon the idea of conducting a small fieldwork study. The project needed to understand how the eDiaMoND technology might fit into the ordinary workflow of the clinical Breast Screening Unit environment. I was keen to find a case study where a workplace study could be applied in the production of requirements. My supervisor, who owned the project's requirements work package, was keen that insights from fieldwork be introduced into the project wherever possible. Added to this there was another DPhil student, CC, who was at that time engaged in an ethnographic study of Mirada, one of the project's commercial partners. CC and Mirada both seemed keen that she should contribute to this study.

The project was at this time already part way through. From a technical perspective there were three envisaged phases for the project: phase 0 sought simply to do some technical proofs of concept; phase 1 was aimed at producing a basic technical infrastructure for the project; and phase 2 was intended for the development of applications. The notion of clinical workflow was for the project an application level concept, for this reason our study was part of the phase 2 requirements process; it became known as the Work Flow Work Package (WFWP). As inputs to the WFWP we would have the existing corpus of project documents which set out at a relatively high level, ideas about what the eDiaMoND system was to do. In return our output, along with a number of other similar work packages, was to become part of the project's phase 2 requirements document. It would consequently be scrutinised by the project's board of investigators.

The basic structure of the WFWP was thus determined. CC and I would work as a pair, and then when we were finished, integrate our findings, along with everyone else, into the main phase 2 requirements document. During our collaboration, CC would visit a number of clinics to do fieldwork, and I would engage in system modelling activities. We would meet regularly to discuss our progress and to guide each other's work. This much was a fairly clear synthesis of our practical circumstances, and the MGF approach.

However, there were also more open aspects to our work. For example, it was far from clear what sort of modelling language would be best. The process was begun using use case modelling as this had been used elsewhere in the project, and indeed many of the initial documents we were working from took this form. Later in the process other kinds of modelling were tried, including CSP and UML activity diagrams. However, all three of these modelling formats were presented in the same way; regardless of the type of modelling used, each document contained a model, and a textual description of the elements of that model.

Neither was it clear from the outset exactly how we were going to present our findings. We had been asked to develop some use cases, but we also wanted to present some stories from the domain. In the end we developed a set of use cases but in the description of each provided some relevant domain information. This description was to come in three parts, the first would provide a description of the normal case scenario for that use case; the second, would provide some stories detailing some of the current arrangements that might be relevant to this new system; and the third to provide some analysis of why the system ought to be that way, often drawing on the evidence presented in the second part.

## 6.2. Approach to studying the WFWP

eDiaMoND was a real project and as such the work we were to do for it was important. However, as a case study of an approach to the use of fieldwork in requirements engineering it was also important that I was able to study how that approach worked in practice. So, just as CC treated the clinic as her domain of study, our work became my domain of study. Although I was engaged in doing the work of modelling, I was also able to treat our meetings as an opportunity for collecting data. All of our meetings, including those where we set up our work, were recorded, and once the project was over, transcribed (a total of 23 hours over 16 occasions). All the documents we physically used were also kept and filed, so that the working annotations they contained could be preserved. Similarly, in the negotiations which followed the WFWP in order to produce the phase 2 requirements document, meetings were recorded and documents were kept. In sum, there were approximately 60 hours of meetings, over 31 occasions, which produced 15 document drafts. Once the project was over, these materials provided a rich and powerful resource for reflecting on our work.

There had been a number of reasons why I had wanted to establish a case study based on at least two participants. The first was to make the situation seem as realistic as possible. It is often the case that there is a division in skills; sociologists are taught how to study various aspects of society, and technologists are taught how to build system specifications. This division can be seen in classic work such as the ATC study [163]. Since this is likely to be a real world constraint, it seemed reasonable to introduce such a division into our work. Similarly, it was important to show that the technique might be plausible for larger groups of workers. By showing that the work could at least be split between two people, it would be possible to consider the scalability of the approach.

However, there were also analytic reasons why this division of roles was desirable. One of the aims of the study was to investigate how observations of rhetorical value might be selected, and how this rhetoric would be developed.[25] This rhetoric would clearly need to involve insights from both the domain and the system. By splitting responsibilities so that one person had access to the domain, and the other access to technical concerns, any synthesis of these concerns would therefore have to be the result of dialogue between the two. This would therefore render that activity at least partly available for study. So, while it was clear than many aspects of this work would occur silently in the heads of the participants, by dividing roles in this way at least some aspects of the task might be available for analysis.

---

25      Requirements elicitation, was described by in previous chapter as an activity which seeks to identify information which may be of rhetorical value to the requirements process.

Appendix A provides a chronological summary of events related to the WFWP, and the way in which each of them was recorded.

## 6.3. Introduction to an eDiaMoND Requirement

As introduced above, the major part of this chapter will be focused on tracing the production of, essentially, a single requirement from the final phase 2 requirements document [document-15]. By doing this it will be possible to show in detail, how observations of the domain helped to shape and influence the eDiaMoND requirements. The requirement in question is:

5.4.1    The system MUST enable users to choose cases by batch.

In order to show how this requirement was produced it will first be necessary to explain a little more about it. The requirement implies that the system will have two important properties: it must allow the user to choose cases, and it must allow the user to select those cases by batch.

Breast Screening Units are organised to process vast numbers of patients. They do this by inviting "batches", often of around 30 patients, to the clinic so that mammograms can be acquired. These mammograms are processed and then examined by two radiologists (although there are exceptions to this arrangement). When this is completed, the results are collated and letters are written to each patient informing them of the outcome of their screening. There are strict targets for the length of time this process can take. Moreover, since patients in a batch are often from the same geographical area, these letters are sent out all at once, so as to minimise any possible anxiety on the part of the patients. Processing the patients by maintaining the batch ordering is therefore one sensible way for the clinic to organise its work-flow.

The process of reading was of critical interest to the project. The eDiaMoND technology sought to take digital mammograms, store them in a national federated database, and move them around the country on a national network. An important part of this was that radiologists would be able to examine images on a digital workstation. This workstation would therefore be a kind of replacement for the more familiar roller light-box viewer.

When a radiologist would read, they would sit down and do a reading session; for example, over the course of an hour they might examine over 100 cases. The most recent few images of each patient are "hung" on the roller viewer by an assistant, and they wait in the roller, alongside hundreds of other recent cases, until such time as they are read. Sometimes the contents of the roller viewer might be represented on a whiteboard so that radiologists can see at a glance what it contains. To read a case the radiologist needs to have at hand a set of images and the patient's file. When they start a reading session they consequently pick up a pile of patient files, often several of the most urgent batches. They then locate the corresponding images on the roller viewer and examine them in sequence; this is possible as the patient files have the same ordering as the mammograms on the roller. Finally the outcome is marked on the patient record and the radiologist moves onto the next case. The idea that the radiologist might choose the work they are about to read, as is implied in the requirement above, is therefore a very natural one.

The notion that the eDiaMoND system should allow its users to choose the work they were about to read, seemed very natural in the light of understanding how radiologists normally work. However, this was not how the system was initially envisaged. If one examines the documents that were in circulation at the very start of the phase 2 requirements activities, ideas about choosing work were very different. Specifically, the use case "prepare screening session" in [document-1] describes a system which, rather than letting readers choose work, would automatically allocate it to them. This automatic mechanism would be based on a criterion, such as oldest-unread-case-first, which could be set by the system administrator. Though the details of this mechanism had not been fully worked out at this stage, there was nevertheless a clear motivation behind this approach.

One perceived benefit was in the ability to control the length of reading sessions. As outlined in the "prepare screening session" use case, it is proposed that a session be:

> considered a time limited thing based on effectiveness studies; so if it is shown that the optimal reading time before fatigue sets in is 1.5 hours, a Radiologist logs on for a screening session that will finish after 1.5 hours.

But subsequent comments [email-1] were less enthusiastic about this plan. They suggested that radiologists were already aware of the need to take breaks, and that concentration was in any case something that varied between individuals and even over the course of a day.

Although the proposal for time-limiting sessions may appear to be of almost incidental importance to the eDiaMoND project, on closer inspection it actually turns out to be quite significant. One of the key aspects of eDiaMoND was the ability to distribute reading work around the country on a national network. In this way it would be possible to make use of extra reading capacity where it was available. As it was conceived, this distribution of work was to be handled by an administrator. To do this the administrator needed to be able to judge the reading capacity of their own clinic. If the screening sessions were of a fixed length then:

> [t]his would have the added advantage of allowing a BCU Administrator to manage the work-load based on capacity.

The implication was that by having a fixed session length, a roughly constant number of cases would be read in each, and this would make the reading capacity of the clinic easier to calculate. Central to this was the assumption that the reading staff might commit to performing a fixed or regular number of sessions. It is a notion expressed clearly in the "Manage Workload" use case of the same document:

> The BCU Administrator will manage the workload vs capacity within the system. On a regular basis she will log on to the work management console and review how many cases are outstanding for first or second reading and arbitration. She will balance this against her manual records of staff availability (i.e. number of Radiologists performing reading and number of hours available per Radiologist) and average statistics for the BCU for time required to read a case.

The notion of "being allocated work", rather than "choosing work" thus acts as a kind of inversion of current arrangements. At present, radiologists have both autonomy and responsibility. They are responsible for understanding the current reading workload and for its completion, but at the same time they have a certain amount of freedom to fit the work of reading around their other clinical duties. The radiologist is a professional, able to judge where her time may be most appropriately spent for the benefit of the clinic. By contrast [document-1] presents a system where it is the clinic administrator who controls the workload, and to a certain extent, the reading of work.

Of course the idea of retaining the existing arrangements within the clinic in no way precludes the idea of sharing work at a national level. Indeed it is relatively easy to see solutions which would allow readers to retain their existing sense of autonomy. To present just one possibility, radiologists could form relationships with other clinics using conventional methods, but then use eDiaMoND to access their reading work as and when necessary; this would allow them to manage these relationships according to their own professional commitments. For these reasons the final phase 2 requirements document presents a system which must allow the reader to choose work for themselves.

Although the two proposals are clearly quite different, this is not to suggest that there was significant conflict over this change in thinking. Although [document-1] presents one view in relation to the allocation of work, as its author pointed out at the time [transcript-7, p3], this was very much the kind of thing that the WFWP was intended to investigate.

Indeed, it would be easy to misinterpret the motivation behind the presentation of this example. Although the notion of "allocating work" vs "choosing it" was significant to the system, this was in no sense *the* defining issue for the eDiaMoND project, merely one amongst many ideas that changed through its course. Fieldwork was involved in this change of attitude, however, the example is not chosen in order to demonstrate the power of fieldwork for systems design. Indeed, there were a great many bits of information which were used in the formation of the requirement in question. The purpose of choosing this example is to show how, out of all the possible facts which could have been elicited, these particular facts were brought together to form an argument which ended up making a small but important contribution to the project. The purpose of doing this will be to reflect on how the method at hand played a role in these activities.

## 6.4. Tracing the Development of an eDiaMoND Requirement

Our main starting point for the WFWP was [document-1], a requirements document written by the lead application developer in Mirada (AD) to outline the clinical aspects of the system. Our task was in the first instance to take this high-level document and produce use cases at "next level down." As mentioned above, this initial document presented a system based around allocation, rather than one which would allow clinicians to choose work. This assumption thus became a part of my initial modelling work. For example, the following use case diagram from [document-2] contains features familiar from [document-1]. It describes a system where an administrator would control the way in which work is allocated to readers, such as by ordering the cases by the oldest unread.

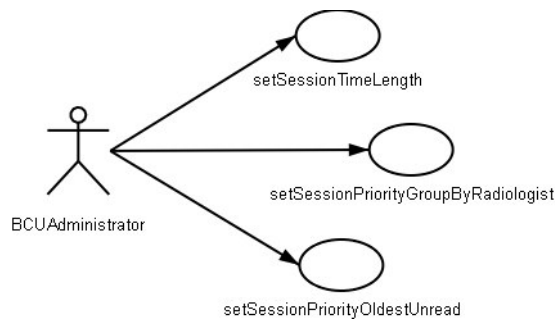

Indeed, this assumption seemed to be quite widely held within the project. For example, the following exchange occurred between myself and the project manager (PM) relatively early in the WFWP process [transcript-9, p1].

> **CH:** How do screening sessions work? Is there one big queue of work and if there are two readers do they each just take the next case off the front of the queue? Or is there a queue for particular people? Or is there a mixture of both, as that effects interruptions. Does the first to log on take the first 100 cases off the front of the queue?

**PM:** from a trust perspective you'd probably assign a particular person to do a particular reading from a particular date or clinic.

The assumptions represented in [document-1] were thus representative of a wider view within eDiaMoND and for this reason became embedded in my first attempt at modelling in [document-2]. It was this document which formed the basis of my first proper working meeting with CC. It was "proper" in the sense that it followed CC's first two days of fieldwork, one in Oxford, and one in London. It was also "proper" in the sense that I had prepared something for us to look over which CC had read before the meeting.

Proceedings began with a discussion of the activities each of us had been engaged in since the last meeting. This also included more political concerns, assessing our own position within the project and what this meant for the work at hand. We then moved on to discuss any particular matters arising that either of us felt were pressing. On this occasion CC wanted to raise some of the difficulties that seemed to face us both as we began our task. For example, that her initial experiences at Oxford and London had highlighted that the actual responsibilities associated with a particular job title tended to differ between locations. Following this we began to go through the models presented in [document-2]. Although we had never discussed a format for our meetings, this format of working became a stable feature. I would prepare a document and CC would read it before the meeting; the meeting itself would begin with administration and then any matters arising, followed by a more systematic walk through of the document.

The meeting seemed to be useful, as CC herself commented at its end:

> **CC:** I can see already that there are ways of working together, to fill in gaps, and ask things. I like that you're trying stuff and I can see it, and I can say that's not good or whatever, we don't have enough time to do it in detail, but I am optimistic that something is happening.

As a result of our discussions I made a number of amendments to our models. For example, CC had described how, if a patient had moved house recently, clinic administrators might need to order her records from another hospital; when these records arrived, her case would then be "fast-tracked" to make up for the delay. As a consequence I added an additional use case to the above diagram to represent the idea of fast-tracking a case. This was just one of many examples where observations from the domain would inform the models we were building.

With the WFWP now under way, I met with the project manager (PM), to show her how our work was proceeding. Rather as with CC, after an initial chat, I began to walk her through the various models I had documented. One of PM's suggestions was that I meet with a member of the Grid technology team (GT) at IBM, one of the project's major partners. This occurred a couple of days later and turned out to be quite valuable.

I had seen the WFWP as predominantly associated with Mirada, as they had provided us with the initial use cases. The idea of sending an ethnographer into clinics to uncover details about the clinical workflow must, at least in part, have been a result of CC's existing presence within the company (albeit in order to write an ethnography about them, rather than to write an ethnography *for* them). This had made sense to me as Mirada were developing the screening application which would sit physically within the clinic. I had not seen the WFWP as being particularly linked to the rest of the project. It became clear during my meeting with GT, that the core development team were more interested in the WFWP than I had imagined. Their work on the previous phases of the project had resulted in some "generic" work-flow, work-list, and database querying Grid infrastructure. Through our conversation it became clear that some of the requirements we might propose within the WFWP would end up being implemented using these technologies. IBM were thus going to be scrutinising our work closely to see how these technologies could be made to implement the resulting WFWP requirements. Knowing the way in which the WFWP deliverables were going to be read, and the purposes to which they were to be put, sharpened my sense of the requirements that our work had to pursue. However, the meeting with GT also helped to affirm that my view of the project made sense, since the technologies he talked about seemed to fit naturally with the initial requirements that Mirada had provided us with. Immediately following this meeting I went through my own document and attempted to mark each of its use cases to indicate whether it might pertain to the work-flow, work-list, or query technologies that GT had outlined.

Following these meetings with PM and GT, the models I focused on producing were of a different form. The use case format, that I had both inherited and begun to elaborate, allowed one to present the individual functions of a system from an external perspective, something that is very valuable when wishing to communicate requirements. However, this seemed somewhat unsatisfactory as it gave no sense of how functionality might be used in a real situation. Following this a more sequential form of modelling was sought and new models were consequently produced. One particular example was the batching model. This attempted to show how a batch would be treated as it passed through the clinical workflow from acquisition to the provision of results.

Neither CC nor I were working on this project full time, and a couple of weeks passed before we met again to discuss a document containing these new models [document-4]. This document, of course, continued to represent a system based on allocation. However, following this meeting, its successor [document-5] a week later, presented a very different view of the system, one based around the choosing of work. The meeting seemed to have been instrumental in this change of thinking. One particular incident which seemed significant was a story that CC told, based on her experiences in the domain. She and I discuss the notion of how a system might allocate work, and during this discussion she reports on an observation she made at a particular Breast Screening Unit. [transcript-11, p7]

> I think I've seen a system in Cambridge where their radiographic assistant writes on a whiteboard how she has loaded the roller... So that you can see where the cases are... and then if you're the first reader then you just cross out if you've done those particular rows, and so the second reader can see "oh these have been read once, and these have not been read even once, OK, I'm now going to do this" and then ticks it off as well...

This particular story outlines an approach to the organisation of work in the Cambridge clinic. Here, a whiteboard is used to represent the available work, so that radiologists can see what needs to be done, and make an appropriate choice of which work to do. This story is something that I orient to quite strongly during the meeting itself. Moreover, immediately following the meeting I scribble a note on the front of [document-4].

> Interestingly, [this] document describes a prescriptive system eg. "allocate this batch to that reader" [while] the domain is much more self-organising eg. people collaborate to complete a batch.

I go on to jot down questions, asking whether or not this "self-organising" property is important, and whether the flexible or public-display aspects of a whiteboard need to be replicated within the system. This discovery, that the domain is based on the choosing of work, rather than allocation, seems to have been instrumental in my subsequent modelling work. This change of thinking was significant, as the notion that work should be chosen rather than allocated, was one which persisted throughout the remainder of the project. Indeed, once the notion of choosing work is introduced into the models presented in [document-5] it remains a feature of every document subsequently produced. It is then captured by a set of requirements in the final phase 2 requirements document [document-15].

This incident is an example where a domain observation seems to have had a notable impact on the direction of the project, one which can be traced through to its final requirements document. Since this case study is in part concerned with a methodological examination, it will be of interest to explore how out of all the domain observations that could have been given, this one was chosen, how it came to be produced at exactly this point during our work, and why this observation seemed to have been instrumental in changing the course of the modelling work. In what follows, the content of this meeting will thus be examined in much greater depth.

## 6.5. Production of a Domain Observation

The meeting in question, the second of its kind, proceeded much like the first working meeting between myself and CC. We began by discussing the work and meetings we'd each attended since our last session, and also touched on our political position and that of the WFWP within the project as a whole. We also talked about more administrative matters, such as how we were to proceed with our work. But, like the previous meeting, the majority of our time was spent looking at a document [document-4], and of this, most of our time was spent examining the aforementioned "batching" model[26].

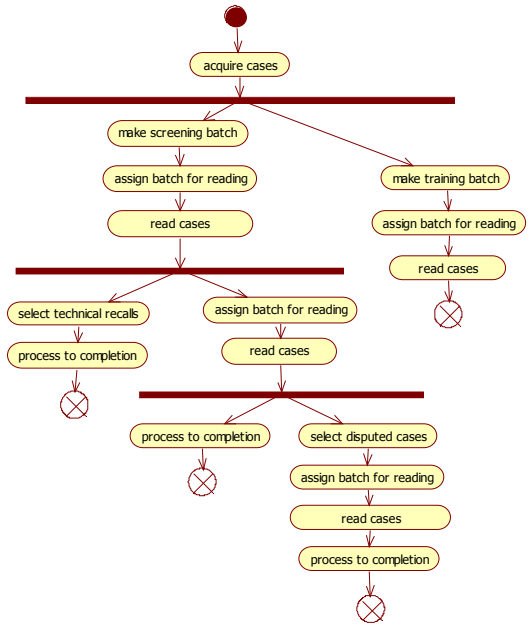

Although this model was different in form from the use case models presented in the first three documents, it nevertheless continues to represent a system based on the allocation of work to readers. This can be seen in the textual description of the "assign batch for reading" model element (italicised as original):

**Assign Batch for Reading:** Batches are then assigned, perhaps to a particular radiologist, or perhaps to a group of radiologists. These batches would make themselves apparent to radiologists, for example by appearing in a list. It may be useful to append a note to each batch, for the radiologist, indicating its priority, or contents. *How and why are individuals chosen to do a read?*

---

26      The original model was created in CSP, however, the model presented here is a semantically identical version that matches later models drawn using the more widely known UML activity diagram notation.

One of the critical features of this description is the italicised question at its end. This kind of expression of uncertainty was common. Models often had an exploratory nature, they would try and define a certain aspect of the system, but such definitions would often end with questions about those aspects.

On this occasion CC had not been able to read the document before the meeting, so after our initial discussion of administrative matters, we adjourned the meeting for half an hour so that she could look over the document, before resuming. For this reason the model was almost certainly very fresh in CC's mind as our walk-through of it commenced. As our discussion reached the "assign batch for reading" activity, CC immediately expressed a difficulty [transcript-030709].

**Fragment a [Audio-030709-1100 0:09:30]**

```
a1  CH  ok s so that that takes us to the next sort of thing which was
        assigning these batches f for reading

a2  CC  yeah

a3  CH  uh: (5)

a4  CC  yeah I didn't understand what you said about

a5  CH  yeah so I have this (2) ah::m I've in a sense treated all of
        the kind of reading activities [ in this one kind of chunk
        here

a6  CC                                 [ yeah
```

CC's use of the past tense perhaps here indicating that this was something which she had already noted as problematic during her read through. However, following this, conversation returns to our previous topic of how technical recalls and assessments might be processed. It is a further 10 minutes before we actually discuss the assign batch event in earnest.

**Fragment b [Audio-030709-1100 0:19:55]**

**b1 CC** so: where are we

**b2 CH** well we're kind of (1.0)

**b3 CC** assign batch for reading?

**b4 CH** assign batches for reading (11.0)

    ((analyst pauses to review the model))

**b5 CH** yeh so I suppose (.) I mean you can imagine what this is going
         to be like there's going to be som:e list of of kind of
         batches [ available

**b6 CC**         [ yeh

**b7 CH** and maybe there's only one batch in that list and it's what's
         happened kind of recently or over the last couple of days

**b8 CC** yep

**b9 CH** there are two or three kind of things in the list and um
         there's some note that says you know this is this is important
         this is the one that's just happened or you can just tell by
         the date hhh um::

**b10 CC** yep

**b11 CH** but how does um:: (.) how does somebody come to read those how
          is that one person knows that they're the person that has to
          read this and not that un un (..) um: for example ^hhh um::
          (.) assign batch for reading there are several readings that
          go on how does somebody know to do the first batch and not the
          second batch and (.) you know (.) the first reading and not
          the second reading

My explanation of the event is essentially a description of the system. I invite CC to

"imagine" what the system might be like. I describe the system as presenting a list of batches which have perhaps been produced from recent screening sessions. I try to describe this by drawing on an example of the system in action; I try to animate the model by describing the system it is supposed to represent within a realistic context.

CC appears to be an active participant, despite the brevity of my description, CC

makes affirmative interjections on no less than three occasions. It is fairly rapidly then, that I come to ask the question that I had previously noted in the document's text. I ask how it is that someone comes to read one batch rather than another, and moreover, how a reader comes to do a first reading rather than a second. If one accepts the assumption that the system is to allocate batches to readers, then these questions are essential to knowing how the system ought to do this. However, this question was perhaps not entirely straightforward for CC to answer.

```
B11 CH  ...the first reading [ and not the second reading

b12 CC                       [ yeh

b13 CC  this is I will I will need to find out [ a bit more about this

b14 CH                                         [ ok

b15 CC  what I see now (at the) clinics where I've been that I've been
        they haven't had a hell of a lot a lot of radiologists or
        readers

b16 CH  yes

b17 CC  which means that it's quite obvious who's going to do it
        because there's only one or two

b18 CH  heh

b19 CC  and

b20 CH  but but there are two people an and (I mean) who does the
        first who does the second

b21 CC  yeh I think they: vary it

b22 CH  ok

b23 CC  but um: just like oh:: um: do you want to do the first today
        or shall I do the second more like quite informal [ that's the
        impression

b24 CH                                                    [ yes yes

b25 CC  I have but I will do more fieldwork on that
```

CC suggests that current arrangements appear, at least to her, to be quite informal. However, she does offer to find out more on this; perhaps it is only the case that matters *seem* informal, perhaps in reality there is actually a careful structure to proceedings which could be elaborated with further ethnographic observations. Throughout our meetings I would ask quite a lot of questions. Sometimes it was possible for CC to answer these immediately, occasionally she might have to go back to her fieldnotes, but at least in these early stages it was common for a question to require CC to revisit the domain for further fieldwork.

We continue our discussion for a further minute and a half. I mention how I had studied some of the Edinburgh research [ethno-1], and whether or not this might imply that radiologists with different kinds of skills might be deliberately paired together for readings. CC also raises a potentially significant point. In the model I present, it is the administration who allocate work to radiologists. CC points out that the administration are not currently involved in the allocation of work. Of course the reality is, as we later establish, that allocation doesn't currently occur at all, but this is not brought out at this stage. Instead I continue to push for some kind of further response to my initial question.

**Fragment c [audio–030709–1100 00:22:56]**

```
c1 CH  it's just you know uh uh what are the mechanisms that are
       going to need to be provided [ in order to enable [ people to
       select and [ give work to the right people umm ((coughs))

c2 CC                                [ yes              [ sure

                   [ sure

c3 CH  because you have this kind of um situation where suddenly all
       of the work uh becomes visible to everybody you know you give
       people a pile it's kind of one (.) one way of doing it (.) and
       you can leave a pile in a certain place for a group of people
```

```
            to do you can be quite configurable when it's all (.) when
            when [ the records are physical

c4 CC       [ yeh
```

As with the previous occasion, I try to illustrate my point with an example. However, unlike the previous occasion, this time my example is not of the imagined system in use, but of the domain itself. For example, I suggest that in the clinic, because the records are currently physical, it would be possible to "leave a pile [of work] in a certain place for a group of people to do." It is in response to this that CC provides the aforementioned account of how reading work is organised in Cambridge.

```
c5 CH   but uh um

c5 CC   I think (1) I've seen a system in er in Cambridge where they
        their radiographic assistant writes on a white board how she
        has loaded the roller so basically where are what cases (.) an
        but this would this would evaporate in a digital system
```

CC continues to talk about the information which is presented on the whiteboard for a further 30 seconds or so, and I ask her a question regarding the size of the roller viewers. But our discussion then returns to the subject of the whiteboard.

```
                    Fragment d [audio-030709-1100 00:24:17]

d1 CC   they have this kind of diagram for (.) uh: first reader second
        reader erm::

d2 CH   a diagram?

d3 CC   well they on the whiteboard it's kind of very much like bom
        bom bom bom and then (1) if you're the first reader then you
        just cross out if you've done that particular (1)[ number of
        rows and then the second reader can see oh these have been
        read once ^hh er these have not been read even once ^hh ok i'm
        now going to do this and then ticks it off as well so
```

```
d4  CH                                                        [ ok

d5  CH   ok (1) (but that) that's really interesting though [ u:::m
         because that that I mean um um I'm I'm sure you (.) I'm sure
         you're sort of (1) you  point to examples where it's done
         differently and it's just organised differently but tha that's
         a really good example of how people are kind of (.) organising
         the work and you can imagine how a digital system will have
         some sort of list of ((coughs)) list of work that needs to be
         done but in s some way is kind of parallel to that

d6  CC                                                        [ mmm
```

CC describes the Cambridge approach where a whiteboard is used to represent the available work. She goes on to elaborate this story, gesticulating with her arm to indicate the lines of a grid, as she says "bom bom bom bom." My response to this observation seems strongly positive, giving particular emphasis in the phrase "that's really interesting." Contrary to my previous description of the domain (line c3) which seems to suggest that work might be allocated, following CC's observation I indicate clearly that I understand work is instead chosen.

As mentioned earlier, the documents that followed this, including the final requirements document all describe a system which allows work to be chosen rather than allocated. The first of these was [document-5], which I presented at our next meeting. During this meeting, by talking through the revised model, we return to the topic of allocation vs. choice of work. I begin by reflecting on the events of the previous meeting [transcript-12, p7].

> CH ... at the time I saw it as a series of small changes, but after I realised it was a complete reversal of the whole concept, I had written a system that was entirely prescriptive someone assigns you some work and you go and do it. But in actual fact it is the other way around you claim some work for your own and then you go and do it.

> CC that's right, because the hierarchy in the clinic is not that the administrator has the authority

CC's contribution to this conversation is particularly valuable. She suggests that one reason why it is important that work is chosen is because it is the radiologists who have the authority within the domain, rather than the administrators. It is the radiologists who have the overall responsibility for the performance of the clinic, and so it would be odd if it were the administrators who were allocating or deciding what work it was that they should do. This argument played a significant role in justifying the idea that work should be chosen rather than allocated, when it came to assembling the final phase 2 requirements document.

## 6.6. Analysis

It is clear that the domain observation which CC provided was important to the progress of the WFWP and furthermore, contributed to the requirements of the project as a whole. Since the identification of domain observations such as this was clearly a major aim of the approach taken, it is therefore now appropriate to consider how and why this observation came about. The kind of answer one might get to such a question will depend upon the kind of materials considered, and the way in which they are viewed. In this section a "close" sequential analysis of the foregoing episode will be considered. To do this the material will be broken up into six sections, and considered in the order in which they occur. First is CC's foreshadowing, that she didn't understand this particular part of the model during her initial read through. Second will be the question that I ask of CC, and why this was relevant to the WFWP at that particular moment in time. Third will be CC's deferral, and why it may have been reasonable for her to do this. Fourth will be my prompt for an answer, and why this could have been seen as necessary. Fifth will be CC's domain story, and why this was a relevant response to my prompt, and finally we will return to my positive response to this domain story.

CC had read through the model we were to consider immediately before the meeting. Her immediate response when I mentioned the event assignBatchForReading in discussion for the first time was that she didn't understand it (line a4). The fact that CC chooses to flag this issue at the outset seems significant, and this is therefore the first matter of analytic concern. There is a sense in which this flagging is perhaps indicative of an expectation by CC that she *ought* to be able to understand the model. Coming to a mutual understanding whether that be of the domain or the system, seemed to be a central theme in our approach. It was therefore entirely appropriate for the fact that CC had not reached a satisfactory understanding of this event, to be treated as problematic. Indeed as lines b5-10 show, I give a detailed rendering of how this part of the system might work, something which might be seen as an attempt to address CC's apparent lack of understanding. In this sense CC's claim foreshadows much of our later talk.

Following this early expression of concern, I provide a detailed rendering of how the system might appear, during which CC makes regular affirmative noises. I then go on to ask a question about the system. It is this question which will form the second topic for analysis. Naturally, the first issue to examine here is exactly what was meant by this question. Both in the document we were examining [document-4], and in our discussion (line b11), I ask about the mechanism through which a reader is chosen to do a particular reading on a particular case. When considered out of context, these questions may seem ambiguous: it is hard, for example, to discern whether they refer to the domain and its current arrangements, or to a system of the future. However, when considered in context, both questions do appear to relate more strongly to the latter. The question in [document-4] comes at the end of a paragraph which explains how a *system* might assign batches for reading. Similarly, my question to CC (line b11) immediately follows my explanation of how a new *system* might work (lines b5-10). Moreover, CC's response to this question (lines b12-13), does not appear to treat the question as ambiguous; rather, she treats the question as a matter for further investigation.

The question was very much relevant to ideas about the system that were current at that time. The model we were considering proposed broadly that work should be allocated to readers, and my question asked more precisely how this should occur. Moreover, there would seem to be good reason to think that a straightforward answer to this question might be forthcoming. For example, previous use case models had presented a requirement to be able to allocate batches of cases to a named radiologist. Evidence from Edinburgh University studies [ethno-1] had shown that readers often had a skill for identifying particular kinds of lesion. This had raised the possibility that knowledge of these skills might be taken into account when allocating work (for example, perhaps by allocating a pair of readers with complementary skills, to examine the same work). These proposals had occurred before the start of any fieldwork investigations, and it remained unclear how such features might be used in practice. Indeed, even before this, the documents which preceded the WFWP spoke about the allocation of work to readers, but gave little detail about how this allocation might occur. My question certainly sought to further define the current model at hand, but this came very much within a context of a long-running unresolved issue of relevance to the WFWP.

Later in our discussion CC produces a domain story that turns out to be quite valuable to our work. It is therefore interesting to consider why, in the first instance she suggests that my question ought instead to be a matter for further investigation; this will be the third issue for analysis. Before CC makes this recommendation, she does talk a little about the current working arrangements, and in sum, says that they are "quite informal" (line b23). However, this observation is of course only her first impression. She had at this stage done relatively little fieldwork, and up until now there had been no reason to give any particular attention to this specific aspect of the clinic. Thus, it is entirely possible that if she were to investigate this concern more closely then a whole range of sensibilities, practices, or rules, might emerge. If such things were a part of the current arrangements then they would certainly be relevant to the WFWP, and to providing an answer to my question. By treating it as a matter for further investigation this future possibility is allowed for.

As an aside, it would be easy to think that because CC was concerned with fieldwork and reporting on the domain's current arrangements, that she had no place commenting on what was essentially a matter which concerned the requirements of the future system, and that my question was hence misplaced in encouraging her so to do. Certainly it was the case that our roles were divided: CC visited the domain and returned with observations, while I proposed ways in which the system might operate. But it would be wrong to think that CC did not routinely comment on the suitability of these proposals, this was very much a part of our work. The fact that she chose to make the above question a matter for "further investigation" is certainly not indicative of an unwillingness to comment on matters relating to the system. The need to postpone pending further fieldwork was a very reasonable answer to my question. Indeed, especially in these early stages this was something that occurred quite regularly.

Moreover, it is reasonable to think that in fact at this stage in the discussion CC may not yet have viewed the information that she later provides about the Cambridge clinic as relevant. The model we were considering presented a view of work that was different to the one observed by CC. However, this did not in itself present a problem. As CC indicates in line c5, just before she does in fact provide her domain observation, the situation she describes would probably "evaporate" in a digital system. CC and I were both very clear that the system we were investigating would change the way people worked in the clinic, and that as a consequence it was not in itself a problem that the model we were examining differed from the domain. For this reason, it was not necessarily relevant for CC to provide her domain observation at this point in our discussion. Rather, there was every reason to imagine that through further fieldwork by CC, or through better explanation on my part, our difficulty might be resolved.

The following turn (line c3) is perhaps the most curious in the sequence and presents the fourth subject of analysis. Following CC's deferral and description of current arrangements as informal (b13-25), I produce my own description of the domain. This seems at first to be a counter-intuitive thing for me to do, since it is evidently CC who is the expert on the domain. Moreover, the turn is made even more interesting because it immediately precedes CC's much valued domain story. For these reasons it will be worth considering why this turn was both a reasonable response to CC's deferral, and relevant to the question I had introduced. In particular it is proposed that my turn be considered as a kind of prompt.

The first sense in which my response may be thought of as a prompt lies in the overall aim of our work. The primary role of the WFWP was to elicit information about clinical workflow. For CC this elicitation was direct, as she would regularly go into the domain and focus on observing particular activities. For me this elicitation was less direct. Sometimes CC would provide information about the domain at relevant moments while we were examining the models. However, at other times I would explicitly ask a question. In this sense, and as a product of the way in which responsibility had been divided between us, I was trying to elicit information from CC. My domain story is a prompt insofar as it sought to sustain our discussion of my initial question, by suggesting a possible answer to it. In this sense the prompt is analogous to the kind used in interview methods.

Our work was, of course, iterative in nature and this was in many respects advantageous. For example, if I were to ask a question, then it was possible for CC to investigate, and then return with relevant observations. While this is definitely a benefit when a topic is not well understood, postponing an answer certainly acts to slow the process of enquiry. Deferring a question is thus acceptable, but only where absolutely necessary. My response here might therefore be seen as a prompt, that perhaps seeks just to double-check that no further immediate progress on this topic may be made.

However, it is possible to see this utterance as more than this. It is relevant to consider how this prompt relates to my state of knowledge about the domain at that point in the process. It may be the case that my prompt represents nothing more than an attempt to elicit further details from CC, and thus that the content of my prompt is comparatively unimportant. Conversely, it may be the case that the domain story I produce actually does represent my view of the domain at that time. If this were the case then it would raise an interesting symmetry between my model and my view of the domain, as both would be based around the notion of allocating work to readers. This symmetry will be returned to later in the analysis. For the moment, if it were the case that I held this view, then this may be significant in relation to the production of the prompt itself. If I did indeed believe that the domain was based around allocation, then CC's impression that matters were organised "informally" might seem almost paradoxical. My subsequent prompt could thus be relevant as an attempt to explore this possible discrepancy.

The fifth turn for analysis contains CC's domain story itself (lines c5-d3). When seen as a prompt, my previous turn can be seen as being quite successful, as it does in fact result in CC speaking further about the domain. It will be particularly interesting to consider how these subsequent turns are related. My description of the domain suggests that work can be allocated by leaving piles of physical case notes in a particular place for a particular group of people to complete. This description is notable in a couple of regards: unlike my earlier description, this one is about the domain rather than the system; and, that the story talks about allocating work to readers in a way that was plainly inconsistent with CC's observations of the domain. It seems reasonable to imagine that this inconsistency was immediately apparent to CC, and it is at this point that she produces her domain description. Having done so, the inconsistency between our views will have been apparent to me too, and in my subsequent response it is CC's view of the domain that I give preference over my previous one. This penultimate turn therefore seems to render our exchange as a demonstration followed by a correction, rather than, as might have appeared at the beginning of the episode, a question followed by an answer.

In my final turn (line d5) I orient quite strongly to CC's domain story. This correction is of course far from being merely pedantic detail, and in my following talk I begin to describe the system in a way very different to my earlier offering (line c3). In this final piece of analysis it will be important to unpack how this newly described system (line d5) is a relevant response to CC's previous domain story (lines c5-d3). I begin my response by emphasising how interesting CC's story is, and I reiterate in very broad terms some aspects of her quite detailed story. However, immediately following this I start to talk about how a system could have a kind of list which might fulfil a role similar to that of the whiteboard grid that she had just described. I describe how this new system might be seen as a "parallel" to the whiteboard grid of CC's story. By implication, and as it is described in subsequent documents, this is a system which, like the whiteboard, would allow readers to choose their own work, rather than have it allocated to them. In my talk I show how CC's domain story is of relevance to our ideas about the system, and since refining such ideas is essential to our overall objective, illustrating this relevance would therefore seem to be an appropriate following turn.

It therefore seems clear that CC's story is relevant because it somehow provides the impetus for a new, and ultimately "successful", view of the system. However, there is a further possible respect in which CC's domain story might be relevant. More than suggesting a new view of the system, it may further, and simultaneously create a difficulty for the previous view. There is of course no reason why this *has* to be the case, after all the new view might just be superior to the old one, rather than necessarily invalidating it. However, in later discussions between myself and CC we do discuss how this old view was in fact inappropriate. For example, in the previous quote from [transcript-12] CC points out that it would be odd for administrators to be allocating work to readers, as readers are typically the more senior people within the domain. In my earlier description of the domain, I suggest that work is allocated rather than chosen. The fact that my initial model of the system is also based on allocation raises the possibility that I had used a similar "parallel" argument here too. CC's domain story would therefore be important because it would immediately show the fallacious nature of the premise to this argument. Though it is impossible to be certain from the evidence available, this provides a further possible reason why my response was so strong to CC's domain story.

## 6.7. Discussion

This chapter has been concerned with the application and evaluation of an approach to using ethnography to inform a requirements exercise. It has considered this task by presenting just a single example and examining it in detail. The episode was chosen because it was possible to show a clear link between activities during a meeting, and a documented change in ideas about the system. The example was one of a handful of episodes that were highlighted for analysis because of this straightforward link between activities within a meeting, and evidence of change in documents containing models and later requirements. Of course much of the work we undertook was more complex in its structure, and perhaps harder to trace, for example, where changes to a model were perhaps the result of activities that spanned multiple meetings. Nevertheless, aside from being able to identify a clear documentary link to the final project requirements, there is nothing at all unusual about the chosen episode; it would have been possible to draw the same analytic conclusions from many other episodes within the data. The fact that a similar analysis could have been given to a great many episodes, means that this analysis can legitimately form the basis of a more general picture of that approach.

The previous chapter introduced an approach called Model Guided Fieldwork (MGF). This discussion section will return to a number of the issues presented there, in light of some of the practical experiences presented here. One of the first things that this chapter provides is an empirical demonstration of the practical plausibility of the MGF approach. Both the events presented in the episode above, and the MGF exercise as a whole, made a positive contribution to a real-world project. This in itself provides evidence that the approach is worth at least examining in more detail.

The previous chapter provided a theoretical discussion of MGF. However, it was noted that discussion of theory is necessarily limited. In particular it was noted that the actual practices which might be used to work with models and fieldwork observations, would need to be the subject of empirical study. Although this case study was comparatively short and the approach itself was unfamiliar to those involved, some initial reflections on these practices are certainly possible.

Models always appeared to take as a starting point the accepted notions of how the system ought to be. In the first instance this was supplied to the WFWP by existing documents within the project. But as the work progressed, these accepted ideas were ones that CC and myself had developed. For example, during a meeting we would often come to an agreement as to how a particular part of the system ought to function; following such a meeting I would then update the models to reflect this agreement. This is very much what happened following the episode presented above.

In the most approximate of senses these models could be seen to move in abstraction from broad high-level sketches, to more detailed lower-level descriptions. However, this was by no means a straightforward or linear progression, and seemed to be guided by some quite complicated concerns. In the episode above, my initial question could be seen as an attempt to elaborate a section of the model which was in some sense unsatisfactorily detailed. The model was clearly being developed in order to contribute to a process of requirements, but in its existing state the model seemed somewhat vague. Models are abstract representations. As such one model will of course describe a whole family of possible systems. If these models are to be of use for requirements engineering then it is clear that the range of systems represented by the model must be such that those systems would be acceptable to the project's stakeholders. The requirements engineer, in trying to develop such a set of models, is therefore constantly in a process of imagining the range of systems implied by those models, and trying to judge their acceptability to stakeholders. In this instance, the model presented a system which would seek to allocate work to readers; evidently a whole range of very different systems might fulfil this model, and it would seem somewhat unlikely that *all* of these might be equally acceptable. In this regard my question addressed a concern with the adequacy of the existing documents for the purposes of developing acceptable requirements. The question hence seeks to progress the overall aims of the WFWP.

Though there are evidently concerns regarding the overall direction and progress of the work-package as a whole, updates to the model were also driven by the more immediate concern of trying to elicit the right information with which to make such progress. In this regard the models were also used for exploratory purposes. An accepted model of the system might be altered to represent a new piece of functionality. It might be changed to explore an alternative, and thus existing features might be altered. Or it might be annotated with a question, as was the case in the episode above, so as to elicit information with which to make an update. In all these regards the models were used in a thoroughly interactive way, and were useful insofar as they played a role in our meetings.

Models then formed a vital part of meetings between myself and CC. They were a guide for our discussions. The models acted like an agenda for the meeting, forming an ordering of matters for our attention. Adding something to the model was therefore tantamount to raising it for discussion. By working in this way our discussions remained closely related to matters of relevance to the system, even when they were about the domain. Moreover, if our discussions strayed off at a tangent, the model provided a reference point with which to bring ourselves back on track.

The models were evidently an important part of structuring our work, and a resource for steering our discussions toward matters of relevance to the system. However, as outlined in the previous chapter, they were ultimately a disposable resource intended only to act as "scaffolding". The aim of MGF was, of course, to be of use to the requirements process. The episode above shows how the elicitation of a domain observation occurred, and that the observation was instrumental in the development of a requirement. The following will consider more closely how the requirement ended up resulting from that observation.

In the above analysis, a sequential relevance is highlighted between CC's domain story and my subsequent description of a possible system property. This sequential relevance is reinforced by the content of our talk; I urge CC to imagine how my proposed system might be seen as "parallel" to the existing domain arrangements which she had described. This concept of exploring systems which were conceptually parallel to existing arrangements was a motif which appeared repeatedly within our work. However, although it seemed to be useful as a way of choosing options to explore, or perhaps a heuristic for generating ideas, this was not in the final analysis how the requirement considered above was justified to the rest of the project.

The fragments presented as part of the episode above focus on the practices involved in establishing the relevance of a particular fragment. They show in particular the role of models in this process. As the evidence shows, this discursive background work was critical to establishing the requirement in question. However, the justification which subsequently became important did not emerge until the following meeting. It is only at this stage that CC outlines more about how the clinic is organised. In particular she talks about the "hierarchy" of people in the clinic. It becomes clear that it is the radiologists who have the power and responsibility within the clinic, and it therefore seems more appropriate to have a system where these people drive the work, rather than have less senior people allocate it to them.

At this stage, while there was a clear justification for a system based around allocation, and while there was a model reflecting this, the requirement itself has not yet actually been written. As mentioned above, the models were used both to document features of the system on which CC and I agreed, and to propose properties ready for exploration. The models consequently ended up having something of a mixed texture, some parts were a matter of firm agreement, others were more speculative. As a result, once the modelling had been completed, it was necessary to distil the experiences of creating them into a more precise deliverable. In the case of the requirement considered here, that distillation process seems almost trivial. The justification discussed above clearly implied a requirement for the system: that it must allow radiologists to choose their own work.

So, from end to end the work completing in the episode above took the following form:

1. models were created representing existing ideas about the system;

2. an observation was produced which was relevant to this model;

3. the observation inspired a change to the model;

4. a justification of this change was then developed;

5. the justification then implied a requirement.

A number of interesting points arise from this. Firstly, the justification (4) was probably not based on facts which were unfamiliar to either myself or CC. The fact that radiologists were seen as the more senior people within the domain was probably something that we both already understood to some extent. However, the combination of model (1) and observation (2) seemed to bring this concern to our attention in a way which had not previously occurred. Secondly, it is interesting to note that the requirement's justification (4) came before (5), the requirement itself. It is the justification which highlights a part of the model as being something that must be subsequently expressed as a requirement. This is something that will be returned to later, in relation to the previous chapter's rhetorical definition of requirements. Thirdly, while myself and CC had found the domain observation extremely valuable, the rest of the project team were much less interested in these observations than I had initially imagined. Ultimately it was the requirement (5) and its justification (4) which turned out to be of greatest interest to the project at large.

This third point was borne of experiences with providing a deliverable from the MGF work, something that the previous chapter also highlighted as a topic of interest. It was noted there that the format of the deliverable itself would need to be something that was tailored to the local circumstances of the project. In this instance, eDiaMoND had been structured into a number of work packages, each produced a work package report, and then these were integrated into a final requirements document.

Our work package report was structured around use cases but contained a significant amount of domain information and analysis. Use cases were viewed positively by the project as a whole, so it made sense to express our requirements in this way. Each use case was then written up in three parts. The first gave a scenario-style description of the use case; the second was ethnographic, containing a collection of domain observations of relevance to the use case; and the third presented an argument which justified the use case, drawing on the evidence in the preceding section. We hoped that this format would be satisfactory in a number of regards. By using use cases as a structure, the document was clearly oriented around matters of technical concern. But by including mini-ethnographies we hoped both to justify our analysis and provide the reader with the ability to re-analyse our work and draw their own conclusions.

This provided a comparatively lengthy summary of the WFWP and was not greeted with the enthusiasm we had hoped for. It is difficult to discern precise reasons for this. It could be that the length and detail presented were off-putting. However, what did seem certain was the lack of interest in our document's ethnographic detail – despite the work that had been done to present its relevance to the requirements.

Nevertheless, our work was influential in the negotiations which followed. This second phase caused me to further distil our work package report. The descriptive use cases were reduced into precise requirements statements of a more traditional form. The final requirement document was then the subject of negotiation, during which justifications could be given and the findings from our fieldwork study invoked. In this context our work seemed much more successful. Once these more precise requirements had been expressed, the developers seemed very willing to engage in debate about them and to be persuaded by our justifications. Our work was thus influential on the project, something which can be seen by comparing our work package report with the final requirements document.

The previous chapter proposed a rhetorical definition of requirements. In light of the experiences above, this is now a topic that it is worth returning to. It was claimed that requirements have an essential rhetorical property; that while requirements certainly describe system properties, what makes them a requirement is their justifiability. The above fragment corroborates this. In the instance above, the empirical evidence suggests that justification may even arise prior to the requirement itself. This may simply be a product of the particular way in which we worked, or a result of the kind of empirical evidence that I gathered. However, at the very least it lends weight to the notion that requirements are essentially rhetorical.

The justification given for the requirement above revolved around the seniority of the reader, the importance of their ability to balance their reading responsibilities with other tasks, and how the responsibility for completion of the clinic's workload was a shared responsibility as well as a personal one. The ability to choose work was thus a feature which would be valued by the readers, in the sense that it would allow them to arrange their activities in a way that they judged to be most effective. The justification of this requirement thus invokes the radiologist as a professional who is best placed to decide what is best for the organisation of reading work.

The previous chapter showed how a requirement might be involved in a variety of different rhetorical structures (p95-104), for example to explore its technical feasibility or its relative value compared to other requirements. Certainly it seems relatively easy to see how the requirement in question could, and indeed would need to be, justified in these terms. However, discussion of these seem conspicuously absent from the empirical evidence. It is likely that this is due to the way in which the empirical study was set up. One of the reasons why the work of the study had been split between two people, myself and CC, was to render the activity of developing models and guiding fieldwork observable. By splitting responsibility for fieldwork from modelling, any relationship between the two would be manifest in our discussion. However, this means that the empirical record will emphasise some aspects of our work more than others. For example, for our requirements to be successful they would clearly have to be technically feasible. Because ensuring that the models represented systems which might reasonably be constructed by the project was a silent part of my role as analyst, it was not made available for analysis in the same way as other issues.[27]

One final concern raised in the previous chapter was the format of the modelling used *during* MGF. This was a challenge which remained throughout the case study. Initially, the use case modelling format was chosen. Existing requirements had been expressed in this form, and the diagrams seemed to be popular within the project. However, while very good for documenting the system, experiences seemed to suggest that it was not a good format for exploring new functionality. In particular the use case format lacked a sense of sequentiality and this made it hard to work out how one piece of functionality might relate to another. Following this a much richer kind of modelling with CSP was tried. CSP was widely taught in Oxford and was favoured by many on the project. However, CSP is textual in nature and it was later felt that it would be worth exploring an equivalent graphical format. For this reason the UML activity diagram was explored and some of the previous CSP models were re-drawn. Overall, this last format seems the best of the three. Sequentiality seemed to be a useful tool in exploring the models, as one was naturally prompted to imagine "what next," and their graphical nature was also beneficial as it provided an at-a-glance sense of the model's content. But aside from this, the most important feature of the modelling format was that it was readily understood by those involved.

---

[27] This is not to suggest that it was not manifest at all, meetings with AD and GT were of course very technical in nature.

Another important part of the documents we were using were the textual descriptions which accompanied the models. These described each element of the model in detail. During meetings these were typically used to tell stories about the system that model represented. In this regard the model and its textual description were being rendered as a kind of scenario, describing an instance of it in action. An example of this can be seen in lines b5-9 of the above episode. This raises the possibility that perhaps, rather than a textual description of the model elements, models should instead be accompanied by a set of scenarios. This might make the documents easier to read for a range of different audiences and less dependent on the analyst to present them.

The models themselves were intended to be disposable, and indeed this is how they were treated. Their circulation within the project was also quite limited; although they were shown to various people, this was always done in person. The models were never intended to stand independent of their author. The models were, by nature, simply sketches. It would have been easy to have misinterpreted these models as holding more weight than they were intended to carry. For example, the models often had a texture which was not apparent to the casual reader. As mentioned above, some parts of the model were a matter of firm agreement between myself and CC, whereas others were entirely speculative proposals. Without a sense of the ongoing status of the enquiry the models would have been virtually meaningless. But this did not in itself seem to be a problem, as it had always been clear that the models were not intended as a deliverable.

However, there was a sense in which the models could also be difficult to interpret even within the WFWP. In the episode above, I presented a model which was later revised as a result of discussion. In the analysis it was noted that the rationale for that model was not readily apparent. It is possible that I had based the model on an incorrect view of the domain, but it was noted that this was difficult to prove from the available evidence. Interestingly, the analytic difficulty faced here in determining a rationale for a system model is probably almost exactly the same difficulty faced by CC. If it had been possible for CC to identify this false view from the model itself or from an explicit statement of the model's premises, she would almost certainly have corrected it sooner. However, the model did not allow for this; no attempt was made in the model document to outline the reasoning behind what was presented. Instead any reasoning had to be determined merely from what was said during our meetings.

This raises the possibility, in the interests of making the process more systematic, that greater support could be built into the process for expressing some of the assumptions behind a model. In the instance above it might have been easier for CC to have understood my model of the system, if I had also produced a model of the domain to represent the world I intended to transform with it. Any errors in my assumptions about the domain would thus be made readily apparent.

Of course there are a number of concerns relating to the introduction of domain modelling. The first would be the increased workload, as the models had been intended to be quite light-weight. A traditional concern with domain modelling is the notion of "paralysis by analysis," expressing the possibility that the analyst may get bogged down in trying to represent an unending quantity of irrelevant domain detail. However, there are many respects in which this proposal to combine domain modelling with system modelling could side-step these traditional concerns. In the first instance the scope of the domain model, and hence the details which it would need to present, would be very tightly circumscribed. The domain model would only need to represent those details which were necessary to support a specific system model. The models could also be quite approximate and rough in character. For instance, rather than do them at the same time as the system model, the analyst could produce them on demand, as a sketch, whenever a fieldworker had a query that she wished to investigate.

Another possible confusion is how such a domain modelling exercise would relate to the issue of positivism, as outlined in Chapter 3. It would be easy to imagine that engaging in this kind of domain modelling necessitates a positivist view of the domain. This would be, however, fallacious. The models need not be taken as absolute representation of the domain, rather, they are simply a representation of the analyst's assumptions about the domain. Moreover, they are not a model of the domain intended for arbitrary usage, their purpose is limited. A particular domain model is local and its applicability is situated in the justification of a particular system model. In this regard the proposal does not suffer from the kind of concerns that were expressed with respect to domain modelling based on fieldwork described in the previous chapter.

## 6.8. Conclusion

In conclusion this chapter has three main outcomes. Firstly it has demonstrated the practicality of the MGF approach. It has shown the value of using models in the way prescribed by the method, of using early system models as a basis for fieldwork investigation. It has shown how they provide a way of both exploring ideas and documenting agreements. Perhaps most importantly, models can provide a resource for managing discussion of fieldwork by acting as a structure or agenda for working meetings. It has shown that, although the models really did "guide" fieldwork, the fieldwork equally well guided modelling.[28] The case study also suggests some basic guidelines for future MGF studies. Sequential forms of modelling seemed to be valuable; a graphical element also seemed to provide a helpful at-a-glance readability; textual description of models is also critical in providing detail for discussion; and models should be treated as a sketch-like, disposable resource. For the future it may be worth experimenting with the use of scenarios, either instead of models, or instead of their accompanying textual descriptions. It may also be worth trying a limited form of domain modelling to help open the analyst's assumptions to greater scrutiny.

The second outcome of the chapter has been to show the nature of contribution which fieldwork may make to requirements engineering. Fieldwork was firstly instrumental in inspiring requirements. The episode detailed in this chapter shows how fieldwork was used with a "parallel" heuristic to explore the appropriateness of possible system properties. However, equally as important was its second use as evidence in the subsequent justification of the resulting requirements. Fieldwork consequently makes a rhetorical contribution to the requirements process, in the sense outlined by the previous chapter.

---

28      "Model Guided Fieldwork" might equally well have been called "Fieldwork Guided Modelling", although the former seems to provide a more satisfactory emphasis.

The final outcome of the chapter has been to show the work necessary in bringing this contribution about. There was a significant amount of work involved in identifying observations from the fieldwork which were relevant to the requirements process. Doing this involved coming to a thorough understanding of aspects of both the technology and the domain; which, in the case of the work presented here, involved myself and CC coming to a mutual understanding of each issue. Another point which this case study made clear was that it is not enough to simply present observations, or even a partial analysis, as a finished deliverable. Observations do not simply "speak for themselves". Models must be distilled into clear requirements, traceably linked to supporting observations. This therefore provides a distinct second phase to MGF, following the proto-modelling activities.

In sum, this chapter shows that it is possible for fieldwork to go beyond merely "informing" development, and play a decisive role in the practical engineering of real requirements. However, the work involved in achieving this should not be underestimated. The WFWP deliverable was based on just 6 full days of fieldwork, although it drew on numerous other interviews, documents and previous studies. However, the modelling work which it contributed to was the result of around 14 hours of analytic discussion, spread over 9 meetings (and this doesn't include the subsequent time that was spent distilling this work into a requirements document). Fieldwork can be a time consuming activity and it is easy for this to become the main focus of effort; one of the things that MGF does is to bring that focus back onto analytic activities which make a precise contribution to the requirements process. It is clear that fieldwork can and should be made to speak to issues of relevance to technology, but it must be recognised that the effort involved in making it do so is by no means trivial.

# 7. A Discussion of Approaches to Justifying Ethnomethodological Fieldwork as a Requirements Elicitation Method.

This discussion chapter is intend to draw together the work of the thesis and consider in brief some of the issues which it raises. Two main lines of discussion will be followed. In the first a number of objections to the use of fieldwork encountered either in the literature or in the practical work this thesis has conducted will be addressed. The second will look more deeply at how fieldwork has been justified in this thesis so far, within the literature, and what might be involved in strengthening these justifications.

The first form of scepticism which might be raised regarding fieldwork is one which has already been examined extensively by the thesis. Chapter 3 has shown that fieldwork, and particularly ethnomethodological fieldwork, is a very distinct activity which avoids the kind of theorising or generalisations that are often associated with engineering. This divide between ideographic and nomological emphasis has led some to suggest that fieldwork may be philosophically irreconcilable with engineering; it is an argument that is carefully dismissed by chapter 4. The idea that the only valid kind of knowledge is the kind expressed in theories is a scientistic position. As outlined there, studies such as Vincenti's [117] show that while theorised knowledge is important to engineering, this is just one of many equally valid epistemic forms. Moreover, and as discussed in the methodology chapter, Garfinkel and others have shown that a theorised view of social action is inherently inadequate. Rules are, for example, inherently treated reflexively by their users. Perhaps more than anyone, Suchman's classic critique [54] helped to bring the idea that there was more to social action than could be expressed within a theorised view of it, to a technical audience. Perhaps for this reason, although Potts and Newstetter [9] raise a concern within an academic context, it was not a complaint that the author has ever encountered in practice.

Though very much advocates of the approach, Jirotka and Wallen nevertheless highlight further challenges for the use of fieldwork [10] (cited in the introduction, p8). On the one hand they point out that fieldwork studies "tantalisingly suggest" how a future domain should be; on the other they point out that they "cannot" conclude this. It would be easy to read this as implying that elicited information *should* say something about how the technology ought to be. Indeed many requirements elicitation techniques can produce some information which has this quality. Surveys and interviews, for example, *do* allow participants to express how they think a technology should be. However, there are other forms of knowledge that do not. Scientific measurements, for instance, only reflect how the world currently is, and say nothing about how it should be.

To interpret the inability of fieldwork findings to say how a system ought to be as a fault is therefore somewhat misleading. After all, other kinds of knowledge such as scientific measurements would not be brought into question in this way. Interestingly, the issue of how an "is" can be used to justify an "ought" is one of long philosophical interest. Hume suggests that to do so is a mistake, Searle for example disagrees [164]. However, it isn't necessary to become caught up in this philosophical debate. Fieldwork does not in itself need to say anything about how a system ought to be, so long as it can be *used* in a way that allows it to do so; so long as it has some rhetorical value to the requirement process. The matter of how such a thing might be possible will be returned to below.

Another complaint, one that sometimes arises in practice, is that fieldworkers are "siding" with the "users" of a system; this may be in contrast to other significant groups within the project, such as the domain's management, or the project's technologists. Since it is very often the case that the focus of a fieldwork study will be the potential users of a system, it is easy to see how this charge could arise. However, this accusation of political bias seems unfair. There may be good practical reasons why fieldworkers have been asked to study a group of potential users. There may be several different communities which a project must satisfy and while some may be well represented within the project, others may be less so. For instance, a domain's management may have good representation on a technological project, perhaps by virtue of controlling its budget; by contrast, the people who may end up using a system may be less well represented, or may have more difficulty articulating the issues which will be important to them. Fieldwork studies of the users' work can address this lack of representation. If they consequently raise issues which may benefit these users, this is a result of what they have been implicitly asked to do. Moreover, if a charge of political bias can be sustained then this is certainly not the result of anything methodologically inherent.

A similar form of scepticism alluded to in the literature [72] (p178) concerns the possibility that using fieldwork will lead to an undesirable conservatism. This is again mistaken, as it is not the case that fieldwork has any inherent agenda relating to its use for requirements or design. The way the information is used is up to those involved in the project. There is a concern that if information about current practices is gathered, then this will limit the possibilities for change. However, if it turns out to be possible to rule out certain possibilities because of information gathered through fieldwork, it must be the case that this information allows a case to be established which members of the project find convincing. In this case, ruling out an undesirable possibility seems to be a positive influence rather than negative one.

Of course, while some issues may be clear, sometimes it will be difficult to make predictions about the nature of future working practices. Some choices may therefore be more risky than others. One way to mitigate this risk would be to consider evaluating a prototype system.[29] However, another might be simply to choose the least risky alternative. The management of risk is a central issue in engineering, where great attention is paid to the reliability of products, both before they are released and throughout their working lifetime [165]. Changing a design may introduce risk and in the pursuit of quality, engineering might sometimes be seen as conservative. Considering work practice is no different. It may be hard to know how a new technology will relate to the working practices of a community. Being conservative in its requirements and design is one strategy to addressing this risk. However, this is not an inherent part of fieldwork, rather it is just one possible engineering strategy for its use.

Both of the above concerns seem to place ethnomethodologists in conflict with other members of a project. These troubles may perhaps be symptomatic of a deeper difference in understanding. One of the most significant tensions over the use of fieldwork takes place over the status of "work practice" itself. On the one hand sceptics suggest that present work practices are unimportant because they will be changed by the introduction of technology. On the other, advocates propose the importance of understanding work practice to requirements and design. The tension between these two positions appears to stem from their two very different interpretations of what "work practice" might consist in. The introduction of technology is often accompanied by a desire to change working practice, for example, to make it more "efficient" in some regard. For the sceptics "work practice" is a transient, perhaps almost arbitrary phenomenon, something that can be changed readily and unproblematically. It is a position which coincides with the aforementioned goal. Of course it is possible to read ethnography in this way, as simply scenic description. Ethnography frequently describes actual events which are at the time of reading, of course, in the past. It would be easy to understand these as simply stories which paint a picture of the past, but which could just as well be quite different in the future. However, this is not how ethnomethodologists understand the work practices that they so carefully detail.

---

[29]     From an ethnomethodological perspective prototyping could be seen as akin to Garfinkel's breaching experiments.

Ethnomethodologists understand "work practice" to be something much more fundamental. Rather than use the term "work practice," ethnomethodologists often describe their studies as being of a domain's "moral order." In some respects this term better reflects the fundamental status of what is being observed during fieldwork. Ethnomethodologists are not simply looking at a sequence of actions, they are studying what makes those particular actions the *right thing to do in that particular circumstance*. It is what Anderson refers to as a "practical logic" [72] (p178). To read them as scenic details is thus to miss their value; they provide access to the professional competencies of the domain. The moral order of a domain consists in what is seen through those professional competencies as being the correct way of going about whatever it is which it is that domain's proper business to do. For ethnomethodologists, "work practice" is understood to reflect something that is essential to whatever makes that domain what it is; "work practice" is merely the tip of a moral iceberg. "Work practice" and "moral order" are two sides of the same coin. In using the former rather than the latter, ethnomethodologists avoid a term which may seem unfamiliar. However, in choosing a more accessible term, they risk the aforementioned "scenic" misinterpretation.

When fieldwork investigations are understood as unpacking the moral order of a domain, rather than as scenic descriptions of past events, it becomes easier to see why ethnomethodologists feel that they have a powerful resource to provide to requirements and design. By unpacking the moral order of a domain, an ethnomethodological study is able to show why a particular piece of work is regarded as being right or proper or adequate (or whatever it is usual to judge work as being) for a particular circumstance. Moreover, it is able to show why the *use of a particular method, or approach*, is considered right or proper or adequate (or whatever it is usual to judge methods of working as being) for a particular circumstance. Ethnomethodological studies help to explain why a piece of work and its method of production are seen as adequate by those in the domain. If one sees ethnography as mere scenic description, then its use to design seems limited. If however one sees it as providing access to what is seen as right, adequate, or valuable within the domain then it is easier to see how this could be a useful resource for considering technological change.[30]

The second part of this discussion chapter concerns the justification of fieldwork as an engineering tool. The objections above, as the discussion shows, are relatively straightforward to address; they are often based on a misunderstanding of fieldwork or its relationship to requirements engineering or design. However, although these responses are useful, they do not in themselves form a satisfactory justification for attempting to use fieldwork for requirements engineering in the first place. A justification of this kind has thus far been avoided by the thesis. The introduction chapter motivates an interest in fieldwork by talking about its previous applications. This implicit claim, that fieldwork can be a useful tool for requirements engineering, is further supported by the eDiaMoND chapter which carefully details particular empirical experiences with its application. Any justification which depends on specific empirical examples is *a posteriori* in nature. With this kind of justification it is always possible to be sceptical by suggesting that the next circumstance of application will be different and could be unsuccessful as a result. It will be interesting to consider whether a stronger kind of justification might be possible.

A reasonable place to begin when considering how fieldwork might be justified is with the justifications that have been used within the literature. However, these often take a similar approach to the one adopted here; they make a claim about the utility of fieldwork studies and then justify this claim by making reference to evidence from a particular case study in which its use was successful. [64] proposes the use of fieldwork in the engineering of requirements, but this proposal is warranted by its successful application in the dealing room case study which the paper goes on to outline (and which is discussed in the literature review, p26). Similarly, [58] suggests that when doing Contextual Design it is important to consider current work practices, so that users can successfully make a transition to the new system (*ibid*, p7). But this, too, is ultimately backed up only with examples (of successful projects which did allow evolution from current work practices and failures which did not). Some have identified faults in existing methods or conceptions and then shown how fieldwork can address these shortcomings. [54], for instance, was instrumental in the turn toward fieldwork methods (discussed in the literature review, p23). Once again though, this is done with reference to a particular case study. Other work might even be seen as alluding to the difficulties of providing a strong justification for the use of fieldwork. [10], discussed above, seems to imply that fieldwork may "tantalisingly" hold an analytic value which is difficult to fully articulate.

---

[30] Chapter 4 also discusses the ethnomethodological notion of "moral order," but in relation to requirements engineering's commitment to investigate notions of "value."

Rather like this thesis, the kinds of justification that have been given for fieldwork ultimately depend on the showing of successful case-studies of their use. It is interesting to consider whether a stronger kind of argument could be given, perhaps *a priori* rather than *a posteriori*. This type of argument would depend only on first principles, on statements which are agreed to be true by definition. The fact that other authors do not attempt such a style of justification seems significant and this raises the issue of what kind of justification is it *reasonable* to give for the use of a requirements method.

In the model guided fieldwork chapter requirements elicitation was defined as being the acquisition of information which may have some rhetorical value for the requirements process, and specifically, value in justifying requirements which that process produces. Proving that the use of fieldwork is valuable will therefore involve proving that it can be put to such rhetorical uses. But this is difficult. Of course one may make a compelling argument for the importance of fieldwork, for example, one may argue to a project's stakeholders that their venture will not be successful unless they account for the practices of their system's users. However, there is no reason why they should accept such an argument. For example, by manipulating their notion of "success" it would always be possible to defeat such a justification.

Developing an *a priori* argument would therefore mean methodologically fixing what project "success" should mean, which seems unsatisfactory. But that a project's stakeholders must decide for themselves what kind of requirements rhetoric is valid on a case-by-case basis, is not to suggest that arguments for requirements methods aren't possible. It is very much the role of requirements engineering research to propose methods, and present arguments for those methods, while recognising that these arguments act as only as guidance to the requirements engineer and to requirements stakeholders. Of interest is to consider what kind of argument might be appropriate.

Arguing for the use of a kind of empirical enquiry seems difficult because by definition before one actually applies that method to a particular situation, it is impossible to know what will be discovered, and therefore it is impossible to know before the event, whether in retrospect knowing that information will have been worth the effort of doing the enquiry. Justifications of elicitation methods are consequently not absolute, but are more like guidance. In general, to justify a method of enquiry is to suggest that following it might yield something useful. However, as a professional discipline, requirements engineering is more likely to be concerned with whether one could be considered negligent in *not* conducting a particular kind of enquiry; that a form of enquiry is sufficiently likely to yield something useful that one would be negligent in not pursuing it.

To use a method of enquiry one must have some kind of belief in the likely outcome of its use, in the value that this might bring. In this regard every method needs some kind of "meta-narrative" which justifies the effort expended in applying it. The notion of meta-narrative is one used by Lyotard [30] to discuss various views of society and forms of enquiry about that society. It has also been used by Bickerton and Sidiqqui [29] to categorize requirements methods. However here it is used slightly differently. The aforementioned authors are taken as showing that methods of worldly enquiry can be usefully discussed in terms of a stable set of beliefs about the world they investigate. However, it is just this observation which will be used here, Lyotard's subsequent taxonomy, which is informed by broadly sociological philosophies, will not.

Meta-narrative is intended in quite a simple sense. For example, a requirements engineer might interview stakeholders and ask them what they want a system to do. Interviews could be said to have a meta-narrative which broadly asserts that to some extent stakeholders know what they want, and are able to articulate this when asked. Although requirements engineers know from experience that this is not always completely true and that sometimes interviews don't yield usable information, the fact that they often do means that *not* asking the stakeholders what they want would probably be considered negligent. The meta-narrative is an argument, or set of assumptions, that justifies the use of a method.

So of interest here is to consider what a meta-narrative of fieldwork might look like. Fieldwork provides knowledge of the practices that are used in a domain to produce adequate work. Its meta-narrative is therefore that this knowledge of practices is important to technology and to the success of a project. In what follows a number of proposals are made which could constitute a meta-narrative for ethnomethodological fieldwork as a requirements elicitation method.

Firstly, it is important to assert the notion and existence of the "practices" which fieldwork seeks to investigate. As indicated above, understanding this term properly is critical to the proper advocacy of fieldwork. The writings of Garfinkel [5] provide for this, and discussion is provided in the methodology chapter.

Secondly, it would be important to show that fieldwork is the right way of studying these practices; that it is necessary to observe them in their naturally occurring circumstances, and that this provides something distinct from and in addition to asking questions about them. Clearly it is sometimes the case that participants can discuss their practices, but there is no reason why this should always be the case, or that these discussions would be adequate for the purposes of requirements engineering. This is quite natural when one considers that these practices may be learnt through practical experience rather than through the discussion of theory. The say-do problem [166] is often cited as evidence of this kind of knowledge, but ethnomethodological studies provide a richer illustration. A meta-narrative for ethnomethodological fieldwork would assert that practices should be studied through observation of practical actions that occur under and in response to, natural mundane circumstances.

Thirdly, it is necessary to assert that the success of a project as a whole will be in some significant part determined by the ability of participants to produce adequate work and to do so in a practically adequate way, following the introduction of the new system. There may be many parties who are qualified to suggest what might make a practical action adequate, but one significant perspective is that of those who already have hands-on experience of the practical circumstances in which this work will take place. This notion of success is clearly additional to what stakeholders may say about the success of other aspects or goals for the project.

Fourthly, that the success of future working practices will be related to the requirements of the new system. So, in a future situation after the introduction of a new technology, the ability of participants to produce work in a practically adequate way will be in some part related to the properties and affordances of that system. These properties and affordances are determined by the requirements and design of that system. Thus, it is the job of requirements to outline the criteria necessary to allow future working practices to evolve successfully.

Finally, and perhaps most importantly, that knowledge of current working practices provides a resource for reasoning about future practices. Evidently what participants regard to be a practically adequate action under current circumstances need not be the same as what they might see as practically adequate after the introduction of a new technology. Nevertheless, knowledge of current circumstances provides a basis for reasoning about the future. In what follows three arguments are given to justify this.

Firstly, what is currently considered as a practically adequate action will be the result of a whole raft of practical considerations, regulations, skills, conventions, the nature or needs of customers or suppliers, constraints of time or space, and the properties and affordances of the tools available. Changing the available tools is evidently just one of these considerations and while technology is often associated with a wider change in the domain, many of the practical day-to-day qualities of working life remain. Understanding the present is a good resource for reasoning about the future because it is likely to be similar in some regards. As Sacks [167] (p578) points out, far from revolutionising it, technology is more often than not simply made at home in our existing world.

Secondly, even if a technology is intended to change the domain in ways that are perceived to be quite radical, a knowledge of what is currently regarded as practically adequate can be useful. While engineering is concerned with innovation, it is equally concerned with addressing risk. Understanding what may be changed when a technology is introduced provides a way of understanding, and hence mitigating, the risk that a project may be taking.

However there is a third more detailed argument for studying the present, concerned with how "practice" and "adequacy" should be understood. For ethnomethodologists studying practices is about coming to understand what is considered by participants to be the right thing to do within a particular practical circumstance. This notion of a practical rightness is often referred to as part of the "moral order" of that community. Understanding practices in this way makes them much more than the sort of prescriptive processes that can be described in formal notations. This cuts to the core of an ethnomethodological philosophy of mind [168]. Clearly the practices and moral order of a domain may be different in the future than they are in the present, and this change may clearly be precipitated by a technological intervention. However, as the participants examine any new technology and decide what it is they will need to do to carry out adequate work, it is their present notion of what is right that will be critical to these investigations. Although the moral order will change, this change will come from within the moral order of the present. What the future will see as the correct way of acting will be determined by those considering within current sense of rightness how they should act with a new technology. Properly understanding the moral order of the present is therefore a resource for understanding the moral order of the future. Moreover, understanding the moral order of the present provides a way of examining the practical adequacy of technological proposals, and a basis for distilling requirements to ensure a system fulfils this.

What fieldwork provides then, is access to the moral order of a workplace, and the above argues that this is a critical resource for requirements and design. It is worth emphasising, as an aside, just how distinctive ethnomethodology is in providing this. Contextual Design [58], with its contextual interviews that encourage relatively short periods in the domain and its emphasis on domain modelling, appears by comparison only to encourage the acquisition of relatively scenic details from the workplace.

Although it is obviously necessary to record what steps a task consists in, what is important for requirements and design (as argued above) is to understand *why* those particular steps are considered to be an adequate path to its completion. To do fieldwork so as to achieve the former without the latter would be, metaphorically, to throw the baby out with the bathwater.

Meta-narratives, like those considered here, help to outline the assumptions behind an elicitation method. They are a way of taking a step back from individual arguments about whether a method is universally right or wrong as a method of elicitation. In practice, the convincingness of a meta-narrative (which is to say, the perceived appropriateness of an elicitation method) is likely to be informed by a whole range of factors, from requirements engineers' philosophical commitments, to the practical aims of the participants in a particular project. One such set of factors are those advocated by professional bodies such as the British Computer Society (BCS). They outline both an ethical code of conduct and a set of good practice guidelines [4] (some of which are cited in the introduction, p5). Ultimately it is up to the requirements engineer to explore the extent to which such guidelines are relevant to their circumstances, and by considering its meta-narrative, to consider the extent to which a particular requirements elicitation method may help them to realise them.

In sum this chapter has discussed: some of the practical difficulties and objections raised to fieldwork; an argument for doing fieldwork; what sort of arguments it is appropriate to give in favour of an elicitation method; and furthermore, how these arguments may relate more widely to its professional discipline.

# 8. Summary and Conclusions

This thesis makes contributions in three areas: it provided substantive support to a flagship e-Science project; it presented a practical approach to addressing the challenges of using ethnomethodological fieldwork for engineering requirements; and it sought to clarify the conceptual relationship between ethnomethodological fieldwork and technological development. Each of these areas will be expanded on in turn.

## 8.1. Substantive Contributions

eDiaMoND was a high-profile e-Science project. The work presented in this thesis played a modest but important part in the overall success of that project. This substantive work is therefore the first and most obvious contribution of the thesis.

## 8.2. Methodological Contributions

The second kind of contribution is methodological in nature. Though its potential value is often acknowledged, there is within the literature a lack of methodological advice regarding how to use ethnomethodological fieldwork for engineering requirements. This thesis therefore makes three main contributions of this kind: the first is to identify a critical problem for its use; the second is to propose an approach to address this problem; and the third is to provide a critical evaluation of this approach as applied under realistic circumstances.

One of the critical steps in presenting the MGF approach was to identify a primary difficulty for fieldwork. In common with [1], this thesis takes this difficulty to be an analytic one, relating to the task of identifying domain observations that are in some way *relevant* to the process of development. However this thesis goes on to outline what "relevance" means for requirements engineering. Specifically, an observation is made relevant through its involvement in a justification of a potential system requirement. This is a detailed task: observations of the domain can be extremely detailed; similarly, requirements are also often very detailed. Any justification of the latter that involves the former must simultaneously draw on a detailed knowledge of each. Existing methods seem to provide little support for this activity.

MGF was essentially an attempt to bring detailed domain observations and detailed proposals about a system into close proximity. By so doing, it sought to foster an opportunity for analysis that would yield arguments about why the system ought to be a certain way; arguments that would use fieldwork observations as their basis. Domain observations and system proposals were brought into close proximity by introducing system proto-models into the debriefing meetings. These were disposable models which represented ideas about what properties a system might possess. The aim was that by bringing insights from the domain to bear on these models, they might be developed, and that through this development, fieldwork would remain focused on issues of relevance.

### 8.2.1. Costs and Benefits of Model Guided Fieldwork

The primary benefit of MGF is that it provides a method through which a requirements process can engage with the *real* work involved in a particular domain of interest. By harnessing an ethnomethodological orientation, it provides understandings of that work beyond merely the factual details of what currently occurs. By additionally unpacking the moral order of that domain it provides a way of directly engaging in analysis of how work could be appropriately changed through the introduction of technology, and the requirements that would realize these changes. Moreover, by using MGF as a structure in which to conduct ethnomethodological enquiry, as opposed to other techniques, the requirements engineer is assured that the observations gained through fieldwork will always be of some relevance to the emerging technological agenda of their project.

MGF is not proposed as a fully formed requirements technique. Rather it is an attempt to solve a perceived problem for the application of ethnomethodological fieldwork. Although the thesis presents its use in a real-world situation, this was done less to evaluate its worth, than to come to a better understanding of the problem itself. Nevertheless, there were respects in which the use of models seemed to be successful. For example, it was possible to observe how models could act as an agenda for a debriefing, as a resource for keeping discussions on matters of relevance to the system. It was also possible to see that the models did influence the direction of fieldwork, and conversely that they were themselves influenced by that fieldwork.

Of course the use of MGF brings with it particular costs. Compared to other requirements methods, MGF introduces the added cost of conducting and analysing fieldwork studies. Doing this not only takes additional time, but requires that a project team have access to quite specialised skills. Of course, these costs can be seen as a way of mitigating the cost of project failure. Compared to other fieldwork-based approaches, and particularly those considered to be "quick and dirty" in nature, MGF appears to shift the focus of effort toward analytic activities and away from purely observational ones. However, this is perhaps an unavoidable necessity if fieldwork is to play a significant role in shaping a technological agenda and in this sense, a greater emphasis on analysis may actually be a desirable outcome.

## 8.3. Conceptual Contributions

The third kind of contribution made by this thesis concerns the conceptual relationship between ethnomethodological fieldwork and technological development. These contributions can be summarised by five main arguments. The first related to claims that had been made about requirements engineering itself. If it is inherently necessary to view requirements engineering as a positivist pursuit, then clearly ethnomethodological fieldwork cannot be a requirements engineering method. Chapter 4 argued against this. It shows that positivism is totally unsuitable as a philosophy of requirements engineering.

The second of these arguments was also made in chapter 4. It presents that case that if one wished to think of requirements engineering in philosophical terms, then this philosophy would look much more like ethics than a philosophy of science. Requirements researchers necessarily position themselves so as to advise practitioners what proper requirements practice ought to be. In so doing the academic field of requirements engineering might be seen as parallel to the field of normative medical ethics, which provides approaches for doctors to determine what the right action might be in any one particular circumstance.[31]

Many ethnomethodologists seem convinced that their perspective holds the key to designing better technology. Indeed, there are many documented examples where this perspective has raised issues that have been extremely valuable to real development projects. However, besides simply citing these examples, actually arguing that ethnomethodology needs to be a part of systems development has proven difficult. The following two arguments both tackle this problem.

So, the third argument presents an alternative way of advocating the use of ethnomethodological fieldwork. Typically this has been done by stressing the importance of understanding work practice. However, "work practice" may be understood differently between ethnomethodologists and non-ethnomethodologists. By understanding that ethnomethodologists also mean "moral order" when they say "work practice" it is possible to give an alternative explanation of the value of ethnomethodological studies. Ethnomethodologists uncover essentially what is considered "right" within a particular circumstance, within a particular locale. By stressing this, it becomes much clearer that this may be a resource for reasoning about what might be considered "right" in the future, and what the "right" technology might be to enable that future.

---

[31] One could extend this analogy and suggest that the software engineering that follows requirements, could be considered to be parallel to the medical science associated with any intervention prescribed during a clinician's initial consultation with the patient.

The fourth argument shows that in any case, the extent to which it is possible to justify a method of requirements elicitation is necessarily limited. One cannot know in advance of using such a method what it will yield, and thus one may only ever proceed on the basis of past experience or best judgement. It goes on to suggest that it may be more helpful to think of elicitation methods as having meta-narratives that capture their assumptions, rather than as being in need of absolute justification. Understanding them in terms of meta-narratives is thus a resource for evaluating how to fit a particular elicitation method to a particular requirements engineering problem.

The fifth argument concerns requirements engineering itself. Though it may be possible to provide appropriate justifications for using ethnomethodological fieldwork in requirements engineering, it may be less clear how such fieldwork might relate to the task of defining requirements. Other forms of requirements elicitation, like interviews, sometimes yield candidate requirements. Fieldwork, however, does not; instead it yields observations of activity within a domain. By arguing that requirements have an essential rhetorical property, it becomes much easier to see how fieldwork might contribute to requirements engineering. Fieldwork observations are thus evidence in the justification of requirements.

It is claimed that these five arguments go some distance toward situating ethnomethodological fieldwork as a requirements elicitation method. However, they also point toward a particular view of how software engineering ought to engage with the world. The introduction to this thesis considered the question of why software engineering should take ethnomethodological fieldwork seriously. It may now be possible to address this question more fully.

When a technology is proposed, it is done so in relation to some activity in which it is intended to assist. The technology is important by virtue of the importance that the world places on the task in which it is assisting. The challenge for requirements engineers is to determine what properties that system needs to have in order that it might be successful. However, it is not their opinion that is important, under this view it is the world for which that technology is intended that should determine whether a technology has ultimately been successful. This consequently motivates an interest in the study of what success would constitute for a particular community.

Different kinds of argument might be used to justify why a requirement may be necessary for the success of a project; this is captured in the aforementioned notion of rhetoric. However, the ethnomethodological view implies one particular kind, one based around the practical success of the technology.[32] Naturally the introduction of a proposed technology will change the practical way that people complete the tasks with which that technology has been associated. Practical success is therefore achieved when these changes result in the successful completion of those tasks in a way that can be considered to be at least as "good" as in the previous approach. More precisely, "good" is taken in relation to the *moral order* of that domain. This ethnomethodological term captures both what the right way of practically completing a task is, and indeed what would be absolutely the wrong way of completing it. If a technology implies that a task should be done using a different practical action, then it is those people who have the practical skills to carry out such actions who also potentially have the skills to identify the consequences, whether positive or negative, of a proposed change.

_____________________

[32] The validity of the ethnomethodological view itself obviously depends on arguments such as those made in [5].

What ethnomethodological fieldwork provides is the opportunity for a requirements engineer to begin to understand the moral order of a particular domain. It is proposed that by coming to such an understanding of what practical success means for this particular location, the requirements engineer will be able to consider the practical consequences of possible changes, and thus be able to determine what properties the system must have in order to achieve this practical success.

Of course this may not be easy. Practical success may depend on fine-grained details of work. It may also be the case that these details are only tacitly known. Previous ethnomethodological studies have demonstrated both these points. It may also be the case that there are different groups within a domain, each of whom have a perfectly legitimate, but different, understanding of what a task consists in, by virtue of engaging with a different aspect of it. The requirements engineers may consequently need to understand aspects of all of these different and interdependent moral orders, and moreover, how they relate to the overall success of the task in which a technology is intended to assist. This may evidently be a complex task, and all the ethnomethodological perspective can provide in conjunction with approaches like MGF is the information with which to begin to try and engineer solutions.

## 8.4. Further Work

This thesis provides a basis for a number of potential future research activities. MGF is, in its current form, just the kernel of a method for using ethnomethodological fieldwork as a requirements engineering tool. The case study presented here shows that this method has significant potential, however, clearly there is much more that could be done to develop it. Perhaps the most helpful way to do this would be to apply the approach in a wider range of case studies. Since each project will have its own organisational structure and approach to development, these experiences would provide a way of refining MGF. The aim of such case studies should be to work toward a more prescriptive statement of how to apply the approach. This is something that would be essential if the approach were ever to be disseminated more widely.

Within such a programme of research it would also be appropriate to try some obvious variations on what was done within this thesis. For example, on p170 it is suggested that a limited form of domain modelling may enhance the process of analysis. Other variations could deal with the scalability of the approach. Although eDiaMoND was significant in size, MGF was only applied within a single work-package. It may also be valuable to see how the approach might work in situations where multiple and related work-packages were being simultaneously tackled by different teams of analysts and fieldworkers. It would also be interesting to investigate different divisions of labour. Although there were good reasons why a pair was used in chapter 6, this decision was also influenced by the particular circumstances at hand. Coordinating fieldwork and analysis within a larger group may require greater methodological support.

Finally, this thesis has proposed some theoretical advances. In particular that the task of engineering requirements can be seen as one of investigating and working with different *moral orders*. It has been proposed that this notion may be of theoretical utility to requirements engineering in general as a way of broadening the theoretical dialogues available to the field. Exploring these possibilities constitutes a further research task. However, part of this will be to develop accounts of the notion of *moral order* itself in such a way that these may be accessible to requirements engineering.

## 8.5. Conclusions

In conclusion, this thesis takes the position that there is something *missing* within both the practice and theory of requirements engineering. Case studies such as the London Ambulance Service Computer Aided Dispatch system seem to indicate that problems exist at a practical level. As discussed in chapter 1, the catastrophic failures of that project might well have been avoided if its participants could have used a methodology capable of providing them with a greater understanding of the *real* work that was involved in dispatching ambulances.

From a theoretical perspective, Requirements Engineering, when taken in the most general terms, is occasionally vague on seemingly important issues. For example, resolving tensions over claims of value in competing requirements is often said to be down to the *skill of the requirements analyst.* While this may very well be so, the fact that so little can be said about something so important must surely pose a problem for the pedagogical and theoretical development of the field.

This thesis has focused substantively on the use of fieldwork as a tool for informing the engineering of requirements. As has already been noted within literature, it is an approach that may help to fill the "practical gap" mentioned above. Through proposing the Model Guided Fieldwork approach, this thesis has helped to develop methodology to support its use. However, there is a broader contribution of a more conceptual nature.

It is proposed that the "theoretical gap" mentioned above can be filled when one takes the engineering of requirements to be a form of *moral enquiry*. So, while the resolution of moral concerns may still be down to the skill of a requirements analyst, understanding the nature of these concerns helps to define what such a skill consists in. Ethnomethodology, both as a methodology and a set of exemplary studies, thus provides requirements engineering with a resource which can be drawn on to provide understandings of what these moral concerns might be. In this regard, ethnomethodology is not just an orientation that can be useful for the practical conduct of fieldwork studies, but something that can inform the theory and pedagogy of Requirements Engineering itself.

# Appendix A: Index of empirical materials

| Code | Date | Type | Description |
|---|---|---|---|
| [ethno-1] | | Meeting (2 days) | Visit to Edinburgh / their previous breast screening ethnography |
| [document-1] | 14/5/03 | Document | Initial use case document written by AD. |
| [transcript-1] | 15/5/03 | Meeting (61 mins) | Informal set up meeting with CC |
| [transcript-2] | 16/5/03 | Meeting (39 mins) | Project meeting at Mirada |
| [transcript-3] | 19/5/03 | Meeting (71 mins) | Set up |
| [email-1] | 20/5/03 | Email | From a university researcher (UR); comments on [document-1] |
| [transcript-4] | 27/5/03 | Meeting (75 mins) | Set up |
| [transcript-5] | 28/5/03 | Meeting (113 mins) | Project meeting on requirements |
| [transcript-6] | 30/5/03 | Meeting (115 mins) | Set up |
| [transcript-7] | 2/6/03 | Meeting (98 mins) | Set up |
| [fieldnotes-1] | 4/6/03 | Fieldwork (1 day) | St Georges (11 pages of notes) |
| [document-2] | 17/6/03 | Document | Initial modelling document |
| [transcript-8] | 17/6/03 | Meeting* | First "proper" working meeting with CC |
| [transcript-9] | 20/6/03 | Meeting (105 mins) | Progress meeting with project manager |
| [document-3] | 24/6/03 | Document | Modelling |
| [transcript-10] | 25/6/03 | Meeting (64 mins) | Meeting with GT from the IBM Core Grid team |
| [fieldnotes-2] | 3/7/03 | Fieldwork interview | Churchill Clinic, (9 pages of notes) |
| [document-4] | 8/7/03 | Document | Modelling |
| [transcript-11] | 9/7/03 | Meeting (139 mins) | Second working meeting with CC |
| [document-5] | 14/7/03 | Document | Modelling |
| [transcript-12] | 15/7/03 | Meeting (131 mins) | Working meeting with CC |
| [document-6] | 16/7/03 | Document | Progress report prepared for Mirada meeting |
| [transcript-13] | 18/7/03 | Meeting (75 mins) | Progress meeting with Mirada |
| [fieldnotes-3] | 22/7/03 | Fieldwork (5 days) | Churchill (35 pages of notes) |
| [document-7] | 28/7/03 | Document | Modelling |
| [transcript-14] | 29/7/03 | Meeting (94 mins) | Working meeting with CC |
| [document-8] | 30/7/03 | Document | Modelling |
| [transcript-15] | 31/7/03 | Meeting (41 mins) | Working meeting with CC |
| [transcript-16] | 31/7/03 | Meeting (82 mins) | Meeting with CC and UR |
| [document-9] | 4/8/03 | Document | Final WFWP report, draft 1 |

| Code | Date | Type | Description |
|---|---|---|---|
| [document-10] | 14/8/03 | Document | Final WFWP report, draft 2 |
| | 15/8/03 | Meeting (119 mins) | Meeting with CC |
| [document-11] | 17/8/03 | Document | Final WFWP report, draft 3 |
| [document-12] | 26/8/03 | Document | Final WFWP report, final version |
| | 29/8/03 | Meeting (38 mins) | Project meeting |
| | 11/9/03 | Meeting (145 mins) | Review of WFWP report with project manager. |
| | 3/10/03 | Meeting* | Main project requirements review |
| | 8/10/03 | Meeting (233 mins) | First phase 2 requirements negotiation |
| | 13/10/03 | Meeting (56 mins) | Workflow meeting with AD |
| | 15/10/03 | Meeting (189 mins) | Second phase 2 meeting |
| [document-13] | 22/10/03 | Document | Phase 2 requirements version 1 |
| | 24/10/03 | Meeting (175 mins) | Third phase 2 meeting |
| | 29/10/03 | Meeting (120 mins) | Forth phase 2 meeting |
| [document-14] | 31/10/03 | Document | Phase 2 requirements version 2 |
| | 31/10/03 | Meeting (133 mins) | With CC, to discuss roles and security |
| | 5/11/03 | Meeting (123 mins) | Phase 2 document wrap-up meeting |
| | 6/11/03 | Fieldwork interview (82 mins) | At Churchill, on roles and responsibilities |
| | 20/11/03 | Document | Phase 2 document version 3 |
| | 21/11/03 | Meeting (58 mins) | Phase 2 document discussion |
| | 26/11/03 | Meeting (64 mins) | Pre- requirements review meeting |
| | 28/11/03 | Meeting (416 mins) | Phase 2 requirements review meeting |
| | 8/12/03 | Meeting (76 mins) | Post- requirements review meeting |
| [document-15] | 5/1/04 | Document | Phase 2 document version 4 |

* These meetings were transcribed from MiniDisc, the rest were recorded digitally.